\documentclass[preprint]{elsarticle}

\usepackage{amssymb}
\usepackage{amsbsy,amsthm}
\usepackage{amsmath}
\usepackage{xcolor}
\usepackage{pdflscape}
\usepackage{afterpage}
\usepackage{capt-of}
\usepackage{fullpage}
\usepackage{mathtools}
\usepackage{float}

\usepackage{tabularx}

\usepackage[normal,tight,center]{subfigure}
\usepackage{array}
\usepackage{calc}
\usepackage[thinlines]{easytable}
\usepackage{tabu}

\usepackage{ulem}

\newcommand{\colr}[1]{\textcolor{red}{#1}}
\newcommand{\colb}[1]{\textcolor{blue}{#1}}

\newcommand{\colm}[1]{\textcolor{magenta}{#1}}

\newcommand{\rtab}[1] {Table~\ref{Tab:#1}}

\newcommand{\rsec}[1] {section~\ref{Sec:#1}}

\newcommand{\rfig}[1] {Fig.~\ref{fig:#1}}
\newcommand{\eqn}[1]{Eq.~(\ref{eqn:#1})}

\newcommand{\eqnlabel}[1]{\label{eqn:#1}}
\newcommand{\figlabel}[1]{\label{fig:#1}}
\newcommand{\tablabel}[1]{\label{Tab:#1}}
\newcommand{\seclabel}[1]{\label{Sec:#1}}

\newcommand\ord{\mathcal{O}}

\def\dx {\Delta x}
\def\dt {\Delta t}

\def\kt {\tilde{k}}

\def\Po {\mathcal{P}}

\def\Le {\mathcal{L}}
\def\ka {\kappa}

\newcommand{\U}[2] {{u_{#1}^{#2}}}
\newcommand{\ran}[1] {{\tilde{#1}}}
\renewcommand{\k}[1] {\ran{k}_{#1}}

\def\bcen {\begin{center}}
\def\ecen {\end{center}}
\newcolumntype{L}[1]{>{\raggedright\arraybackslash}p{#1}}
\newcolumntype{C}[1]{>{\centering\arraybackslash}p{#1}}
\newcolumntype{R}[1]{>{\raggedleft\arraybackslash}p{#1}}

\journal{Journal of Computational Physics}

\begin{document}

\begin{frontmatter}



\title{Evaluation of finite difference based asynchronous partial differential equations solver for reacting flows}


\author[label1]{Komal Kumari}
\address[label1]{Department of Aerospace Engineering, Texas A\&M University, College Station, TX, United States}
\author[label2]{Emmet Cleary}
\address[label2]{California Institute of Technology, Pasadena, CA, United States}
\author[label3]{Swapnil Desai}
\address[label3]{Combustion Research Facility, Sandia National Laboratories, Livermore, CA, United States}
\author[label1]{Diego A. Donzis}
\author[label3]{Jacqueline H. Chen}
\author[label4]{Konduri Aditya\corref{cor1}}
\address[label4]{Department of Computational and Data Sciences, Indian Institute of Science, Bengaluru, KA, India}
\ead{konduriadi@iisc.ac.in}
\cortext[cor1]{Corresponding author.}

\begin{abstract}
Next-generation exascale machines with extreme levels of parallelism will
provide massive computing resources for large scale numerical simulations of complex physical systems at unprecedented parameter ranges.
However, novel numerical methods, scalable algorithms
and re-design of current state-of-the art numerical solvers
are required for scaling to these machines with minimal overheads.
One such approach for partial differential equations based solvers involves computation of spatial derivatives with possibly
delayed or asynchronous data using high-order asynchrony-tolerant (AT) schemes
to facilitate mitigation of communication
and synchronization bottlenecks without affecting the numerical accuracy.
In the present study, an effective methodology of
implementing temporal discretization using
a multi-stage Runge-Kutta method
with AT schemes is presented.
Together these schemes are used to perform asynchronous simulations of
canonical reacting flow problems, demonstrated in one-dimension
including auto-ignition of a premixture, premixed flame propagation
and non-premixed autoignition.
Simulation results show that the AT schemes
incur very small numerical errors
in all key quantities of interest including stiff intermediate
species despite delayed data at processing element (PE) boundaries.
For simulations of supersonic flows, the degraded numerical accuracy of
well-known shock-resolving WENO (weighted essentially
non-oscillatory) schemes when used with relaxed synchronization is also discussed. To overcome this loss of accuracy,
high-order AT-WENO schemes are derived
and tested on linear and non-linear equations.
Finally the novel AT-WENO schemes are demonstrated in the propagation of a detonation wave
with delays at PE boundaries.
\end{abstract}


\end{frontmatter}


\section{Introduction}
The fundamental understanding of a multitude of complex
scientific and engineering phenomena
relies extensively on high fidelity numerical simulations of the
governing equations. Of particular interest is the
so-called Direct Numerical Simulations (DNS)
\cite{moinDIRECTNUMERICALSIMULATION1998}
of reacting and non-reacting turbulent flows wherein all turbulence scales are fully resolved.
In the DNS approach the time-dependent governing Navier-Stokes, energy and species continuity equations are solved with high accuracy, and  all dynamically relevant
ranges of unsteady spatial and temporal scales are numerically resolved.
This stringent resolution
criteria imposes prohibitive computational cost and
consequently limits the attainable parametric range of
relevant DNS.
Even for moderate Reynolds number or number of transported species, for example, DNS
requires massive supercomputers with hundreds of
thousands of processing elements (PEs) working concurrently.

In some of the most well resolved DNS of complex flows \cite{visbalHighOrderSchemesDNS, chenTerascaleDirectNumerical2009,attiliFormationGrowthTransport2014, adityaDirectNumericalSimulation2019,savard2019,gruberDirectNumericalSimulations2018, zhangExergyLossCharacteristics2021,bergerDNSStudyImpact2020,kimEffectsDifferentialDiffusion2020,Nivarti2017,desai2021direct,BEARDSELL2021}
high-order finite difference schemes
have been used extensively to approximate the spatial derivatives.
The parallel efficiency of these schemes is high since they use
local stencils extending only to the
nearest grid-points to approximate the
derivatives. However, in a data parallel decomposition of the domain, where multiple PEs are working on
different parts of the computational domain in parallel, the
PEs also
need to communicate across the processor boundaries of the nearest computational domain neighbors
in each direction. These communications proceed in the
form of halo exchanges of processor boundary information
that is stored in buffer/ghost points. For current
state-of-the-art numerical solvers, the PEs synchronize and wait for
these halo exchanges to complete before
derivatives at PE boundaries are computed. If this synchronization is not imposed, data from the latest communication may not be available at the buffer points.
However, if old or delayed data is used at the
boundaries with the standard finite difference
schemes, the resulting solution is inconsistent and
inaccurate \cite{konduri2012async,donzisAsynchronousFinitedifferenceSchemes2014}.
Thus, standard schemes inherently necessitate synchronizations at
PE boundaries and consequently incur
severe penalties due to processor idling, especially when a large number of PEs are used. This is the communication
and synchronization bottleneck that is expected to pose a major challenge in efficiently scaling to next-generation
exascale machines \cite{dongarraInternationalExascaleSoftware2011}.

An efficient way to mitigate this bottleneck is to relax
the strict communication and synchronization
requirements and perform simulations asynchronously.
Essentially the derivatives are computed with the latest
available data which may or may not be at the
current time level and thus a delay is encountered at the PE boundaries.
However, due to
the degraded numerical accuracy of standard schemes with asynchrony,
alternate approaches are needed for asynchronous computations
such that  numerical errors incurred are minimal.
While asynchronous computations have been utilized successfully
for iterative linear
solvers \cite{bert1989,frommerAsynchronousIterations2000,leeRelaxedSynchronizationApproach2016,leeSwitchedDynamicalSystem2015},
some of the early work in asynchronous simulations of partial differential equations (PDEs) is
limited to simple canonical equations in one-dimension.
In \cite{amitai1992,amitaiAsynchronousCorrectedasynchronousFinite1994,mittalProxyequationParadigmStrategy2017} the governing equation is modified \textit{apriori} to
offset the effect of asynchrony on the numerical solution.
However, extension of this approach to higher order and dimension
is very challenging. An alternate approach to
realize asynchronous computing is to
modify the numerical scheme \cite{konduri2012async,donzisAsynchronousFinitedifferenceSchemes2014,adityaHighorderAsynchronytolerantFinite2017,mudigereDelayedDifferenceScheme2014}.
An example of this approach would be the
asynchrony-tolerant (AT) schemes of arbitrarily high orders that were used
for asynchronous
simulations of one-dimensional linear and non-linear equations in \cite{adityaHighorderAsynchronytolerantFinite2017}.
The statistical error analysis of these
AT schemes presented in \cite{adityaHighorderAsynchronytolerantFinite2017}
shows that the error depends upon simulation parameters, such as
the number of PEs and delays at the boundaries. In a more recent
study, the AT schemes were used to perform accurate asynchronous simulations
of Burgers' turbulence \cite{aditya2019} and compressible turbulence \cite{kumariDirectNumericalSimulations2020}
that also exhibited superior scaling to their synchronous counterpart.
Two distinct algorithms, synchronization avoiding (SAA) and communication avoiding (CAA), to efficiently introduce asynchrony at
boundaries in a three-dimensional
flow solver were also presented and verified
in \cite{kumariDirectNumericalSimulations2020}. The ability of the asynchronous method to absorb system noise effects was demonstrated in \cite{aditya2019}.
The stability and spectral accuracy of AT schemes was analyzed in detail in \cite{kumariGeneralizedNeumannAnalysis2021}. To further advance the applicability of AT schemes, multi-physics simulations such
as turbulent combustion, requiring massive computations,
are a natural next choice.

As a first step to evaluate the numerical performance of AT schemes
for computationally expensive and highly nonlinear
turbulent combustion simulations, several canonical reacting
one-dimensional flow problems are performed.
In particular, the effect of data asynchrony is studied
for autoignition, premixed flame propagation and non-premixed autoignition.
Both single-step and detailed chemical mechanisms
with stiff reactions are used to test the
efficacy of the AT schemes. Moreover, for simulations
of flows involving shocks and discontinuities AT-WENO schemes
are derived for the first time. These AT-WENO schemes are then
demonstrated for their numerical performance in simulations of both reacting and non-reacting flow.

The remainder of the paper is organized as follows. In section 2
a brief overview of asynchronous computing and AT schemes is presented along with
time integration using multi-stage Runge-Kutta schemes.
The AT-WENO schemes are derived in section 3 followed by
order of accuracy tests. The governing equations are
included in section 4 and  the numerical test cases and results are
presented in section 5. Finally, sections 6 and 7 comprise
discussions and conclusions, respectively.
The AT schemes used in the simulations and
expressions for the AT-WENO schemes are included in the Appendix.

\section{Asynchronous computing method}

\subsection{Concept}
To illustrate the asynchronous computing method proposed in \cite{konduri2012async,donzisAsynchronousFinitedifferenceSchemes2014}, consider the simple one-dimensional time-dependent heat equation,
\begin{equation}
    \frac{\partial u}{\partial t}=\alpha \frac{\partial^2 u}{\partial x^2},
    \eqnlabel{heat}
\end{equation}
where u$(x,t)$ is the temperature and $\alpha$ is the diffusion coefficient. This equation can be approximated using a first-order Euler and second-order central difference schemes for spatial and temporal derivatives, respectively, as
\begin{equation}
    \frac{{u_j^{n+1}}-u_j^n}{\dt}=\alpha\frac{u_{j+1}^n-2u_j^n+u_{j-1}^n}{\dx^2}+ \ord(\dt,\dx^2).
    \eqnlabel{sync-fde}
\end{equation}
Here $u_j^{n+1}=u(x_j,t_n)$ is the temperature at a grid point $j$ and a time level $n$. This finite difference equation, for a given set of initial and boundary conditions, can be used to evolve the temperature field in a discretized one-dimensional domain (see \rfig{schematic-1d}(a)). Implementation of this numerical method in a serial code is straightforward, as function ($u$) values at the grid points and at necessary time levels would be available in the processing element's memory. On the other hand, in a parallel computing model with a distributed memory setting, the one-dimensional domain is decomposed into $\Po$ sub-domains that are mapped to $\Po$ processes/processing elements (PEs), as illustrated in \rfig{schematic-1d}(b) for $\Po=2$. Now, each PE's memory would contain a subset of function values that belong to the mapped sub-domain. The grid points in each sub-domain can be further divided into two sets: first, the interior points at which the function updates are computed with data available within the PE's memory, and second, the PE boundary points at which the function updates depend on the data from neighbouring PEs. This is facilitated by, as mentioned earlier, the so-called halo exchanges which communicate the necessary function values from a neighbouring PE into buffer or ghost points. Note that the function update using \eqn{sync-fde} at PE boundary points cannot proceed until the halo exchanges are complete. This is often ensured by imposing explicit synchronization that results in significant overheads.

\begin{figure}[h!]
\begin{center}
\includegraphics[trim={0cm 0cm 0cm 0cm},clip,width=0.75\textwidth]{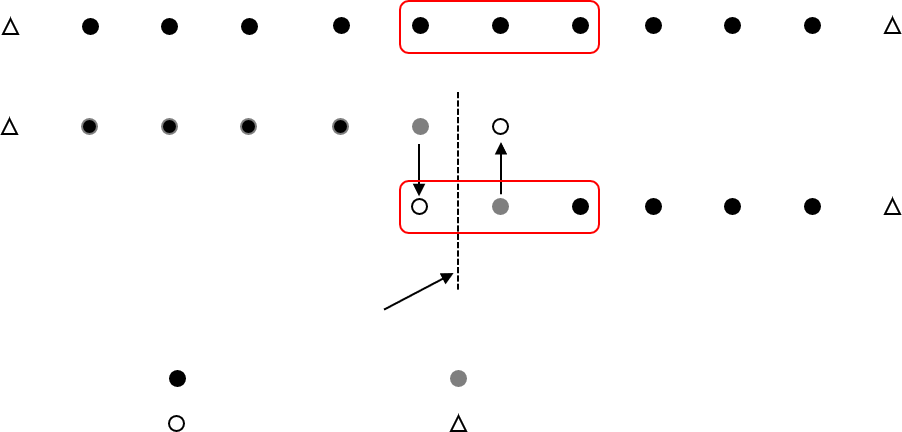}
         \put(-270,130){PE 0}
           \put(-270,85){Synchronization}
             \put(-270,100){Communication}
         \put(-90,67){PE 1}
        \put(-265,45){PE boundary}
         \put(-274,19){Internal point}
         \put(-274,0){Buffer point}
         \put(-162,19){PE boundary point }
         \put(-162,-0){Physical boundary point}
          \put(-135,172){$j+1$}
           \put(-158,172){$j$}
            \put(-200,172){$j-1$}
             \put(-135,67){$j+1$}
           \put(-158,67){$j$}
            \put(-200,67){$j-1$}
            \put(-380,155){(a)}
            \put(-380,115){(b)}
\caption{A schematic of a one-dimensional discretized domain in (a) serial and in (b) parallel (bottom) with two processing elements. The red curve denotes the three-point stencil
about point $j$ and the dashed line denotes the processor boundary.}
\figlabel{schematic-1d}
\end{center}
\end{figure}
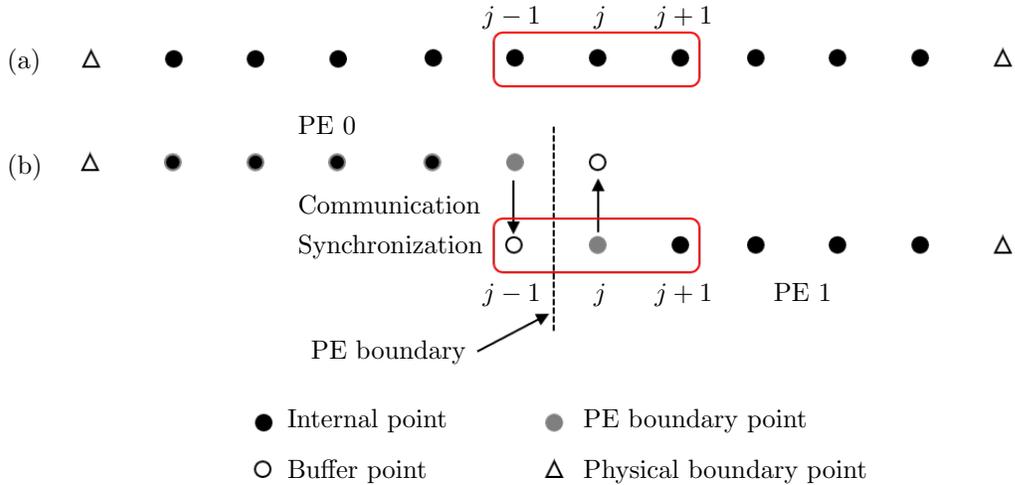

In the asynchronous computing method \cite{donzisAsynchronousFinitedifferenceSchemes2014}, the halo exchanges are initiated, but no explicit synchronization is imposed. Therefore the function value at buffer points may or may not be at the latest time level $n$, which modifies the update equation in \eqn{sync-fde} for PE boundary points (assuming that the points are at the left boundary) to
\begin{equation}
    \frac{{u_j^{n+1}}-u_j^n}{\dt}=\alpha\frac{u_{j+1}^n-2u_j^n+u_{j-1}^{n-\kt}}{\dx^2}+ \ord(\dt/\dx^2,\dt,\dx^2),
    \eqnlabel{async-fde}
\end{equation}
where $\kt$ represents the time level delay resulting from the relaxed synchronization.
If the communication at the latest time level is complete, then $\kt=0$ and the computations at the PE boundary points are synchronous (like the interior points). Otherwise, $\kt>0$ and the update equation would involve a delayed function value, resulting in \textit{asynchronous computations}. Note that depending on the computation and communication time scales, delays at different PE boundaries can be different and can be several time levels older. Let $\Le$ be the maximum allowable delay between the two PEs at a PE boundary. Then, the delay $\kt \in \left\{ 0,1,2,\dots,\Le-1\right\}$. As the arrival of messages at PE boundaries can be modeled as a random process \cite{hoefler2008noise}, the corresponding delay values, $\kt$, take on random values. Let $p_k$ represent the probability of occurrence of a $k$ time level delay, i.e. $\kt=k$, in a simulation. The sum of probabilities is $\sum_{k=0}^{\Le-1}p_k=1$.

It is evident that the numerical properties of standard schemes used asynchronously would not just depend on numerical and physical parameters such as grid resolution ($\dx$), time step ($\dt$) and diffusion coefficient ($\alpha$), but would also depend on simulation parameters such as the number of PEs ($\Po$), stencil size, delay probabilities ($p_k$) and maximum allowable delay ($\Le$).
A statistical framework was developed in \cite{donzisAsynchronousFinitedifferenceSchemes2014} to assess the numerical properties of finite differences implemented with relaxed communication synchronizations. It was shown that asynchronous computations at the PE boundaries would not only affect the stability and consistency of the overall numerical method, but also significantly degrade the accuracy. For the schemes considered in \eqn{async-fde}, the truncation error consists of terms that are functions of $\dt/\dx^2$, $\dt$ and $\dx^2$ to leading order. From stability analysis, which gives the relation $\dt\sim\dx^2$ for a constant CFL number, the leading order term is identified as $\ord{(\dt/\dx^2)}$, which is zeroth-order accurate. Following the analysis described in \cite{donzisAsynchronousFinitedifferenceSchemes2014}, one can obtain the overall accuracy in the domain to be $\ord{(\dx)}$ or first-order accurate. It can also be shown that the error would scale linearly with the number of PEs $P$ and mean delay $\Bar{\kt}$. As standard schemes would result in poor accuracy under relaxed data synchronization, new asynchrony-tolerant schemes were derived  subsequently \cite{adityaHighorderAsynchronytolerantFinite2017}, which are summarised next.

\subsection{Asynchrony-tolerant (AT) spatial schemes}
The basic idea of asynchrony-tolerant (AT) schemes is to use an extended
stencil in space and/or time to recover
loss of accuracy due to the use of delayed data. For example,
the spatial derivative of $u_j^{n}:=u(j\dx,n\dt)$ at
location $j$ and time level $n$ can be computed
using
\begin{equation}
\left.\frac{\partial u}{\partial x} \right|_j^{n}
=\frac{1}{\dx}\sum_{\ell=0}^{\Le-1} \sum_{m=-M_1(\ell)}^{M_2(\ell)}a_{m}^{\ell}
u_{j+m}^{n-\ell},
\eqnlabel{geDer}
\end{equation}
where $\dx$ is the grid size, $M_1(\ell)$ and $M_2(\ell)$ are the
extents of stencil to the left and right of $u_{j}^{n}$ and
$\Le$ is the maximum allowable delay level. For a given set of parameters $\{\Le,M_1(\ell),M_2(\ell)\}$
the unknown weights $a_m^{\ell}$ are obtained by solving a linear
system of equations that is formed by imposing necessary accuracy constraints
on terms of the Taylor series expansion of $u_{j+m}^{n-\ell}$ about $u_{j}^{n}$.
If $\Le=1$ and $M_1(0)=M_2(0)=M$,
this system of equations yields
coefficients for the standard central difference scheme.
For $\Le>1$, AT schemes
can be derived if the stencil parameters $\{M_1(\ell),M_2(\ell)\}$
are appropriately chosen. An example of a second-order AT scheme with
$\dt\sim \dx^2$ at the
left processor boundary, \textit{i.e.} delayed data is used
only at the left stencil points, is
\begin{equation}
   \left.\frac{\partial u}{\partial x} \right|_j^{n}
=\frac{\kt u_{j-1}^{n-\kt-1}-(\kt+1)u_{j-1}^{n-\kt} +u_{j+1}}{\dx}
\eqnlabel{at1}
\end{equation}
where $\kt$ is the delay \cite{adityaHighorderAsynchronytolerantFinite2017}. We note that the coefficients in \eqn{at1} depend
upon delays, unlike the fixed coefficients in standard schemes. Furthermore,
in the absence of delays \textit{i.e} when $\kt=0$, \eqn{at1} reduces
to the standard central difference scheme for the first derivative.
The general methodology to derive different families of AT
schemes of arbitrary order of accuracy is described in   \cite{adityaHighorderAsynchronytolerantFinite2017}. Here the AT schemes from \cite{adityaHighorderAsynchronytolerantFinite2017} are directly used
and are listed in the Appendix without details of the derivation.

\subsection{Temporal scheme: Runge-kutta}
\seclabel{rkder}
For a fully-discrete system, the global order
of accuracy depends on both the spatial and temporal discretizations.
In order to compute this global order, a stability relation of the form
$\dt \sim \dx^r$ is used to express the leading order truncation error term of
the time discretization scheme also in terms of grid size ($\dx$).
For example, for a fourth-order spatial scheme  ($\mathcal{O}(\dx^4)$),
if a second-order temporal discretization scheme ($\mathcal{O}(\dt^{2})$) is used
then the global spatial order is two if $\dt \sim\dx$
and four if $\dt \sim \dx^2$.
One of the most widely used schemes for high-order temporal
discretization is the multi-stage Runge-Kutta (RK) method. An $S$-stage explicit
RK scheme for an equation of the form $du/dt = f(u,t)$ over a time step $\Delta t$
is given by
\begin{equation}
   u^{n+1} = u^{n} + \Delta t \sum_{s=0}^{S} b_s K_s,
\end{equation}
where the stages $K_s$ are computed using
\begin{equation}
    \begin{aligned}
    K_1 &= f(u^n,t^n) \\
    K_s& = f(u^n +\Delta t
 \sum_{i=1}^{s-1}(a_{si}k_i),t^n + c_s\Delta t).
    \end{aligned}
    \eqnlabel{rk1}
\end{equation}
Here $a_{si},b_s,c_s$ are the coefficients of the RK scheme. For
an RK scheme, in advancing from time level $n$ to $n+1$, intermediate stages
$K_s$ are computed. Each of these stages depends upon the previous stages as can
be seen from \eqn{rk1}. Thus, for simulations performed in parallel,
RK schemes require PEs to communicate and synchronize after every
stage in order to update the corresponding values at buffer points.
However, if simulations are performed with asynchronous
communications using AT schemes,
then a fractional delay will be encountered at the intermediate stages and
different AT schemes will be required for each stage.
An alternate method is proposed to effectively use multi-stage RK schemes with
AT that circumvents the need  to communicate at every stage and avoids
fractional delays.

In the presence of delays, each stage $K_s$
is computed using AT schemes at PE boundary points.
As mentioned before, the current stage $K_s$ explicitly depends upon
the previous stages, $K_{s-1},K_{s-2},\dots, K_1$.
Thus, to ensure accuracy all the stages of RK scheme are
computed locally at both PE boundary and buffer points of each PE.
Furthermore, at the old time levels required by the AT scheme, these
stages are also computed at the internal points close to the PE boundary.
In order to perform the additional computations at buffer points,
a larger message of size (= No. of stages in RK
 $\times$ spatial stencil) has to be communicated across processors
 at every time step. However, the PEs no longer communicate at every stage
 of the RK. Thus, there exists a trade-off between communication
 which is expensive and computation
which is relatively cheap.

To illustrate that the RK schemes preserve the order of accuracy when combined with high-order AT schemes, the one-dimensional diffusion equation
in \eqn{heat} is reconsidered. 
Denoting the right-hand-side of
\eqn{heat} by $f(u,t)=\partial ^2u/\partial x^2$ (excluding the diffusion coefficient, $\alpha$),
the temporal discretization
at the $j$-th spatial point using a two-stage
second-order RK scheme (RK2) is given by
\begin{equation}u_j^{n+1}=u_j^{n} + \alpha \Delta t \left(\frac{K_{j,1}}{2} +
\frac{K_{j,2}}{2}\right)
\eqnlabel{rk2}
\end{equation}
where
    $K_{j,1}=f(u_j^n,t^n)$, $K_{j,2} = f\left((u_j^n)^*,t^n+\Delta t \right)$ and $(u_j^n)^* = u_j^n +\Delta t (f(u_j^n,t^n))=u_j^n +\Delta t \alpha K_{j,1}$.
When a standard fourth-order central difference scheme is
used for numerical approximation
of $f(u_j^n,t^n)$, the following $K_{j,1}$ and $K_{j,2}$
are obtained,
\begin{equation}
\begin{aligned}
K_{j,1}=&\frac{-u_{j-2}^n+16 u_{j-1}^n-30 u_j^n+16 u_{j+1}^n-u_{j+2}^n}{12\dx^2}\\
K_{j,2}=&-\frac{\alpha \dt (K_{j-2,1})+u_{j-2}^n}
{12\dx^2}+\frac{4 \left(\alpha \dt (K_{j-1,1})+u_{j-1}^n\right)}
{3\dx^2}
-\frac{5 \left(\alpha \dt
 \left(K_{j,1}\right)+u_j^n\right)}
{2\dx^2}\\
&
+\frac{4 \left(\alpha \dt
(K_{j+1,1}) +u_{j+1}^n\right)}
{3\dx^2}
-\frac{\alpha \dt
 \left( K_{j+2,1}\right)+u_{j+2}^n}
{12\dx^2}.
\end{aligned}
\eqnlabel{ks}
\end{equation}
The Taylor series expansion of \eqn{rk2} on substituting \eqn{ks}
then yields a truncation error of the form
\begin{equation}
\text{TE}_j^n|_{\text{sync}} = \frac{1}{6} \left(-f^{(0,3)}(x,t)\right) \Delta t^2
-\frac{1}{90} \alpha  f^{(6,0)}(x,t) \Delta x^4+\mathcal{O}(\Delta x^6,\Delta t^3,\Delta x^4 \Delta t)
\end{equation}
or
\begin{equation}
\text{TE}_j^n|_{\text{sync}} = (-\frac{r_{\alpha }^2 u^{(0,3)}(x,t)}{6 \alpha ^2}
 -\frac{1}{90} \alpha  u^{(6,0)}(x,t))\Delta x^4 + \mathcal{O}(\Delta x^6),
 \eqnlabel{rks}
 \end{equation}
where $r_{\alpha} = \alpha \Delta t/\Delta x^2$ is the diffusive CFL number and the
subscript ``sync" denotes synchronous. Note that from \eqn{rks}
the global spatial order of accuracy is four.

When a fourth-order AT scheme
is used at the boundary points for computation
of spatial derivatives, we have
\begin{equation}
\begin{aligned}
K_{j,1}=&\left(\kt^2+\kt\right)\frac{\left(- u_{j-2}^{n-\kt-2}
+16 u_{j-1}^{n-\kt-2} -30 u_j^n +16 u_{j+1}^{n}- u_{j+2}^n
\right)}{24\Delta x^2}\\
&-\left(\kt^2+2 \kt\right) \frac{ \left(- u_{j-2}^{n-\kt-1}
+16 u_{j-1}^{n-\kt-1} -30 u_j^n +16 u_{j+1}^{n}- u_{j+2}^n
\right)}{12 \Delta x^2}\\
&+\left(\kt^2+3 \kt+2\right)
\frac{\left( - u_{j-2}^{n-\kt}
+16 u_{j-1}^{n-\kt} -30 u_j^n +16 u_{j+1}^{n}- u_{j+2}^n \right)}
{24 \Delta x^2}
\end{aligned}
\end{equation}
which is used to compute $K_{j,2}=f(u_j^n+\Delta t \alpha K_{j,1},t^n+\Delta t)$
with $f(.,.)$ being evaluated using the fourth-order AT-scheme as well.
Once again, the Taylor series expansion is used
to obtain the truncation error,
\begin{equation}
\text{TE}_j^n|_{\text{AT},\tilde k}=-\frac{5}{24}  \alpha  \kt \left(\kt^2+3 \kt+2\right) u^{(0,3)}(x,t) \frac{\Delta t^3}{\Delta x^2}
-\frac{1}{6} u^{(0,3)}(x,t) \Delta t^2
-\frac{1}{90} \alpha  u^{(6,0)}(x,t) \Delta x^4
 +\mathcal{O}(\kt^3 \frac{\dt^3}{\dx}).
 \eqnlabel{rkat1}
 \end{equation}
The above expression simplifies to
give a global fourth-order truncation error term when $r_\alpha$
is substituted,
\begin{equation}
\text{TE}_j^n|_{\text{AT},\tilde k}=
-\left(\frac{75 \kt \left(\kt^2+3 \kt+2\right) r_{\alpha }^3
 u^{(0,3)}(x,t)+60 r_{\alpha }^2 u^{(0,3)}(x,t)+4 \alpha ^3
 u^{(6,0)}(x,t)}{360 \alpha ^2}\right)\Delta x^4 +\mathcal{O}(\Delta x^6).
 \eqnlabel{rkat2}
 \end{equation}
When the delay is zero \textit{i.e.} $\kt=0$, the truncation error
in \eqn{rkat2} reduces to
$\text{TE}_j^n|_{\text{sync}}$ (\eqn{rks}) which is the synchronous
truncation error at the internal points.
Therefore, RK schemes can be used with the AT-schemes without
affecting the global order of accuracy.
During a simulation
the stencil data at older time levels for
the buffer as well as for internal points are available, and hence
standard synchronous schemes can be used to
compute the sub-stages $K_{j,s}$ at these points for efficient
implementation of RK with AT.
Numerical simulations of linear equations were performed
using this approach and the global order of accuracy was preserved.
Here the order of accuracy tests are shown for RK schemes with
AT-WENO schemes that are derived next (see \rtab{wenoLin}).

\section{Asynchrony-tolerant Weighted Essentially
Non-oscillatory schemes (AT-WENO)}
Despite their advantages, central finite difference schemes are not
always suitable for performing DNS of,
for example, supersonic flows. In the presence of sharp gradients or discontinuities in density, temperature and composition due to detonations, shocks or high pressure flames, the central difference schemes are prone to numerical oscillations and instabilities. A well established numerical technique to accurately simulate flows with piece-wise smooth solutions between discontinuities is the so called weighted essentially non-oscillatory schemes~(WENO)~\cite{shuEssentiallyNonoscillatoryWeighted1998, shuHighOrderWeighted2009}. These schemes aim to achieve high-order accuracy at the smooth regions of the flow and resolve the discontinuities with minimal oscillations by automatically selecting the locally smoothest stencil. WENO schemes are considerably more expensive than regular central difference schemes due to the need to evaluate the same flux functions from multiple candidate stencils~\cite{pirozzoli2011numerical}. However, the numerical method is suitable for implementation in finite difference codes using stencil operations very similar to the central difference schemes. Also, the usual advantages of DNS using finite difference schemes including parallelism and scalability apply to the WENO schemes as well. Apart from DNS~\cite{bermejo2013scaling,desai2021direct}, the advantages of WENO schemes have also been demonstrated in implicit large eddy simulations (iLES)~\cite{mosedale2007assessment,ritos2018physical,ritos2018performance}.

In this section, a brief overview of the standard WENO schemes is
provided followed by a discussion on the effect of asynchrony on the standard WENO schemes (AS-WENO) and derivation of the asynchrony-tolerant WENO (AT-WENO) schemes for accurate numerical simulations when delays are observed at PE boundaries. Consider the one-dimensional ($x$-direction) version of the governing equations:
\begin{equation}
    \frac{\partial Q}{\partial t} +
      \frac{\partial C}{\partial x} +
        \frac{\partial D}{\partial x} = S,
        \eqnlabel{ge1}
\end{equation}
where $Q$ is the solution vector,
$C$ is the vector comprising the convective flux terms,
$D$ is the viscous and molecular diffusion
flux vector, and $S$ is the vector of source terms.
The exact form of terms in $Q,C$ and $D$ is defined in \rsec{gov_eqn}.
In the WENO framework, the terms contained in the $D$ vector do not require any special treatment and are evaluated using central difference schemes (standard
or AT). However, the convective flux terms in $C$ have
to be computed appropriately to ensure both
stability and accuracy in regions near and at discontinuities.
Specifically, the derivative of the convective
flux $\partial C/\partial x$ at
a point $j$ is approximated using the flux at the
edges, \textit{i.e.}, $\widehat{C}_{j\pm\frac{1}{2}}$ of a cell $\mathcal{I}_j=[x_{j-\frac{1}{2}},x_{j+\frac{1}{2}}]$. The flux at
the edges is computed using the WENO approximation procedure
that can be carried out using
interpolation or reconstruction, and
accordingly, requires point values or cell averages
of the fluxes. Note that the hat notation
$(~\widehat{}~)$ is used to denote the variables at cell edges. The derivative is then computed using the relation
\begin{equation}
\left.\frac{    \partial C}{\partial{x}}\right|_j
=\frac{\widehat{C}_{j+\frac{1}{2}} -
\widehat{C}_{j-\frac{1}{2}}}{\dx}
\eqnlabel{cder}
\end{equation}
for a uniform grid and the order depends upon the order
of the numerical flux approximation $\widehat{C}_{j\pm\frac{1}{2}}$. When
using finite-differences, the
WENO approximation directly yields the fluxes at the edges in terms
of the fluxes at the grid points. For stability, appropriate up-winding of fluxes is also required and is achieved through splitting the
flux, for example using the local Lax-Friedrichs flux splitting methodology,
\begin{equation}
\begin{aligned}
    &C=C^{+}+C^{-}, \quad \text{where} \\
    &C^{\pm} = \frac{1}{2}(C\pm\lambda_{max}Q),
\end{aligned}
\eqnlabel{Csplit}
\end{equation}
at every grid-point. The quantity $\lambda_{max}$ in \eqn{Csplit} is the maximum local
wave propagation speed which can be computed as
\begin{equation}
\left.    \lambda_{max}\right|_j=\max\{(|u|)_{j-1},(|u\pm c|)_{j-1},(|u|)_j,
(|u\pm c|)_{j}\},
\end{equation}
where $u$ and $c$ are the local flow velocity and
the speed of sound, respectively. The derivative is then simply evaluated
using
\begin{equation}
    \left.\frac{\partial C}{\partial x}\right|_j
    = \frac{ \widehat{C}^{+}_{j+\frac{1}{2}} -
    \widehat{C}^{+}_{j-\frac{1}{2}} }{\dx}+
     \frac{ \widehat{C}^{-}_{j+\frac{1}{2}} -
    \widehat{C}^{-}_{j-\frac{1}{2}} }{\dx}
       \eqnlabel{wender}
\end{equation}
with an appropriate upwind stencil for approximating both positive and negative fluxes \cite{desai2021direct}.
For computation of derivative in \eqn{wender} using the WENO procedure,
the first step is to approximate the fluxes at
the edges ($j\pm1/2$) through an interpolation or a
reconstruction procedure. For simplicity, the approximation of these fluxes using
an interpolation technique is discussed here in detail.
However, the approximation through
reconstruction that treats grid-point values as cell averages
can also be performed following the steps detailed here.
In order to do so, the primitive function defined in
\cite{shuEssentiallyNonoscillatoryWeighted1998}
is required.

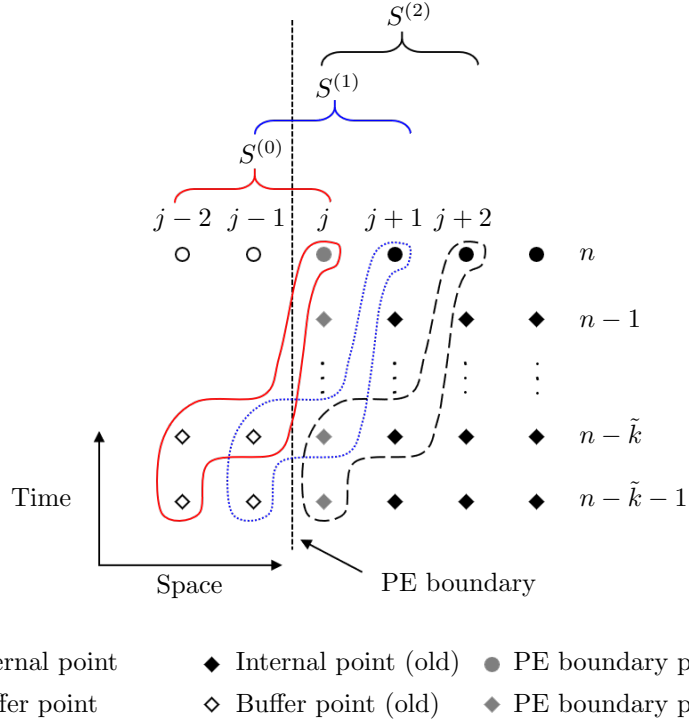
\begin{figure}[h!]
\begin{center}
\hspace{-2cm}
\includegraphics[trim={0cm 0cm 0cm 0cm},clip,width=0.5\textwidth]{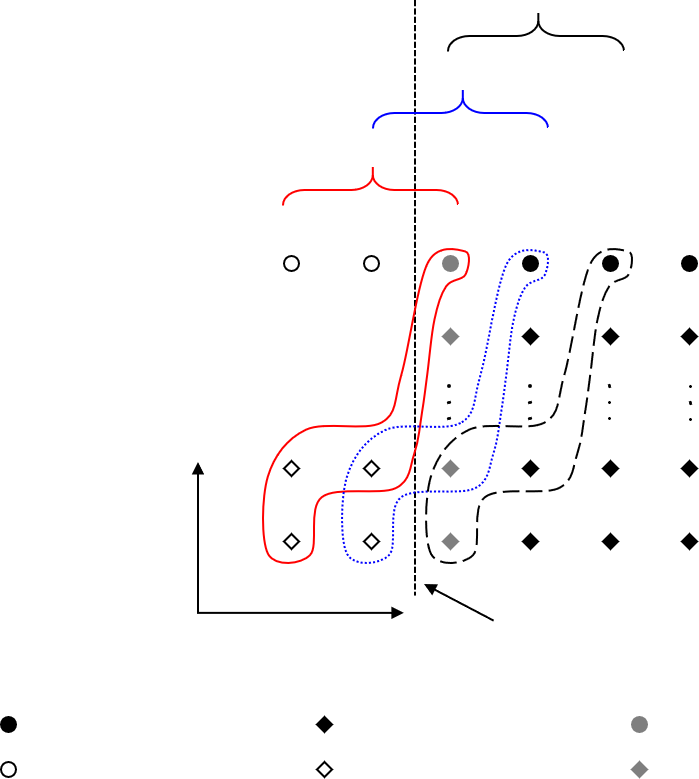}
\begin{picture}(0,0)
        \put(-205,78){Time}
        \put(-150,45){Space}
        \put(-65,46){PE boundary}
         \put(-225,0){Buffer point}
         \put(-225,16){Internal point}
         \put(-120,0){Buffer point (old)}
         \put(-120,16){Internal point (old)}
          \put(-15,0){PE boundary point }
         \put(-15,16){PE boundary point (old)}
          \put(-119,208){$S^{(0)}$}
            \put(-90,233){$S^{(1)}$}
              \put(-63,260){$S^{(2)}$}
               \put(10,171){$n$}
                \put(10,146){$n-1$}
                 \put(10,102){$n-\kt$}
                  \put(10,78){$n-\kt-1$}
             \put(-45,184){$j+2$}
           \put(-71,184){$j+1$}
           \put(-89,184){$j$}
            \put(-123,184){$j-1$}
           \put(-151,184){$j-2$}

        \end{picture}
\caption{Four-point candidate stencils for asynchronous WENO approximation at
the left processor boundary. $S^{(0)}, S^{(1)}$ and $S^{(2)}$ are the
three-point candidate synchronous WENO stencils for approximation at
point $j+1/2$ and the corresponding curves are the AT
candidate stencil with delays on the left.}
\figlabel{sten1}
\end{center}
\end{figure}

For illustration, consider a specific WENO scheme
that uses three
sub-stencils each of which comprises three grid-points denoted
by $S^{(i)}$, $i=\{0,1,2\}$, in \rfig{sten1} to approximate the
value of a quantity $\hat{u}$ at cell edges, say $j+1/2$, and time-level $n$.
Taken together the
larger stencil obtained by combining these $S^{(i)}$
contains five points. Since the approximation order depends upon the number
of points, a fifth-order accurate approximation can be obtained with the full stencil for a smooth function. This is the classical three-fifth order WENO scheme, where the three smaller candidate stencils $S^{(0)}=\{j-2,j-1,j\}$, $S^{(1)}=\{j-1,j,j+1\}$, and
$S^{(2)}=\{j,j+1,j+2\}$ give a degree two polynomial
interpolation of $\hat{u}_{j+\frac{1}{2}}^{n}$ at time level $n$.
The synchronous third-order interpolant in each of the three stencils can be computed using Taylor series expansion or Lagrange interpolation \cite{shuEssentiallyNonoscillatoryWeighted1998,shuHighOrderWeighted2009},
\begin{equation}
    \begin{aligned}
    &\hat{u}_{j+\frac{1}{2}}^{n,(0)}=\frac{3 }{8}u_{j-2}^n-\frac{5 }{4}u_{j-1}^n+\frac{15 }{8}u_j^n+\mathcal{O}(\dx^3)\\
    &\hat{u}_{j+\frac{1}{2}}^{n,(1)}=-\frac{1}{8} u_{j-1}^n+\frac{3 }{4}u_j^n+\frac{3 }{8}u_{j+1}^n+\mathcal{O}(\dx^3)\\
    &\hat{u}_{j+\frac{1}{2}}^{n,(2)}  = \frac{3 }{8}u_j^n+\frac{3}{4} u_{j+1}^n-\frac{1}{8}u_{j+2}^n+\mathcal{O}(\dx^3).
    \end{aligned}
    \eqnlabel{interpS}
\end{equation}

The final approximation is taken as a convex combination of the above three third-order approximations, yielding a higher order interpolant in the
larger stencil. Ideally, if $u(x,t)$ is smooth in the large stencil
$S=\bigcup\limits_{i=0}^{2}S^{(i)}=\{j-2,j-1,j,j+1,j+2\}$, a
fifth-order approximation can be achieved using the following expression:
\begin{equation}
    \hat{u}_{j+\frac{1}{2}}^{n}=\sum_{i=0}^{2}\omega_i\hat{u}_{j+\frac{1}{2}}^{n,(i)},
    \eqnlabel{fullSten}
\end{equation}
where $\omega_i$'s are the non-linear weights
\begin{equation}
    \omega_i=\frac{\alpha_i}{\sum\limits_{i=0}^{2} \alpha_i},
    \quad
    \alpha_i=\frac{\gamma_i}{(\epsilon+\beta^{(i)})^2}.
\eqnlabel{omg}
\end{equation}
The non-linear weights in \eqn{omg} satisfy $w_i\ge0$ and  $\sum\limits_{i=0}^{2}\omega_i=1$ and are related to the ideal or linear weights $\gamma_i$ through the smoothness indicator $\beta^{(i)}$ \cite{shuEssentiallyNonoscillatoryWeighted,shuHighOrderWeighted2009,jiangEfficientImplementationWeighted}. A simple Taylor series expansion yields the ideal weights $\gamma=\{\frac{1}{16},\frac{5}{8},\frac{5}{16}\}$, such that
\begin{equation}
     \hat{u}_{j+\frac{1}{2}}^{n}=\frac{3}{128}u_{j-2}^{n}-\frac{5}{32}u_{j-1}^{n}
     +\frac{45}{64}u_{j}^{n}+\frac{15}{32}u_{j+1}^{n}-\frac{5}{128}u_{j+2}^{n}
     +\mathcal{O}(\dx^5).
\end{equation}
We now focus on the effect of data asynchrony on each of the individual approximations.
Consider a point $j$ at the left processor boundary
with a delay $\kt$ encountered at
each of the buffer points. The order of accuracy of the interpolants in the first two stencils is affected by this delay,
\begin{equation}
    \begin{aligned}
    &\hat{u}_{j+\frac{1}{2}}^{n,(0)}=\frac{3 }{8}u_{j-2}^{n-\kt}-\frac{5 }{4}u_{j-1}^{n-\kt}+\frac{15 }{8}u_j^n+\mathcal{O}(\dt,\dx\dt,\dx^3)\\
    &\hat{u}_{j+\frac{1}{2}}^{n,(1)}=-\frac{1}{8} u_{j-1}^{n-\kt}+\frac{3 }{4}u_j^n+\frac{3 }{8}u_{j+1}^n+\mathcal{O}(\dt,\dx\dt,\dx^3).
    \end{aligned}
    \eqnlabel{asweno}
\end{equation}
More precisely, there are terms in the truncation error that depend upon $\dt$.
On relating $\dt\sim\dx^r$, the interpolants with delays reduce to first order
when $r=1$ (convective CFL) and to second order when $r=2$ (diffusive CFL). The convex combination of
the three interpolants with ideal weights $\gamma_i$ also has a leading order truncation term that
scales as $\mathcal{O}(\dt)$, thereby degrading the order even in the
full stencil, \textit{i.e}
\begin{equation}
     \hat{u}_{j+\frac{1}{2}}^{n}=\frac{3}{128}u_{j-2}^{n-\kt}-\frac{5}{32}u_{j-1}^{n-\kt}
     +\frac{45}{64}u_{j}^{n}+\frac{15}{32}u_{j-1}^{n}-\frac{5}{128}u_{j+2}^{n}
     +\mathcal{O}(\dt,\dx\dt,\dt^2,\dx^2\dt).
\end{equation}
Clearly, the standard interpolants cannot be used in the presence of delays.
A similar degradation in order was observed when standard finite difference schemes are used with asynchrony~\cite{donzisAsynchronousFinitedifferenceSchemes2014}
for computation of spatial derivatives. Following the general methodology to overcome this loss of
accuracy in \cite{adityaHighorderAsynchronytolerantFinite2017},
the asynchrony-tolerant WENO schemes (AT-WENO) can be derived. For this,
the AT-interpolant for each of the substencils in the presence of delays is computed.
When multiple time-levels are used, the interpolation is two-dimensional \text{i.e.} in space and time, and thus, requires a greater numbers of points to achieve a given order of accuracy. Since the Taylor series expansion is used to obtain the coefficients,
a relation of the form $\dt\sim\dx^r$ is also needed to identify the
overall low-order terms (in space) that should be eliminated for achieving the desired order of accuracy.
A schematic representation to
identify low-order terms which need to be subjected to constraints is illustrated in Figures 2 and 3 in \cite{adityaHighorderAsynchronytolerantFinite2017}.
For example, for a third-order accurate approximation,
when $\dt \sim \dx$, six equations corresponding
to terms of the order $\{1,\dt ,\dx , \dt^2, \dt\dx , \dx^2\}$
are needed, and therefore, each candidate stencil comprises six points.
Similarly, if $r=2$ only four equations $\{1,\dt ,\dx , \dx^2\}$ are required, and therefore, a
four-point stencil is sufficient to achieve a third-order approximation.
Since the stencil now depends upon the delays observed at the processor boundaries, the stencil itself is random because of the random nature of the delays.
One can represent this stochastic stencil as $\tilde{S}$. For the first candidate stencil (as shown in \rfig{sten1}),
$\tilde{S}^{(0)}=\{u_{j-2}^{-\kt+n-1},u_{j-2}^{n-\kt},u_{j-1}^{n-\kt},u_j^{n}\}$ with delay $\kt$ and $r=2$, the AT interpolant is
\begin{equation}
    \tilde{u}_{j+\frac{1}{2}}^{n,(0)}=\frac{7}{8} \tilde{k} u_{j-2}^{-\tilde{k}+n-1}+\frac{1}{8} \left(3-7 \tilde{k}\right) u_{j-2}^{n-\tilde{k}}-\frac{5}{4} u_{j-1}^{n-\tilde{k}}+\frac{15}{8} u_j^n +
    \text{TE}^{(0)}.
    \eqnlabel{atS0}
\end{equation}
The leading order truncation error in \eqn{atS0} term depends on the
delay $\kt$ and is equal to
\begin{equation}
    \text{TE}^{(0)}=\frac{-5}{4} \tilde{k} \dot{u}^{\prime}(x,t)\dt\dx + \frac{5}{16} u^{\prime \prime
    \prime}(x,t)\dx^3 +\dots \sim \mathcal{O}(\dx^3),
    \eqnlabel{teATS0}
\end{equation}
where $\dot{u}=\partial u/\partial t$ and $u^{\prime}=\partial u/\partial x$ and
$\dt\sim\dx^2$ gives the overall spatial
order of accuracy to be $\mathcal{O}(\dx^3)$. When $\kt=0$, \eqn{atS0}
reduces to the standard third-order synchronous interpolant in \eqn{interpS}.
For the second stencil $\tilde{S}^{(1)}=\{u_{j-1}^{-\kt+n-1},u_{j-1}^{-\kt},u_{j}^{n},u_{j+1}^{n}\}$,  with delays considered only at the buffer point ($u_{j-1}$),
the third-order interpolant is
\begin{equation}
\tilde{u}_{j+\frac{1}{2}}^{n,(1)}=
    \frac{1}{8} \tilde{k} u_{j-1}^{-\tilde{k}+n-1}+\frac{1}{8} \left(-\tilde{k}-1\right) u_{j-1}^{n-\tilde{k}}+\frac{3}{4} u_j^n+\frac{3 }{8}u_{j+1}^n+\text{TE}^{(1)},
\eqnlabel{atS1}
\end{equation}
where the leading order truncation error term
\begin{equation}
  \text{TE}^{(1)}= -\frac{1}{16}u^{\prime \prime
    \prime}(x,t)\dx^3 +\dots \sim \mathcal{O}(\dx^3)
\end{equation}
is independent of delay $\kt$, unlike \eqn{teATS0}.
Lastly, for stencil $\tilde{S}^{(2)}=\{u_{j}^{n},u_{j+1}^{n},u_{j+2}^n\}=S^{(2)}$, since all the points are within the PE,  the standard interpolant \textit{i.e.} $\tilde{u}_{j+\frac{1}{2}}^{n,(2)}=
\hat{u}_{j+\frac{1}{2}}^{n,(2)}$ can readily be used. The full stencil is comprised of
seven points $\tilde{S}=\bigcup\limits_{i=0}^{2}\tilde{S}^{(i)}=
\{u_{j-2}^{-\kt+n-1},u_{j-2}^{n-\kt},
u_{j-1}^{-\kt+n-1},u_{j-1}^{n-\kt},u_{j}^{n},u_{j+1}^{n},
u_{j+2}^n
\}$. With this choice of stencil, at most, a
fourth-order accurate approximation can be obtained for the full
stencil since the degrees of freedom are sufficient only to
eliminate seven low-order terms. Extension to higher orders would
require extending the stencil in both space and time. This
is explained in great detail in
\cite{adityaHighorderAsynchronytolerantFinite2017} for AT finite difference schemes.

On evaluation of the ideal weights
for this larger stochastic stencil $\tilde{S}$ \textit{i.e.}
\begin{equation}
\tilde{u}_{j+\frac{1}{2}}^{n}=\sum_{i=0}^{2}\gamma_i\tilde{u}_{j+\frac{1}{2}}^{n,(i)},
\eqnlabel{idw}
\end{equation}
yields
\begin{equation}
    \gamma=\left\{0,\frac{1}{2},\frac{1}{2}\right\}
\end{equation}
which, in turn, gives a fourth-order truncation error term in \eqn{idw}. These
ideal weights eliminate the first stencil $\tilde{S}^{(0)}$ and
reduce the full stencil to a four-point synchronous stencil in the absence of delays. If the
second stencil is modified $\tilde{S}^{(1)}=\{u_{j-1}^{-\kt+n-1},u_{j-1}^{n-\kt},u_j^{n-\kt},u_{j+1
}^n\}$, the ideal weights are no longer positive,
 $   \gamma=\left\{-\frac{3}{2},-\frac{5}{2},5 \right\}. $
Further changing the third stencil to
$\tilde{S}^{(2)}=\{u_{j}^{-\kt+n-1},u_{j}^{n-\kt},u_{j+1}^{n-\kt},u_{j+2
}^n\}$ such that all
three stencils now have similar stencil structure
and thus use four points as shown in \rfig{sten1}, the ideal weights
obtained are
\begin{equation}
    \gamma=\left\{\frac{3}{8},\frac{5}{4},-\frac{5}{8} \right\}.
    \eqnlabel{gammAT2}
\end{equation}
This exercise shows that
irrespective of the choice of asynchronous stencil with four points,
at least one of the ideal
or linear weights is non-positive for integer values of $\kt$.
This non-convex nature of ideal weights leads to instabilities
and oscillations \cite{shiTechniqueTreatingNegative2002}. The procedure to
deal with the non-positive weights has been described in \cite{shiTechniqueTreatingNegative2002} and it involves splitting
all the weights into positive and negative parts,
\begin{equation}
\begin{aligned}
    &\tilde{\gamma}_i^{+} =\frac{1}{2}\left( \gamma_i +\theta |\gamma_i|\right) \\
    &\tilde{\gamma}_i^{-} =\tilde{\gamma}_i^{+} -\gamma_i
\end{aligned}
\eqnlabel{splitGam}
\end{equation}
for $i=\{ 0,1,2\}$. The split ideal weights are then scaled by the
parameters
\begin{equation}
    \sigma^{\pm}=\sum_{i=1}^{3}\tilde{\gamma}_i^{\pm};
    \quad \gamma_i^{\pm}=\tilde{\gamma}_i^{+}/\sigma^{\pm},
\end{equation}
which are used to write the split polynomial interpolations
\begin{equation}
    \left(\hat{u}_{j+\frac{1}{2}}^{n}\right)^{\pm}=\sum_{i=0}^{2}\gamma_i^{\pm}\hat{u}_{j+\frac{1}{2}}^{n,(i)},
    \eqnlabel{fullSplit}
\end{equation}
which is equivalent to replacing
the ideal positive weight $\gamma_i$ in \eqn{omg} by the
corresponding scaled split weight $\gamma^{\pm}_i$
\cite{shiTechniqueTreatingNegative2002}. However, this
procedure does involve a series of additional computations to
compute the derivative of the flux accurately.


For this particular choice of stencil with delayed data at the
left boundaries, the reconstruction polynomial
can also be derived using the primitive function \cite{shuEssentiallyNonoscillatoryWeighted1998}. For example, for a stencil
of the form $\tilde{S}^{(i)}=\{u_{j-1+i}^{-\kt+n-1},u_{j+i}^{n-\kt},u_{j+1+i}^{n-\kt},u_{j+1+i}^{n}\}$ where $i=\{0,1,2\}$, one obtains the following AT approximations,
\begin{equation}
\begin{aligned}
\tilde{u}^{n,(i)}_{j+\frac{1}{2}}=
        &\frac{1}{6} \left(3 i^2-12 i+5\right) \tilde{k} u_{i+j-2}^{-\tilde{k}+n-1}+\frac{1}{6} \left(-3 i^2 \tilde{k}+12 i \tilde{k}-5 \tilde{k}+3 i^2-6 i+2\right)
        u_{i+j-2}^{n-\tilde{k}}\\&+\frac{1}{6} \left(-6 i^2+18 i-7\right) u_{i+j-1}^{n-\tilde{k}}+\frac{1}{6} \left(3 i^2-12 i+11\right) u_{i+j}^n.
\end{aligned}
\eqnlabel{recon}
\end{equation}
The reconstruction polynomials in \eqn{recon} reduce to the
traditional candidate reconstruction polynomials in \cite{jiangEfficientImplementationWeighted,shuEssentiallyNonoscillatoryWeighted1998} when $\kt=0$ for three-fifth order WENO scheme with ideal
weights $\{1/10, 6/10 ,3/10\}$.

As is evident from \eqn{omg}, the non-linear weights $\omega_i$ that
are integral to the WENO formulation depend upon the smoothness indicators $\beta_i$
defined in \cite{shuEssentiallyNonoscillatoryWeighted1998} as
\begin{equation}
    \beta^{(i)}=\sum_{\ell=1}^{k}\dx^{2\ell-1}\int_{x_{j-\frac{1}{2}}}^{x_{j+\frac{1}{2}}}
    \left( \frac{d^{\ell}}{dx^{\ell} }p^{(i)}(x)\right)^2dx
\eqnlabel{bet1}
\end{equation}
where $p^{(i)}(x)$ is the polynomial of degree $k$ (=2, in the example considered here) for
stencil $S^{(i)}$. For a one-dimensional polynomial interpolation
\eqn{bet1} can be easily evaluated to give a smoothness indicator for each stencil.
However, for a simpler extension to an asynchronous stencil, Simpson's 3/8 rule is used to
compute $\beta^{(i)}$,
\begin{equation}
\begin{aligned}
    \beta^{(i)}=\frac{\dx}{6}
    &\Bigg(
    \sum_{\ell=1}^{k}\dx^{2\ell-1}
    \Bigg(\left. \frac{d^{\ell}}{dx^{\ell} }p^{(j)}(x)\right|_{x=x_{j-\frac{1}{2}}}\Bigg)^2
    +
    4 \sum_{\ell=1}^{k}\dx^{2\ell-1}
    \Bigg(\left. \frac{d^{\ell}}{dx^{\ell} }p^{(j)}(x)\right|_{x=x_{j-\frac{1}{6}}}\Bigg)^2
    \\
    &+
    4 \sum_{\ell=1}^{k}\dx^{2\ell-1}
    \Bigg(\left. \frac{d^{\ell}}{dx^{\ell} }p^{(j)}(x)\right|_{x=x_{j+\frac{1}{6}}}\Bigg)^2
    +
    \sum_{\ell=1}^{k}\dx^{2\ell-1}
    \Bigg(\left. \frac{d^{\ell}}{dx^{\ell} }p^{(j)}(x)\right|_{x=x_{j+\frac{1}{2}}}\Bigg)^2
    \Bigg),
    \end{aligned}
    \eqnlabel{beta}
\end{equation}
such that the derivatives of the interpolant are computed
numerically at each of $\{ x_{j-\frac{1}{2}},x_{j-\frac{1}{6}},x_{j+\frac{1}{6}},x_{j+\frac{1}{2}}  \}$ using all the points in the stencil $S^{(i)}$. This yields the smoothness indicator expressed
in the form
\begin{equation}
    \begin{aligned}
        &\beta^{(0)}=\frac{1 }{3} \left(4 \left(u^n\right)_{j-2}^2+\left(11 u^n_j-19 u^n_{j-1}\right) u^n_{j-2}+25 \left(u^n\right)_{j-1}^2+10 \left(u^n\right)_j^2-31 u^n_{j-1} u^n_j\right) \\
        &\beta^{(1)}=\frac{1 }{3} \left(4 \left(u^n\right)_{j-1}^2+\left(5 u^n_{j+1}-13 u^n_j\right) u^n_{j-1}+13 \left(u^n\right)_j^2+4 \left(u^n\right)_{j+1}^2-13 u^n_j u^n_{j+1}  \right) \\
        &\beta^{(2)}=\frac{1 }{3} \left( 10 \left(u^n\right)_j^2+\left(11 u^n_{j+2}-31 u^n_{j+1}\right) u^n_j+25 \left(u^n\right)_{j+1}^2+4 \left(u^n\right)_{j+2}^2-19 u^n_{j+1} u^n_{j+2}\right) \\
    \end{aligned}
    \eqnlabel{betaS}
\end{equation}
which is consistent with $\beta^{(j)}$ in the literature. Extending this
to the asynchronous stencil, the
smoothness indicator of the following
form is obtained when a four-point stencil is used for all
three candidate stencils,
\begin{equation}
    \begin{aligned}
         \tilde{\beta}^{(0)}=\frac{1}{3} &
         \Bigg(
        \left(10 \tilde{k}^2-11 \tilde{k}+4\right) \left(u_{j-2}^{n-\tilde{k}}\right)^2+\Big(-20 \tilde{k}^2 u_{j-2}^{n-\tilde{k}-1}+\tilde{k} \left(11 u_{j-2}^{n-\tilde{k}-1}+31 u_{j-1}^{n-\tilde{k}}-20 u_j^n\right)-19 u_{j-1}^{n-\tilde{k}}\\
        &+11 u_j^n\Big)
        u_{j-2}^{n-\tilde{k}}+10 \tilde{k}^2 \left(u_{j-2}^{n-\tilde{k}-1}\right)^2+25 \left(u_{j-1}^{n-\tilde{k}}\right)^2-31 u_j^n u_{j-1}^{n-\tilde{k}}+\tilde{k} u_{j-2}^{n-\tilde{k}-1} \left(20 u_j^n-31 u_{j-1}^{n-\tilde{k}}\right)+10 \left(u_j^n\right)^2
         \Bigg)
    \end{aligned}
    \eqnlabel{betaAT0}
\end{equation}
\begin{equation}
    \begin{aligned}
        \tilde{ \beta}^{(1)}=\frac{1}{3} &
         \Bigg(
         \left(4 \tilde{k}^2-5 \tilde{k}+4\right) \left(u_{j-1} ^{n-\tilde{k}}\right) ^2+\Big(-8 \tilde{k}^2 u_{j-1} ^{n-\tilde{k}-1}+\tilde{k} \left(5 u_{j-1} ^{n-\tilde{k}-1}+13 u_j ^{n-\tilde{k}}-8 u_{j+1} ^n\right)-13 u_j ^{n-\tilde{k}}\\
         &+5 u_{j+1} ^n\Big) u_{j-1} ^{n-\tilde{k}}+4 \tilde{k}^2 \left(u_{j-1} ^{n-\tilde{k}-1}\right) ^2+13 \left(u_j ^{n-\tilde{k}}\right) ^2-13 u_{j+1} ^n u_j ^{n-\tilde{k}}\\
         &+\tilde{k} u_{j-1} ^{n-\tilde{k}-1} \left(8 u_{j+1} ^n-13 u_j ^{n-\tilde{k}}\right)+4 \left(u_{j+1} ^n\right) ^2
         \Bigg)
    \end{aligned}
    \eqnlabel{betaAT1}
\end{equation}
\begin{equation}
    \begin{aligned}
        \tilde{ \beta}^{(2)}=\frac{1}{3} &
         \Bigg(
         \left(4 \tilde{k}^2-11 \tilde{k}+10\right) \left(u_j^{n-\tilde{k}}\right)^2+\Big(-8 \tilde{k}^2 u_j^{n-\tilde{k}-1}+\tilde{k} \left(11 u_j^{n-\tilde{k}-1}+19 u_{j+1}^{n-\tilde{k}}-8 u_{j+2}^n\right)-31 u_{j+1}^{n-\tilde{k}}
         +11 u_{j+2}^n\Big) \\
        &u_j^{n-\tilde{k}}+4 \tilde{k}^2 \left(u_j^{n-\tilde{k}-1}\right)^2+25 \left(u_{j+1}^{n-\tilde{k}}\right)^2-19 u_{j+2}^n u_{j+1}^{n-\tilde{k}}+\tilde{k} u_j^{n-\tilde{k}-1} \left(8 u_{j+2}^n-19 u_{j+1}^{n-\tilde{k}}\right)+4 \left(u_{j+2}^n\right)^2
         \Bigg).
    \end{aligned}
    \eqnlabel{betaAT2}
\end{equation}
Each of the $\tilde{\beta}^{i}$ listed in \eqn{betaAT0}, \eqn{betaAT1}, \eqn{betaAT2}
reduce to the corresponding $\beta^{(j)}$ expressions in \eqn{betaS} when
$\kt=0$. With the approximations and smoothness indicators in
the candidate stencil known along with the ideal weights, the
derivative of the flux can then be computed using the standard WENO procedure.

\begin{figure}[h!]
\begin{center}
\hspace{-2cm}
\includegraphics[trim={0cm 0cm 0cm 0cm},clip,width=0.5\textwidth]{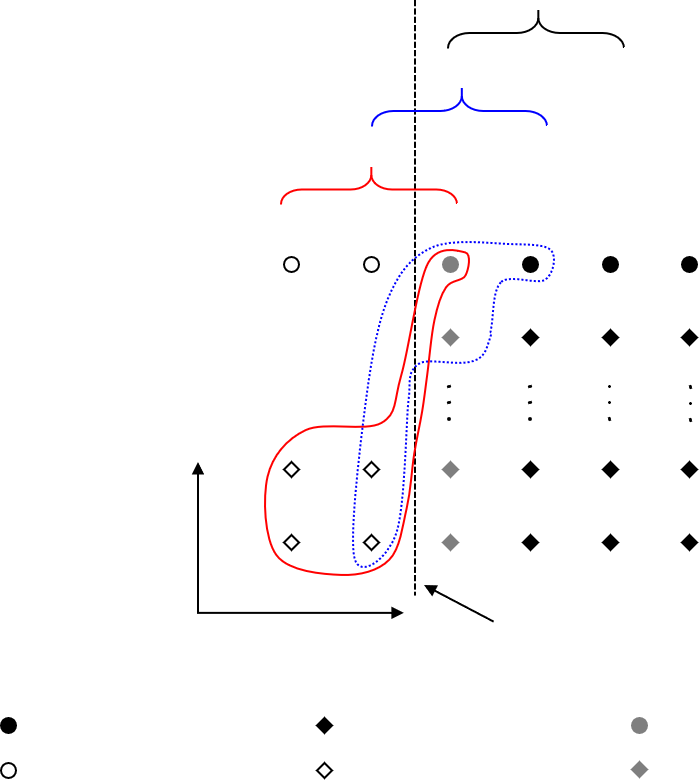}
\begin{picture}(0,0)
        \put(-205,78){Time}
        \put(-150,45){Space}
        \put(-65,46){PE boundary}
         \put(-225,0){Buffer point}
         \put(-225,16){Internal point}
         \put(-120,0){Buffer point (old)}
         \put(-120,16){Internal point (old)}
          \put(-15,0){PE boundary point }
         \put(-15,16){PE boundary point (old)}
          \put(-119,208){$S^{(0)}$}
            \put(-90,233){$S^{(1)}$}
              \put(-63,260){$S^{(2)}$}
               \put(10,171){$n$}
                \put(10,146){$n-1$}
                 \put(10,102){$n-\kt$}
                  \put(10,78){$n-\kt-1$}
                   \put(-71,184){$j+1$}
                   \put(-45,184){$j+2$}
           \put(-89,184){$j$}
               \put(-123,184){$j-1$}
           \put(-151,184){$j-2$}
        \end{picture}
\caption{Updated stencil for WENO at the left processor boundary.}
\figlabel{stenNew}
\end{center}
\end{figure}

For the choice of stencil discussed above, if the delay is zero
then instead of achieving fifth-order approximation in the
larger stencil, only fourth-order accuracy is obtained.
This is explained as follows. The ideal weights $\{\gamma_0,\gamma_1,\gamma_2\}$
in \eqn{idw} satisfy
$\sum_{i=0}^{2}\gamma_i=1$ and thus, two additional
constraints need to be imposed to find a unique solution when seeking a combination of smaller stencils. Upon upgrading from a
third-order approximation in the smaller stencil $\tilde{S}^{(j)}$ to a fourth-order
approximation in the full stencil $\tilde{S}$, only two lower order
truncation error terms of the form $\{ \dt  \dx, \dx^3\}$ can be eliminated.
Thus, a unique solution for $\gamma_i$ is obtained.
However, such an approximation will
never recover fifth-order accuracy in the absence of delays since the
$\dx^4$ truncation error term is non-zero.
Consequently, when there is no delay,
the large stencil $\tilde{S}$ will have at most four points and the
approximation is only fourth-order accurate. For example, the ideal weights \eqn{gammAT2} yield non-zero
coefficients only for points $\{j-2,j-1,j,j+2\}$ in the full stencil when $\kt=0$. Moreover, this stencil is spatially discontinuous.

The reduction in the number of points in the larger stencil when the delay is zero and a corresponding lower order accurate approximation can be
overcome by selectively
eliminating  additional truncation error terms.
For an asynchronous fifth-order approximation,
terms of the form $\{\dt^2,\dt\dx^2,\dx^4\}$ in addition to $\{\dt\dx,\dx^3 \}$ also need to be eliminated.
Since this yields a system with six constraints and
only three unknowns, a unique solution cannot be determined. However,
it is still possible to achieve a fifth-order approximation in the absence of delays.
%
Addition of one more point in the smaller asynchronous stencil
and using this degree of freedom to eliminate the
$\dt\dx$ truncation term does not
affect the order of the resulting interpolant. However, when the ideal weights
in the large stencil are computed, one of the constraints can now be imposed on the
 $\dx^4$ term. This exercise then yields convex ideal weights that
are exactly equal to the ones obtained when
all three small stencils are synchronous.
While the asynchronous approximation in the
full stencil is still of order four, the synchronous approximation recovers
fifth-order accuracy when the delay goes to zero.
Moreover, a
convex combination of the candidate stencil
with none of the weights being negative is achieved.
Thus, the additional computation involved in the treatment on
negative weights is no-longer required.
This new asynchronous stencil at the left boundary is given as
\begin{equation}
    \begin{aligned}
    &\tilde{S}^{(0)}_L=\{ u_{j-2}^{n-\kt -1}, u_{j-2}^{n-\kt},u_{j-1}^{n-\kt-1},
u_{j-1}^{n-\kt},u_j^n\}, \\
&\tilde{S}^{(1)}_L=\{ u_{j-1}^{n-\kt -1}, u_{j-1}^{n-\kt},
u_{j}^{n-\kt},
u_{j}^{n},u_{j+1}^n\}\\
&\tilde{S}^{(2)}_L=S^{(2)}=\{ u_{j}^n,u_{j-1}^n,u_{j+2}^n\}.
    \end{aligned}
    \eqnlabel{stenF}
\end{equation}
and is shown in \rfig{stenNew}.
For this stencil the third-order interpolation of the form
\begin{equation}
    \begin{aligned}
    &\tilde{u}_{j+\frac{1}{2}}^{n,(0)}=
    -\frac{3}{8} \tilde{k} u_{j-2}^{-\tilde{k}+n-1}+\frac{5}{4} \tilde{k} u_{j-1}^{-\tilde{k}+n-1}+\frac{1}{8} \left(3 \tilde{k}+3\right) u_{j-2}^{n-\tilde{k}}+\frac{1}{4} \left(-5 \tilde{k}-5\right) u_{j-1}^{n-\tilde{k}}+\frac{15}{8} u_j^n\\
     &\tilde{u}_{j+\frac{1}{2}}^{n,(1)}=
\frac{1}{8} \tilde{k} u_{j-1}^{-\tilde{k}+n-1}+\frac{1}{8} \left(-\tilde{k}-1\right) u_{j-1}^{n-\tilde{k}}+\frac{3}{4} u_j^n+\frac{3}{8} u_{j+1}^n\\
   &\tilde{u}_{j+\frac{1}{2}}^{n,(2)}=
\frac{3}{8}8 u_j^n+\frac{3}{4} u_{j+1}^n-\frac{1}{8}u_{j+2}^n
    \end{aligned}
\end{equation}
is obtained when a constraint is imposed on the truncation error term $\dt\dx$ for the
first two stencils, $\tilde{S}^{(0)}$ and $\tilde{S}^{(1)}$. The ideal weights are then computed to be
\begin{equation}
\gamma=\left\{\frac{1}{16},\frac{5}{8},\frac{5}{16}\right\}.
\end{equation}
While the interpolation procedure for approximating the
fluxes at the edges is explained in detail,
approximation using reconstruction, which is equivalent to the WENO approximation for the first derivative, is the main relevant WENO procedure when designing finite volume or finite difference
schemes to solve hyperbolic conservation laws~\cite{liu2009positivity}. Similar constraints
in terms of the order-of-accuracy and elimination
of specific terms can also be used to derive a
reconstruction function with an asynchronous stencil using a primitive
variable. For the choice of stencil listed in \eqn{stenF}, by treating point values
as cell-averages  the third-order reconstruction
approximation at
the left boundary assumes the following functional form
\begin{equation}
    \begin{aligned}
    &\tilde{u}_{j+\frac{1}{2}}^{n,(0)}=
-\frac{1}{3} \tilde{k} u_{j-2}^{-\tilde{k}+n-1}+\frac{7}{6} \tilde{k} u_{j-1}^{-\tilde{k}+n-1}+\frac{1}{6} \left(2 \tilde{k}+2\right) u_{j-2}^{n-\tilde{k}}+\frac{1}{6} \left(-7 \tilde{k}-7\right) u_{j-1}^{n-\tilde{k}}+\frac{11}{6} u_j^n
\\
     &\tilde{u}_{j+\frac{1}{2}}^{n,(1)}=
         \frac{1}{6} \tilde{k} u_{j-1}^{-\tilde{k}+n-1}+\frac{1}{6} \left(-\tilde{k}-1\right) u_{j-1}^{n-\tilde{k}}+\frac{5}{6} u_j^n+\frac{1}{3}u_{j+1}^n
     \\
   &\tilde{u}_{j+\frac{1}{2}}^{n,(2)}=
   \frac{1}{3}{u_j^n}+\frac{5}{6} u_{j+1}^n-\frac{1}{6}u_{j+2}^n
    \end{aligned}
    \eqnlabel{atRecon}
\end{equation}
such that the coefficients reduce to their corresponding
synchronous values when $\kt=0$. The ideal or
linear weights in this case can
then be computed to be
\begin{equation}
    \gamma =\left\{\frac{1}{10}, \frac{3}{5}, \frac{3}{10}\right \}
    \eqnlabel{idRec}
\end{equation}
which provides the fifth-order reconstruction polynomial in the
larger stencil when the delay is zero, and is fourth-order, otherwise.
Finally, the smoothness indicator for stencil \eqn{stenF} is
computed using Simpson's 3/8 rule,
\begin{equation}
    \begin{aligned}
         \tilde{\beta}^{(0)}=\frac{1}{3} &
         \Bigg(
         4 \left(u_{j-2}^{n-\tilde{k}}\right)^2+25 \left(u_{j-1}^{n-\tilde{k}}\right)^2+\tilde{k}^2\Big(4 \left(u_{j-2}^{n-\tilde{k}-1}\right)^2+4 \left(u_{j-2}^{n-\tilde{k}}\right)^2+19 u_{j-2}^{n-\tilde{k}} \left(u_{j-1}^{n-\tilde{k}-1}-u_{j-1}^{n-\tilde{k}}\right) \\
         &+25 \left(u_{j-1}^{n-\tilde{k}-1}-u_{j-1}^{n-\tilde{k}}\right)^2+u_{j-2}^{n-\tilde{k}-1} \left(19 \left(u_{j-1}^{n-\tilde{k}}-u_{j-1}^{n-\tilde{k}-1}\right)-8 u_{j-2}^{n-\tilde{k}}\right)\Big)-31 u_j^n u_{j-1}^{n-\tilde{k}}+10 \left(u_j^n\right)^2 \\
         &+u_{j-2}^{n-\tilde{k}} \left(11 u_j^n-19 u_{j-1}^{n-\tilde{k}}\right)+\tilde{k} \Big(8 \left(u_{j-2}^{n-\tilde{k}}\right)^2+\left(19 u_{j-1}^{n-\tilde{k}-1}-38 u_{j-1}^{n-\tilde{k}}+11 u_j^n\right) u_{j-2}^{n-\tilde{k}}\\
         &+u_{j-2}^{n-\tilde{k}-1} \left(-8 u_{j-2}^{n-\tilde{k}}+19 u_{j-1}^{n-\tilde{k}}-11 u_j^n\right)-\left(u_{j-1}^{n-\tilde{k}-1}-u_{j-1}^{n-\tilde{k}}\right) \left(50 u_{j-1}^{n-\tilde{k}}-31 u_j^n\right)\Big)
         \Bigg)
    \end{aligned}
    \eqnlabel{beta2AT0}
\end{equation}
\begin{equation}
    \begin{aligned}
        \tilde{ \beta}^{(1)}=\frac{1}{3} &
         \Bigg(
         4 \tilde{k}^2 \left(u_{j-1}^{n-\tilde{k}-1}-u_{j-1}^{n-\tilde{k}}\right)^2+4 \left(u_{j-1}^{n-\tilde{k}}\right)^2+13 \left(u_j^n\right)^2+4 \left(u_{j+1}^n\right)^2
         +\left(5 u_{j+1}^n-13 u_j^n\right)u_{j-1}^{n-\tilde{k}}\\
         &-13 u_j^n u_{j+1}^n-\tilde{k} \left(u_{j-1}^{n-\tilde{k}-1}-u_{j-1}^{n-\tilde{k}}\right) \left(8 u_{j-1}^{n-\tilde{k}}-13 u_j^n+5 u_{j+1}^n\right)
         \Bigg)
    \end{aligned}
    \eqnlabel{beta2AT1}
\end{equation}
\begin{equation}
    \begin{aligned}
        \tilde{ \beta}^{(2)}=\frac{1}{3} &
         \Bigg(
       10 \left(u_j ^n\right) ^2-31 u_{j+1} ^n u_j ^n+25 \left(u_{j+1} ^n\right) ^2+4 \left(u_{j+2} ^n\right) ^2+\left(11 u_j ^n-19 u_{j+1} ^n\right) u_{j+2} ^n
         \Bigg) = \beta^{(2)}.
    \end{aligned}
    \eqnlabel{beta2AT2}
\end{equation}
In terms of the leading order truncation error, these smoothness indicators
have the same properties as their synchronous counterparts, and
therefore, preserve the order characteristics of the non-linear
weights in \eqn{fullSten}. Thus, the AT-WENO approximation
in the simulations presented here is carried out
using the
reconstruction polynomials in \eqn{atRecon}, ideal
weights in \eqn{idRec} and the
smoothness indicators listed in \eqn{beta2AT2}.
While only a specific example of AT-WENO is presented here,
similar steps can be followed to derive high-order schemes as well by
extending the stencil in both space and time.
Furthermore, additional low-order truncation error terms
can also be selectively eliminated
in both candidate and full stencil such that the AT-WENO
scheme in the full stencil reduces to
the corresponding synchronous scheme in the absence of
delays.

Before proceeding with the validation simulations in Section 5,
the order of accuracy of the synchronous and the new AT-WENO schemes
is tested on a linear advection equation,
\begin{equation}
\begin{aligned}
    &\frac{\partial u}{\partial t}+c \frac{\partial u}{\partial x} = 0
\end{aligned}
\end{equation}
in a periodic domain $x\in[-1,1]$ for an initial condition $u(x,0)=\sin^4(\pi x+ 0.25)$ and a convective velocity of $c=5$ for
different spatial resolutions. Since the AT-WENO scheme is derived using the
relation $\dt \sim \dx^2$, we retain this power-law for
computing the time step in the simulations to assess the order of
accuracy. The temporal scheme used is RK-3 implemented
using the procedure listed in \rsec{rkder}.
Delays with uniform probability are introduced at every grid-point.
 The $L_1$ and
$L_{\infty}$ norm of the error at $t=1.0$ is
tabulated in \rtab{wenoLin} for the synchronous case, standard
WENO used asynchronously (AS-WENO) and the
asynchrony-tolerant WENO (AT-WENO). As expected, the
order degrades to two when the standard WENO sees delays at the boundaries and the error is orders of magnitude larger than the synchronous error.
On the contrary, the AT-WENO schemes exhibit errors comparable to the
synchronous case and the order of accuracy is close to four.

The analysis was repeated for an inviscid Burgers equation,
\begin{equation}
\frac{\partial u}{\partial t}+ \frac{\partial }{\partial x}\left( \frac{u^2}{2}\right)=0
\end{equation}
in a
periodic domain with an initial condition,
$u(x,0)=0.5+\sin(\pi x)$. The order of accuracy both
before and after the formation of the shock is computed. At $t=0.15$, the solution is
smooth throughout the domain and the order deterioration only occurs when
the standard WENO is used asynchronously (AS-WENO). Both
synchronous and AT-WENO have an order higher than the theoretical
order when the grid is refined.
For $L_1$ and $L_{\infty}$ norm of error after shock
formation, only the smooth region at a distance of 0.05
on both sides of the shock, \textit{ie.} $|x-\text{x}_{\text{shock}}|>0.05$, is considered
\cite{shuEssentiallyNonoscillatoryWeighted}.
Once again, errors in both the $L_1$ and $L_{\infty}$ norms and the order of accuracy for AT-WENO are consistent with that for standard WENO.
Thus, AT-WENO exhibits expected numerical accuracy for both linear and non-linear
equations.

\begin{table}
\begin{center}
\begin{tabular}{|c|c|c|c|c|c| }
 \hline
  Case & N & $L_1$ error & $L_1$ order & $L_{\infty}$ error & $L_{\infty}$ order \\
 \hline
            &   16	&1.95E-01	&-	    & 3.93E-01 &	-   \\
            &   32	&3.87E-02	&2.3   & 9.09E-02 &	2.1\\
Synchronous &   64	&2.79E-03	&3.8   & 5.57E-03 &	4.0\\
 (Order 5)  &   128	&2.65E-04	&3.4   & 1.07E-03 &	2.4\\
            &   256	&1.09E-05	&4.6	& 7.00E-05 &	3.9\\
            &   512	&3.01E-07	&5.8	& 2.18E-06 &	5.0\\
\hline
\hline
&16	                    &3.25E-01	&-	    &6.10E-01	&-      \\
&32	                     &1.42E-01	&1.2	&2.93E-01	&1.0   \\
AS-WENO&64	             &2.84E-02	&2.3	&6.38E-02	&2.2   \\
$p=[0.5,0.5]$    &128   &5.70E-03	&2.3	&1.32E-02	&2.3   \\
 (Order 5) &256                    &1.22E-03	&2.2	&2.90E-03	&2.2   \\
&512                    	&2.82E-04	&2.1	&6.79E-04	&2.1   \\
\hline
\hline
\hline
        &   16  &   2.00E-01  &	-   &	3.86E-01&	-   \\
        &   32  & 	4.15E-02	&   2.3&	1.00E-01&	1.9\\
AT-WENO &   64  &   2.98E-03  &	3.8&	6.70E-03&	3.9\\
$p=[0.5,0.5]$       &   128 &   2.90E-04  &	3.4&	9.54E-04&	2.9\\
 (Order 4)        &   256 &	1.42E-05  &	4.4&	7.04E-05&	3.8\\
        &   512	&   5.79E-07	&   4.6&	2.40E-06&	4.9\\
        \hline
        &16	    &3.01E-01	&-	 &  5.09E-01	&-\\
        &32	    &6.18E-02	&2.3&	1.37E-01	&1.9\\
AT-WENO &64	    &5.20E-03	&3.8&	1.13E-02	&3.6\\
$p=[0.3,0.3,0.3]$   &128	&4.22E-04	&3.6&	1.12E-03	&3.3\\
  (Order 4)       &256	&2.40E-05	&4.1&	7.73E-05	&3.9\\
        &512	&1.26E-06	&4.3&	2.86E-06	&4.8\\
        \hline
\end{tabular}
\caption{Order of accuracy for linear advection equation $u_t+cu_x=0$ with
initial condition $u(x,0)=\sin^4(\pi x +0.25)$ at $t=1.0$ for synchronous, AS-WENO and AT-WENO.}
\tablabel{wenoLin}
\end{center}
\end{table}

\begin{table}
\begin{center}
\begin{tabular}{|c|c|c|c|c|c| }

        \multicolumn{6}{c}{}\\
         \multicolumn{6}{c}{$t=0.15$: Before shock}\\
         \multicolumn{6}{c}{}\\
         \hline
  Case & N & $L_1$ error & $L_1$ order & $L_{\infty}$ error & $L_{\infty}$ order \\
 \hline
	&	32	&	3.03E-04	&	-	&	2.36E-03	&	-	\\
Synchronous	&	64	&	1.14E-05	&	4.7	&	1.08E-04	&	4.4	\\
 (Order 5)	&	128	&	4.38E-07	&	4.7	&	3.71E-06	&	4.9	\\
	&	256	&	1.41E-08	&	5.0	&	2.12E-07	&	4.1	\\
	&	512	&	3.37E-10	&	5.4	&	4.53E-09	&	5.5	\\
\hline
\hline
	&	32	&	7.99E-04	&	-	&	4.77E-03	&	-	\\
AS-WENO	&	64	&	1.44E-04	&	2.5	&	7.21E-04	&	2.7	\\
$p=[0.5, 0.5]$	&	128	&	3.39E-05	&	2.1	&	1.49E-04	&	2.3	\\
 (Order 5)	&	256	&	8.61E-06	&	2.0	&	3.76E-05	&	2.0	\\
	&	512	&	2.05E-06	&	2.1	&	8.93E-06	&	2.1	\\
\hline
\hline
	&	32	&	3.11E-04	&	-	&	2.48E-03	&	-	\\
AT-WENO	&	64	&	1.21E-05	&	4.7	&	1.15E-04	&	4.4	\\
$p=[0.5, 0.5]$	&	128	&	4.85E-07	&	4.6	&	4.14E-06	&	4.8	\\
 (Order 4)	&	256	&	1.82E-08	&	4.7	&	2.07E-07	&	4.3	\\
	&	512	&	6.61E-10	&	4.8	&	4.82E-09	&	5.4	\\
        \hline
     	&	32	&	3.23E-04	&	-	&	2.67E-03	&	-	\\
AT-WENO	&	64	&	1.32E-05	&	4.6	&	1.23E-04	&	4.4	\\
$p=[0.3, 0.3,0.3]$	&	128	&	6.14E-07	&	4.4	&	4.88E-06	&	4.7	\\
 (Order 4)	&	256	&	2.74E-08	&	4.5	&	1.99E-07	&	4.6	\\
	&	512	&	1.25E-09	&	4.5	&	7.42E-09	&	4.7	\\
        \hline
        \multicolumn{6}{c}{}\\
         \multicolumn{6}{c}{$t=0.55$: After shock}\\
         \multicolumn{6}{c}{}\\
        \hline
          Case & N & $L_1$ error & $L_1$ order & $L_{\infty}$ error & $L_{\infty}$ order \\
 \hline
        	&	32	&	6.26E-03	&	-	&	1.20E-01	&	-	\\
	&	64	&	2.43E-04	&	4.7	&	1.14E-02	&	3.4	\\
Synchronous	&	128	&	6.09E-06	&	5.3	&	3.00E-04	&	5.2	\\
 (Order 5)	&	256	&	1.83E-07	&	5.1	&	2.18E-05	&	3.8	\\
	&	512	&	2.25E-09	&	6.3	&	3.79E-07	&	5.8	\\
	\hline
	\hline
		&	32	&	6.49E-03	&	-	&	1.20E-01	&	-	\\
AS-WENO	&	64	&	3.21E-04	&	4.3	&	1.10E-02	&	3.4	\\
$p=[0.5,0.5]$	&	128	&	3.17E-05	&	3.3	&	3.48E-04	&	5.0	\\
 (Order 5)	&	256	&	6.96E-06	&	2.2	&	1.75E-05	&	4.3	\\
	&	512	&	1.70E-06	&	2.0	&	4.14E-06	&	2.1	\\
	\hline
	\hline
		&	32	&	6.26E-03	&	-	&	1.20E-01	&	-	\\
AT-WENO	&	64	&	2.44E-04	&	4.7	&	1.14E-02	&	3.4	\\
$p=[0.5,0.5]$	&	128	&	6.10E-06	&	5.3	&	2.99E-04	&	5.2	\\
 (Order 4)	&	256	&	1.85E-07	&	5.0	&	2.18E-05	&	3.8	\\
	&	512	&	2.43E-09	&	6.2	&	3.78E-07	&	5.8	\\
	\hline	&	32	&	6.26E-03	&	-	&	1.20E-01	&	-	\\
AT-WENO	&	64	&	2.44E-04	&	4.7	&	1.14E-02	&	3.4	\\
$p=[0.3,0.3,0.3]$	&	128	&	6.15E-06	&	5.3	&	2.99E-04	&	5.3	\\
 (Order 4)	&	256	&	1.89E-07	&	5.0	&	2.18E-05	&	3.8	\\
	&	512	&	2.80E-09	&	6.1	&	3.76E-07	&	5.9	\\
	\hline
\end{tabular}
\caption{Order of accuracy for inviscid Burgers' equation $u_t+(u^2/2)_x=0$ with
initial condition $u(x,0)=0.5 + \sin(\pi x)$ before and
after the shock for synchronous, AS-WENO and AT-WENO.}
\tablabel{wenoNLin}
\end{center}
\end{table}

\section{Governing equations}
\seclabel{gov_eqn}
In the following sections the effect of asynchrony is evaluated for a set of canonical reacting flow configurations.  The AT schemes described in Section 3 are implemented in a compressible reacting flow solver with periodic and open boundaries. The governing conservation equations and constitutive laws are presented in this section.
The one-dimensional form of the conservation equations for mass, momentum, total energy and species are
\begin{equation}
    \begin{aligned}
        &\frac{\partial  \rho}{ \partial t}=-\frac{\partial (\rho u)}{\partial x}\\
        &\frac{\partial  (\rho u)}{ \partial t}=-\frac{\partial (\rho u u)}{\partial x} + \frac{\partial \tau}{\partial x}
        -\frac{\partial P}{\partial x}\\
        &\frac{\partial  (\rho e_0)}{ \partial t}=-\frac{\partial [u(\rho e_0+P)]}{\partial x} + \frac{\partial (\tau u)}{\partial x}
        -\frac{\partial q}{\partial x}\\
        &\frac{\partial  (\rho Y_i)}{ \partial t}=-\frac{\partial (\rho u Y_i)}{\partial x} -\frac{\partial ( \rho Y_i V_{i})}{\partial x}
        + W_i\dot{\omega}_i,\\
    \end{aligned}
    \eqnlabel{ge}
\end{equation}
where $Y_i$ is the mass fraction, $W_i$ is the
molecular weight, $V_i$ is the species mass diffusion, $\dot{\omega}_i$ is the
molar production rate of species $i$ and $e_0$ is the specific
total energy
\begin{equation}
    e_0=\frac{u^2}{2}-\frac{P}{\rho} + h.
\end{equation}
$h=\sum_{i=1}^{N_s}Y_ih_i=\sum_{i=1}^{N_s}Y_i\left(
h_i^{0}+\int_{T_0}^{T}c_{p,i}dT\right)
$ is the total enthalpy expressed in terms of
$h_{i}^{0}$ which is the enthalpy of formation of species $i$ at temperature $T_0$ and the isobaric heat capacity
$c_p=\sum_{i=1}^{N_s}Y_ic_{p,i}$. For an ideal gas mixture, $P=\rho{R_u}T/W$ and $c_p-c_v=R_u/W$, where
$W=\left( \sum_{i=1}^{N_s}Y_i/W_i\right)^{-1}$ and $R_u$ is the universal gas constant, are used to compute the pressure and specific heats.
The viscous stress $\tau$ is
\begin{equation}
    \tau=\frac{4}{3}\mu \frac{\partial u}{\partial x},
\end{equation}
and the heat flux and species diffusion
velocities are given by
\begin{equation}
\begin{aligned}
&q=-\lambda\frac{\partial T}{\partial x} +\sum_{i=1}^{N_s}h_iJ_i\\
&V_i=-\frac{D_i^{\text{mix}}}{X_i}
\frac{\partial X_i}{\partial x}
\end{aligned}
\end{equation}
where $J_i=\rho Y_i V_{i}$ is the
species diffusive flux, $D_i^{\text{mix}}$ is the mixture-averaged
diffusion coefficient, and $X_i=Y_i W/W_i$ is the mole fraction.
Barodiffusion and the Soret and Dufour effects are not considered.
CHEMKIN~\cite{kee1996chemkin} and TRANSPORT~\cite{kee1986fortran} software libraries
were linked with the solver and used for
evaluating reaction rates, thermodynamic and mixture-averaged transport properties.
\eqn{ge} can be written in a compact form as in \eqn{ge1} where
\begin{equation}
Q=
    \begin{pmatrix}
    \rho \\ \rho u\\ \rho e_0\\  \rho Y_i
    \end{pmatrix},
    \quad
    C=
      \begin{pmatrix}
    \rho u\\ \rho u^2 + P\\  u(\rho e_0+P)\\  \rho u Y_i
    \end{pmatrix},
    \quad
    D=
        \begin{pmatrix}
    0 \\  -\tau \\ -\tau u+q\\  \rho Y_iV_i
    \end{pmatrix},
    \text{ and}
    \quad
        S=
        \begin{pmatrix}
    0 \\  0 \\0\\  W_i \dot{\omega}_i
    \end{pmatrix}.
\eqnlabel{convEq}
\end{equation}
The derivatives of product terms in \eqn{ge} are
expanded using the chain rule. For example,
$\partial (\rho u Y_i)/\partial x:=(\rho u Y_i)_x =
\rho u (Y_i)_x + \rho u_x Y_i +\rho_x u Y_i$. This essentially allows
computation of derivatives of such terms using AT schemes
without having to retain every product term at multiple time-levels.
Apart from the expansion using the chain rule, the one-dimensional asynchronous solver used in the present study is largely based on S3D \cite{chenTerascaleDirectNumerical2009} which is widely used to perform DNS
of turbulent combustion.

\section{Numerical Results}
In this section, five different flow configurations are
selected to assess the
effect of asynchrony on canonical combustion problems.
For the first case, two types of asynchronous simulations are considered,
\begin{enumerate}
    \item Standard schemes used asynchronously (AS-SFD or AS, AS-WENO as applicable)
    \item AT schemes for asynchronous computation (AT, AT-WENO as applicable).
\end{enumerate}
For the remainder of the cases, only the
the asynchronous simulation
performed using AT schemes is compared with the synchronous simulation. The domain
is decomposed into $P$ processors and delays are introduced
using a random number generator at each processor
boundary similar to \cite{adityaHighorderAsynchronytolerantFinite2017}.
The maximum allowed delay levels are three with the
probability
of non-zero delay gradually increasing from Set-1 to Set-3.
The different probability sets
considered in numerical simulations are tabulated in \rtab{dels} where Set-4 represents a synchronous simulation. The probability for Set-2 and Set-3 is similar to the probability of
delays observed on TACC supercomputers \cite{kumariDirectNumericalSimulations2020}.
A summary of all the numerical experiments performed and their relevance is
listed in \rtab{nums}.

\begin{table}
\centering
\begin{tabular}{ |c|c|c| }
 \hline
  & Probability & Legend\\
  & $[p_0~ p_1 ~p_2]$& \\
 \hline
 Set-1 &  [0.8 0.1 0.1] & \colb{\_\_\_ $\circ$} \\
 Set-2 &  [0.6 0.3 0.1] & - . *\\
 Set-3 &  [0.4 0.5 0.1]  & \colr{- - $\square$ }\\
 Set-4 &  [1.0 0.0 0.0] &\colm{$\dots$ +}\\
 \hline
\end{tabular}
\caption{Probability of simulated delays for different sets used in the
numerical experiments presented in Section 5.
This legend (color and/or symbol) is used
in all the figures from Fig. 4 to Fig. 13.}
\tablabel{dels}
\end{table}

\begin{table}
\centering
\begin{tabular}{|C{2cm}|L{6cm}|L{7cm}|}
 \hline
 Case number & Case name & Relevant processes to be resolved\\
 \hline
   5.1 & Acoustic wave propagation (non-reacting) &
  Pressure perturbation travelling at speed of sound\\
  5.2 & Auto-ignition of $H_2$ (periodic domain)  & Spontaneous ignition
  dominated by reaction term\\
  5.3 & Auto-ignition of $C_2H_4$ (temperature fluctuations at inflow)  &
  Unsteadiness, oscillatory ignition front\\
  5.4 & Premixed flame propagation  &
  Reactive-diffusive balance in reaction zone\\
  5.5 & Non-premixed ignition &  Diffusion controlled reaction front \\
  5.6 & Propagation of a detonation wave & Jump due to shock front followed by a reaction zone \\
 \hline

 \hline
\end{tabular}
\caption{Summary of numerical simulations and their relevance.}
\tablabel{nums}
\end{table}

\subsection{Non-reacting case: acoustic wave propagation}
For this non-reacting case the propagation
of an acoustic wave in air is considered.
The effect of the source of the acoustic wave, for example
from an ignition kernel, is decoupled from its propagation, and
the only focus here is on whether the asynchrony-tolerant framework can
accurately capture waves traversing with the speed of sound. Furthermore,
if an error in gradients due to delayed data at the boundaries
manifest themselves in the form of dilatational modes then
one should observe these effects in the propagation of acoustic waves through
larger errors. In the current problem the acoustic wave is considered
to be a perturbation in the pressure field that is
described  with the following initial condition \cite{baum1993},
\begin{equation}
\begin{aligned}
&u(x,0)=u_0+\mathcal{A}\exp{\left[
- \mathcal{B}\left(\frac{x-x_0}{L}\right)\right]}\\
&P(x,0)=P_0+\rho_0c_0\left(u-u_0\right)\\
&\rho(x,0)=\rho_0+\frac{\rho_0\left(u-u_0\right)}{c_0}\\
&T=P/\rho R
\end{aligned}
\eqnlabel{acIC}
\end{equation}
where $u_0,\rho_0,P_0$ prescribe the uniform mean
values, and ideal gas law is used to compute the temperature field. Here $\mathcal{A}$ and $\mathcal{B}$ determine the
magnitude and stiffness of the acoustic fluctuation
and $x_0$ is the location of the fluctuation peak. This
initial field is shown as a light blue line in \rfig{aAWP} and \rfig{AWP}. Non-reflecting inflow/outflow boundary conditions
\cite{thompsonTimeDependentBoundary1987,poinsotBoundaryConditionsDirect}
are used and the initial fluctuation is allowed to traverse across at least one processing element boundary where it encounters delays. While at the
internal points standard fourth-order
central difference schemes are used, at the physical
boundary points the derivatives are computed using second-order finite
difference schemes. For computation of derivatives at the processor
boundaries, fourth-order AT schemes are used (see Appendix A).

\begin{figure}[h!]
\vspace{2mm}
\begin{center}
\subfigure{\includegraphics[trim={0cm 0cm 0cm 0cm},clip,width=0.31\textwidth]{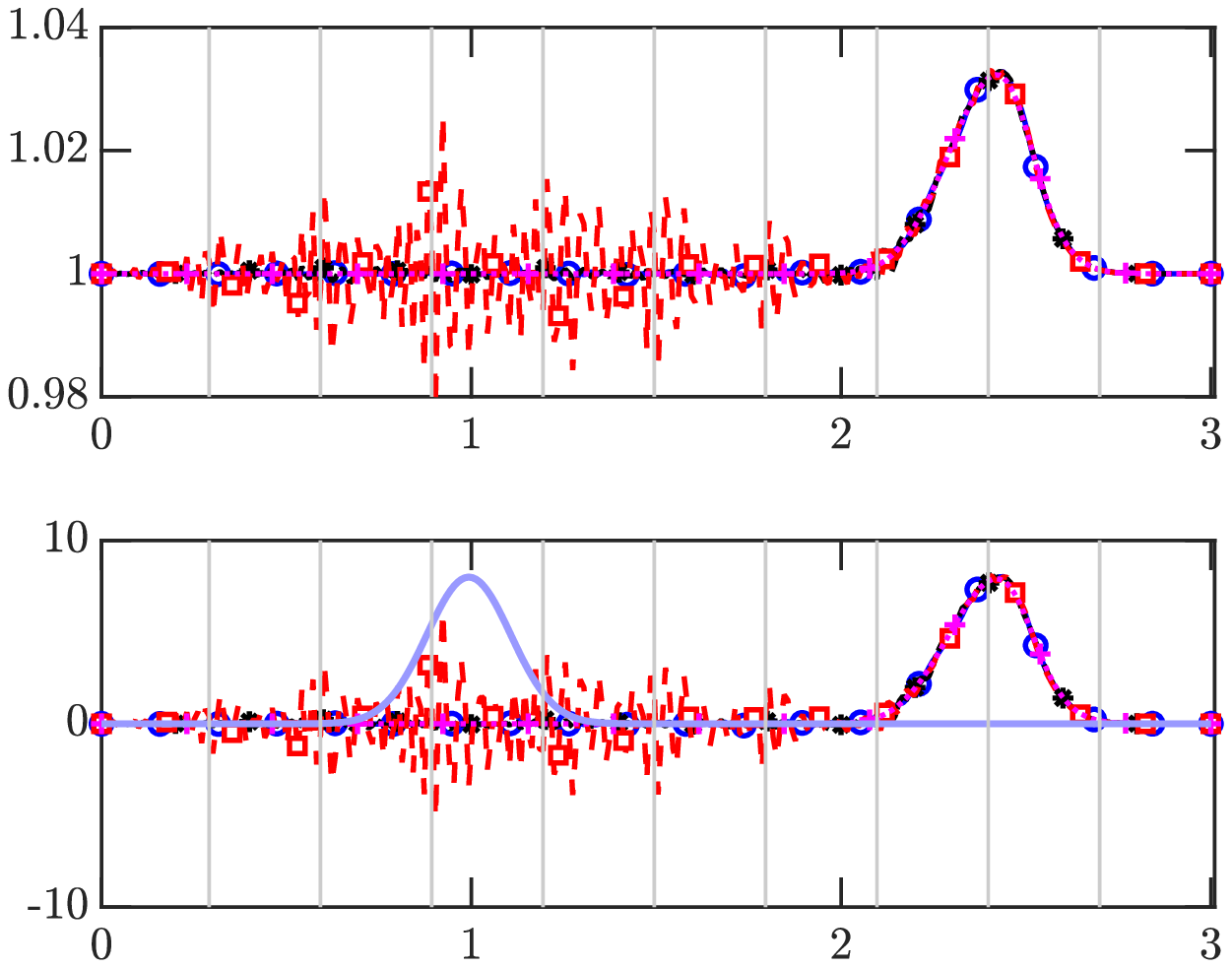}}
\hspace{1cm}
\subfigure{\includegraphics[trim={0cm 0cm 0cm 0cm},clip,width=0.31\textwidth]{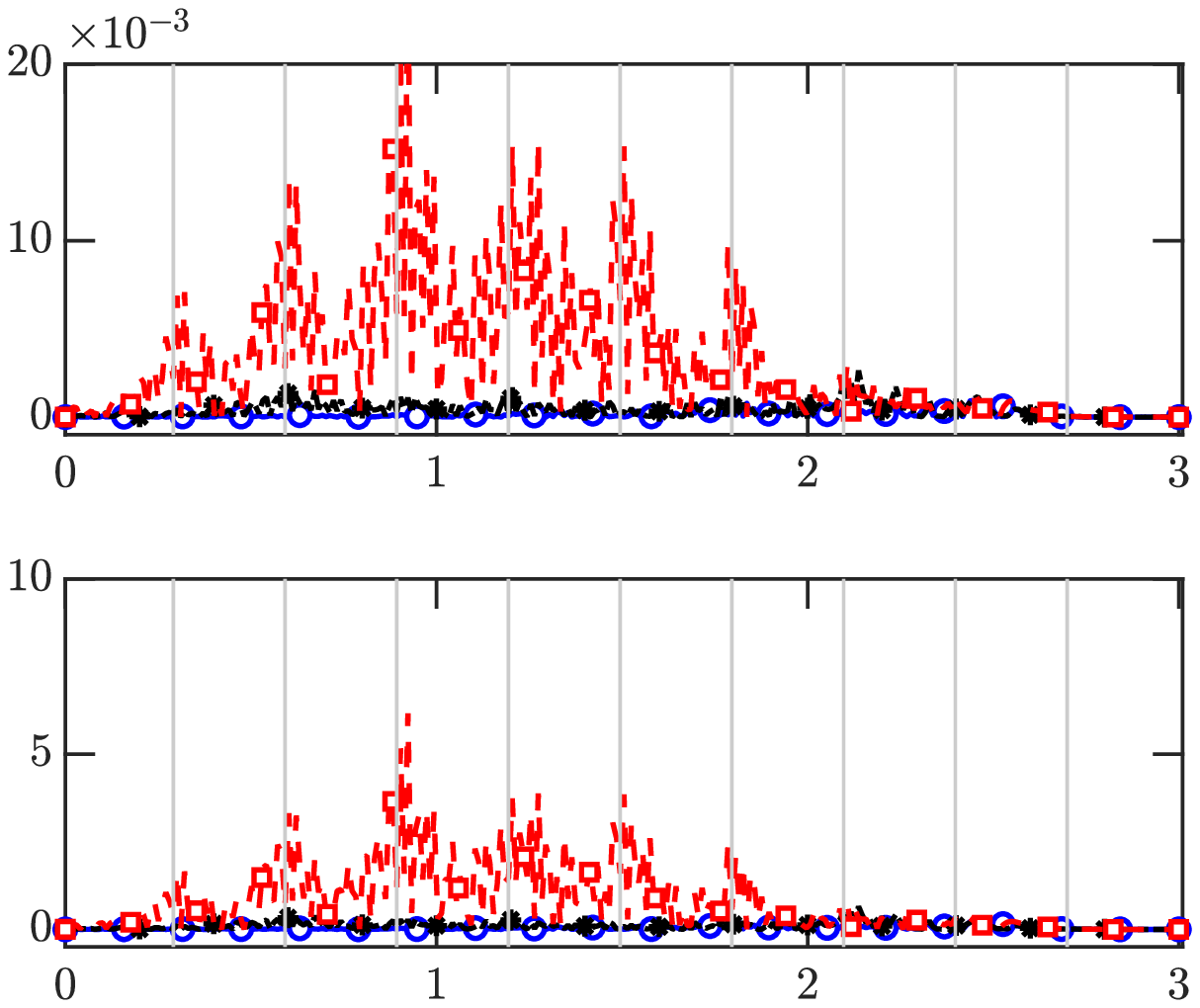}}
\begin{picture}(0,0)
        \put(-95,120){$|P_{AS}-P_{S}|$}
        \put(-340,85){$P$}
        \put(-340,25){$U$}
        \put(-95,57){$|U_{AS}-U_{S}|$}
        \put(-280,-10){$x~(mm)$}
        \put(-85,-10){$x~(mm)$}
        \end{picture}
\caption{AS-SFD: instantaneous profiles and errors in pressure ($atm$) and  velocity ($ms^{-1}$) at $t=4 \times 10^{-6} s$.  The different lines: blue (Set-1), black (Set-2), red (Set-3) and magenta (Set-4) are defined in \rtab{dels}. The light blue line indicates the initial condition and gray vertical lines represent processor boundaries.
AS: Asynchronous with standard finite-difference schemes, S: synchronous.}
\figlabel{aAWP}
\end{center}
\end{figure}

\begin{figure}[h!]
\begin{center}
\subfigure{\includegraphics[trim={0cm 0cm 0cm 0cm},clip,width=0.31\textwidth]{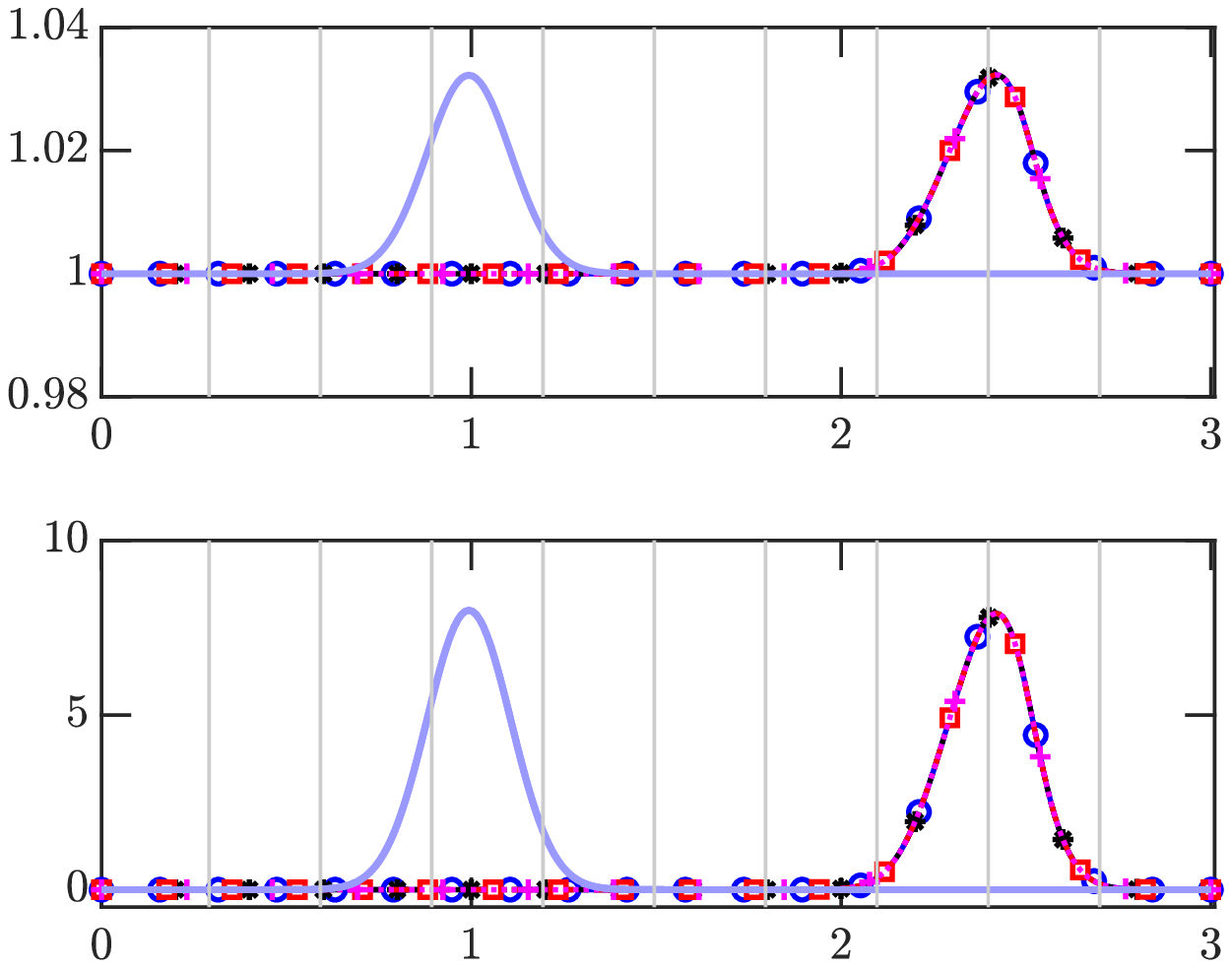}}
\hspace{1cm}
\subfigure{\includegraphics[trim={0cm 0cm 0cm 0cm},clip,width=0.31\textwidth]{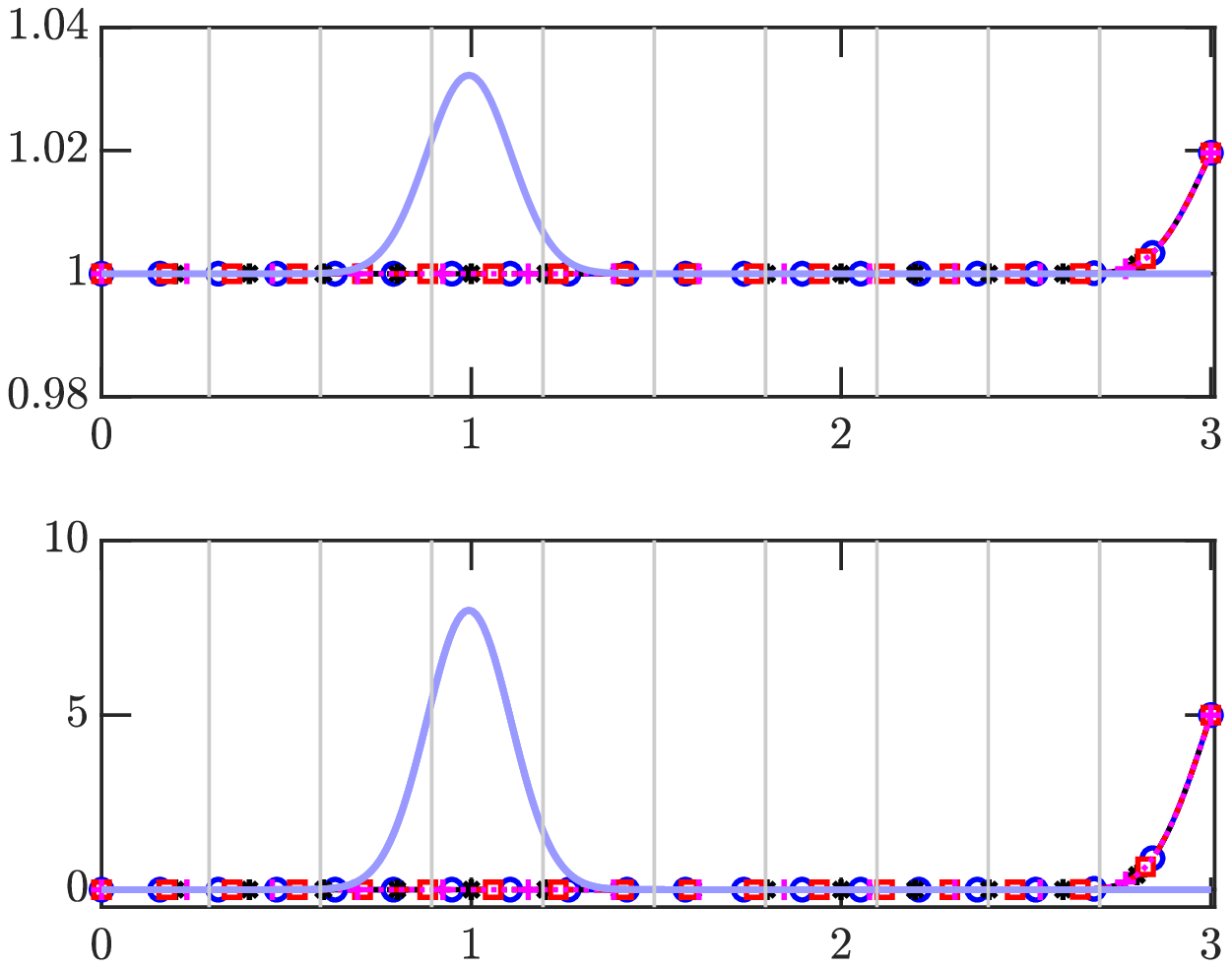}}
\\
\hspace{2mm}
\subfigure{\includegraphics[trim={0cm 0cm 0cm 0cm},clip,width=0.3\textwidth]{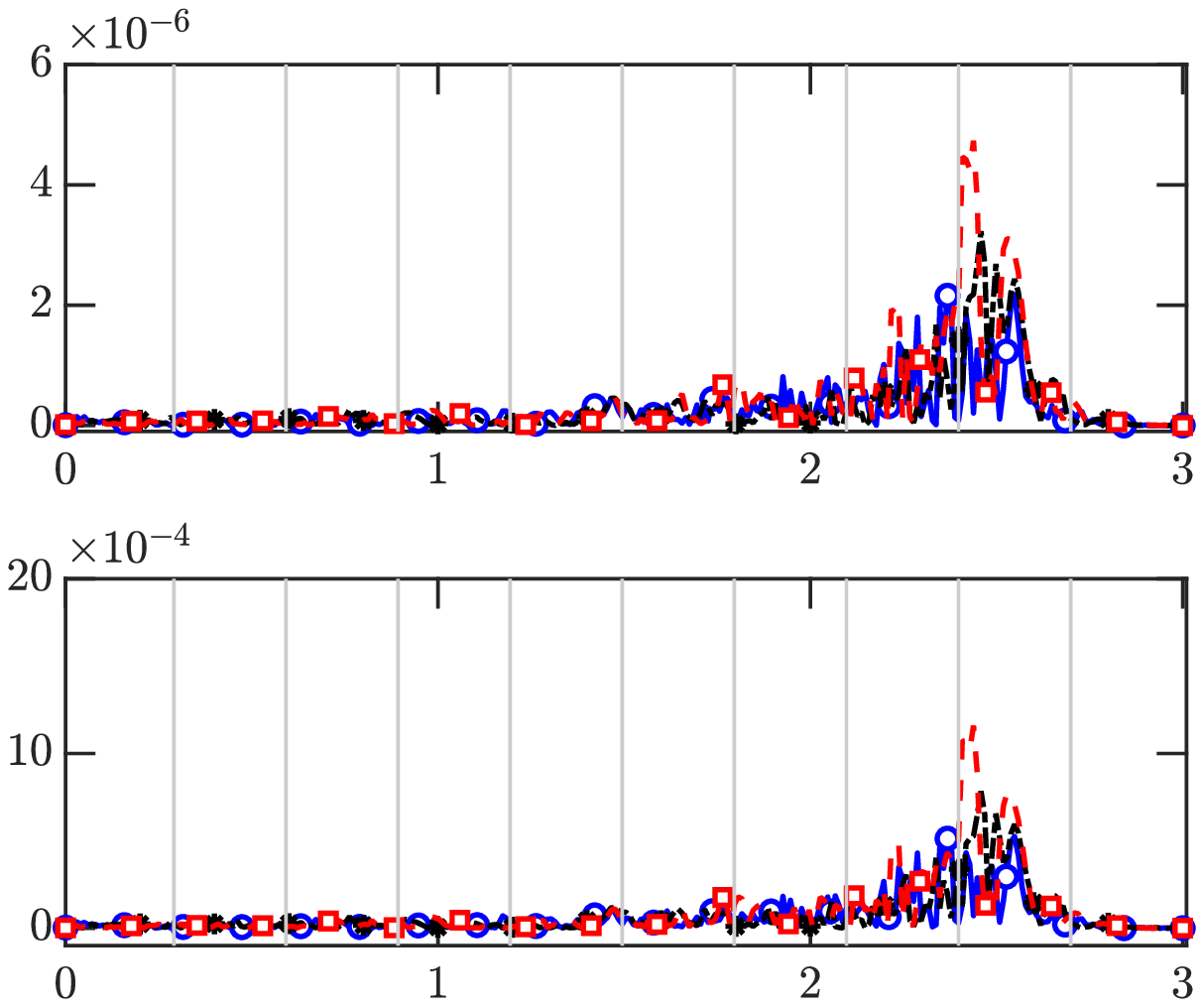}}
\hspace{1.2cm}
\subfigure{\includegraphics[trim={0cm 0cm 0cm 0cm},clip,width=0.3\textwidth]{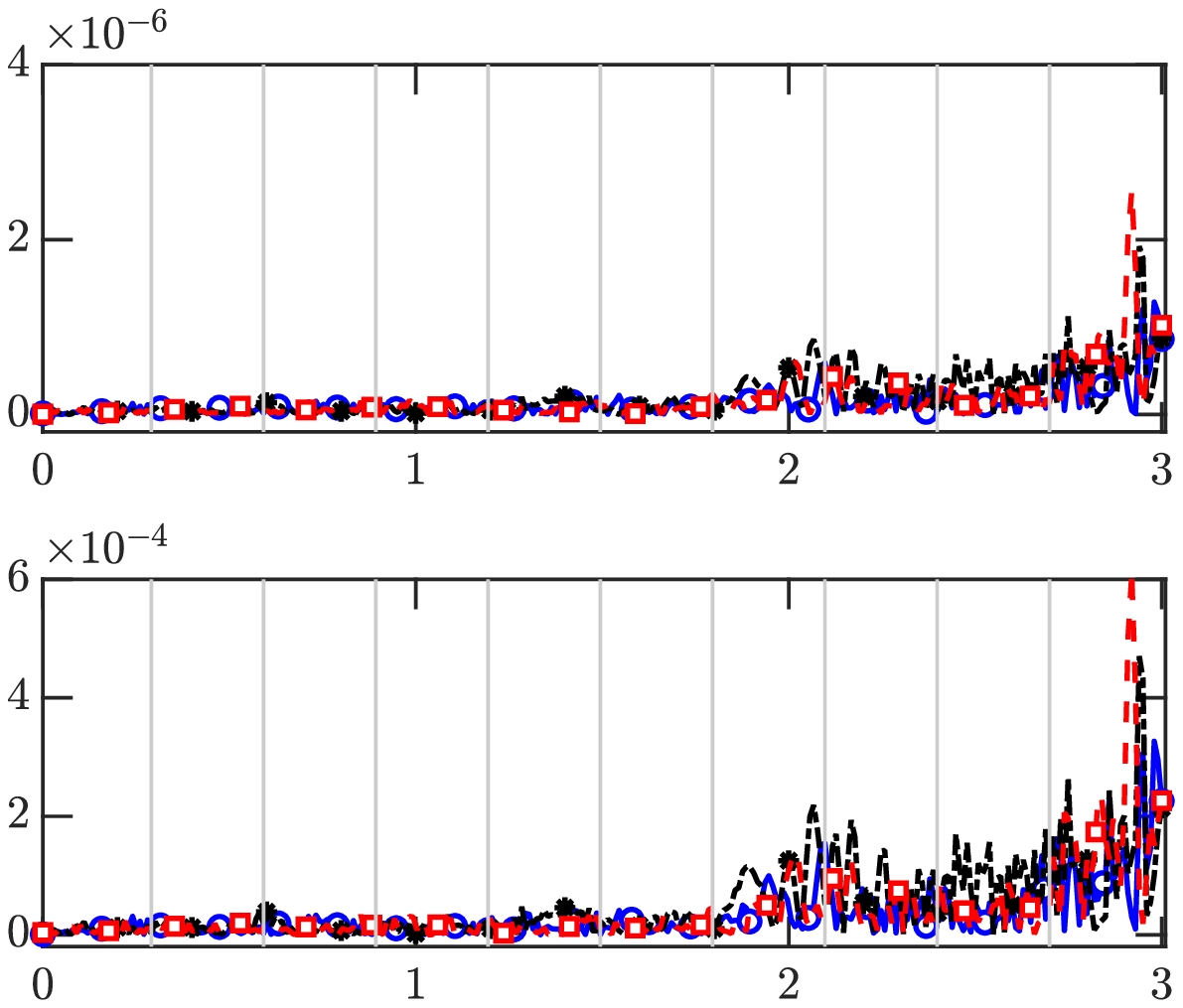}}
\begin{picture}(0,0)
         \put(-380,30){$|P_{AT} -P_S|$}
         \put(-380,85){$|U_{AT}-U_S|$}
        \put(-340,215){$P$}
        \put(-340,155){$U$}
        \put(-250,-10){$x$}
         \put(-80,-10){$x$}
         \put(-275,240){$t=4 \times 10^{-6} s$}
         \put(-100,240){$t=6 \times 10^{-6} s$}
        \end{picture}
\caption{Asynchrony-tolerant (AT): instantaneous profiles and errors in pressure ($atm$) and velocity ($ms^{-1}$) at $t=4 \times 10^{-6} s$ (left) and
$t=6\times 10^{-6} s$ (right). The different lines: blue (Set-1), black (Set-2), red (Set-3) and magenta (Set-4) are defined in \rtab{dels}. The light blue line indicates the initial condition and gray vertical lines represent processor boundaries.
AT: asynchronous with asynchrony-tolerant, S: synchronous.}
\figlabel{AWP}
\end{center}
\end{figure}

When the standard central-difference schemes are
used asynchronously (AS-SFD), there are visible
fluctuations in the pressure field even at early times.
This is evident from \rfig{aAWP} where both instantaneous pressure and
velocity fields at time $t=4 \times 10^{-6}s$ show
large numerical errors and clear deviation from the
corresponding synchronous field. For larger delay values corresponding to Set-3 in \rtab{dels}, the numerical perturbations become significant at
much shorter times and render the simulation unstable eventually. These
errors are amplified even after the acoustic wave leaves the domain.
However, when AT schemes are used for
the propagation of an acoustic wave with delays at processor boundaries,
the solution
remains in close agreement with its
synchronous counterpart and the errors are several orders
of magnitude smaller (see \rfig{AWP}). Similar behavior is also observed at
later times when the wave is close to leaving the right boundary. Moreover, the
errors in  both pressure and velocity are not localized to near processing element boundary
points where AT schemes are used to compute derivatives.

The $L_1$ and $L_{\infty}$ norms of the errors
computed with respect to the synchronous simulation
are defined as
\begin{equation}
\begin{aligned}
&Z_{L_1} =\langle Z_{\text{synchronous}} - Z_{\text{AS or AT}} \rangle \\
&
Z_{L_{\infty}} =\max \left(Z_{\text{synchronous}} - Z_{\text{AS or AT}} \right),
\end{aligned}
\end{equation}
where $Z$ is any quantity, for example temperature, pressure or mass fractions, and $\langle .\rangle$ is the spatial average.
These errors are listed in \rtab{ap} for both AS-SFD and AT simulations at
two different times. The AT schemes
exhibit significantly less numerical error at both times, and the error does not grow monotonically  with time like it does for the AS-SFD.

\begin{table}
\centering
\begin{tabular}{ |c|c|c c|c c| }
  \hline
  & & \multicolumn{2}{|c|}{$t=4\times10^{-6}s$} &\multicolumn{2}{|c|}{$t=6\times10^{-6}s$} \\
  \hline
   \multicolumn{6}{|c|}{AS-SFD}\\
 \hline
  & Case & $U~(ms^{-1})$& $P~(atm)$ & $U~(ms^{-1})$& $P~(atm)$ \\
  \hline
             &1& 3.33e-02 &  1.36e-04   & 3.30e-02 & 1.34e-04       \\
 $L_1$ error &2& 9.87e-02 &  4.00e-04   & 3.78e-01 &  1.52e-03         \\
             &3& 8.01e-01 &  3.26e-03  & 2.28e+01  & 9.31e-02 \\
             \hline
                    & 1& 2.27e-01   &9.43e-04 &   2.72e-01   &1.08e-03   \\
 $L_{\infty}$ error & 2&  6.53e-01  & 2.69e-03 &   2.03e+00  & 8.08e-03     \\
                    & 3& 6.16e+00   &2.57e-02  & 1.45e+02 &  7.56e-01  \\
                    \hline
   \hline
 \multicolumn{6}{|c|}{Asynchrony-tolerant (AT)}\\
 \hline
  & Case & $U~(ms^{-1})$& $P~(atm)$ & $U~(ms^{-1})$& $P~(atm)$ \\
  \hline
             &1& 5.88e-05 &  2.43e-07   & 3.38e-05  & 1.40e-07   \\
 $L_1$ error &2& 6.70e-05&  2.75e-07   & 4.82e-05 &  1.95e-07 \\
             &3& 8.83e-05 &  3.64e-07   & 3.81e-05  & 1.57e-07\\
             \hline
                    & 1& 5.25e-04&   2.18e-06 &  3.27e-04 &  1.29e-06 \\
 $L_{\infty}$ error & 2& 7.84e-04&   3.23e-06 & 4.70e-04  & 1.92e-06 \\
                    & 2& 1.16e-03 &  4.73e-06  &  6.09e-04 &  2.56e-06 \\
                    \hline
\end{tabular}
\caption{$L_1$ and $L_{\infty}$ norms of error in velocity and pressure in acoustic wave propagation simulations.}
\tablabel{ap}
\end{table}

\subsection{Premixed Auto-ignition of $H_2$ using one-step chemistry}
\begin{figure}[h!]
\begin{center}
\includegraphics[trim={0cm 0cm 0cm 0cm},clip,width=0.31\textwidth]{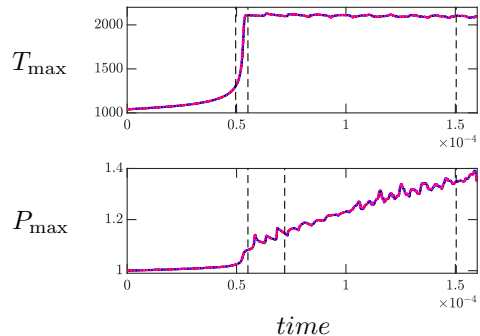}
\begin{picture}(0,0)
\put(-80,-10){$time$}
\put(-180,90){$T_{\text{max}}$}
\put(-180,30){$P_{\text{max}}$}
\end{picture}
\caption{Temporal evolution of the peak temperature ($K$) and pressure ($atm$) during auto-ignition of
a premixture of hydrogen/air in a periodic domain.  The different lines (without symbols): blue (Set-1), black (Set-2), red (Set-3) and magenta (Set-4) are defined in \rtab{dels}.
The dashed line indicates the time at which the instantaneous scalar and velocity
profiles and errors are listed in \rtab{aiL1} and \rtab{aiLinf}.}
\figlabel{tpmaxAI}
\end{center}
\end{figure}

A canonical problem of practical relevance to compression ignition in internal combustion engines
is auto-ignition of a premixed fuel-air mixture at constant volume.
Here auto-ignition of a fuel lean $H_2$/air mixture at
an equivalence ratio of $0.4895$ at  an initial temperature of $400K$ is considered. A hot spot represented by a Gaussian
temperature spike with a peak value of $1040K$, which is above the ignition limit, is introduced at the center of a one-dimensional closed domain and the mixture eventually auto-ignites after an induction period.
This thermal explosion leads to a rapid increase in temperature to nearly 2000K. Following the ignition of the hot spot, reaction fronts  emanate from the spontaneous ignition kernel, one propagating to the left, while the other to the right.  The gas expansion from the heat release during auto-ignition results in an induced velocity from an initially quiescent flow field.
  The time elapsed from the initial hot-spot to a time at which the temporal gradient of temperature or heat release rate is maximum is the so-called ignition delay time $(\tau_{ign})$. The time evolution of
maximum temperature and pressure is shown in \rfig{tpmaxAI}.
It is clear that while the temperature increases slowly initially,
there is an exponential increase in the temperature at around $t=5\times 10^{-5}s$. The
temporal evolution of the peak temperature and pressure is captured accurately by
the asynchrony-tolerant schemes. Following ignition, the two fronts propagate towards
the left and right boundary and are allowed to traverse across at least one
processing element boundary where
delays are explicitly encountered.
For simplicity we consider a periodic
domain that allows us to compute spectral characteristics of
both thermodynamic and hydrodynamic quantities.
Note that the periodic boundary conditions considered here provide a closed volume with compression heating
from the exothermic reactions that lead to a gradual rise in pressure with time.
This is evident from \rfig{tpmaxAI} where the pressure increases from
$1~atm$ to $1.4~atm$ after ignition.

\begin{figure}[h!]
\begin{center}
\subfigure{\includegraphics[trim={0cm 0cm 0cm 0cm},clip,width=0.31\textwidth]{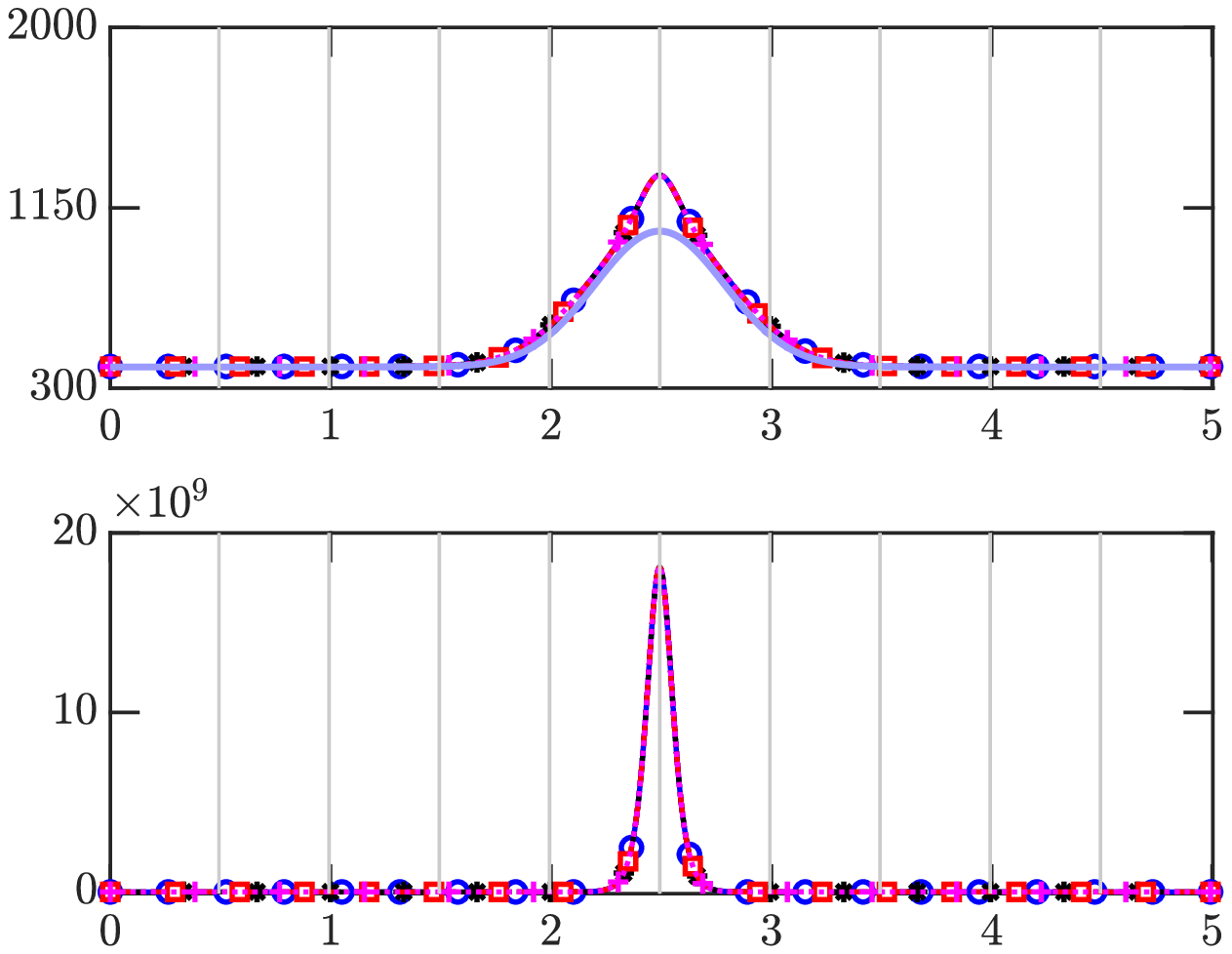}}
\hspace{0.3cm}
\subfigure{\includegraphics[trim={0cm 0cm 0cm 0cm},clip,width=0.31\textwidth]{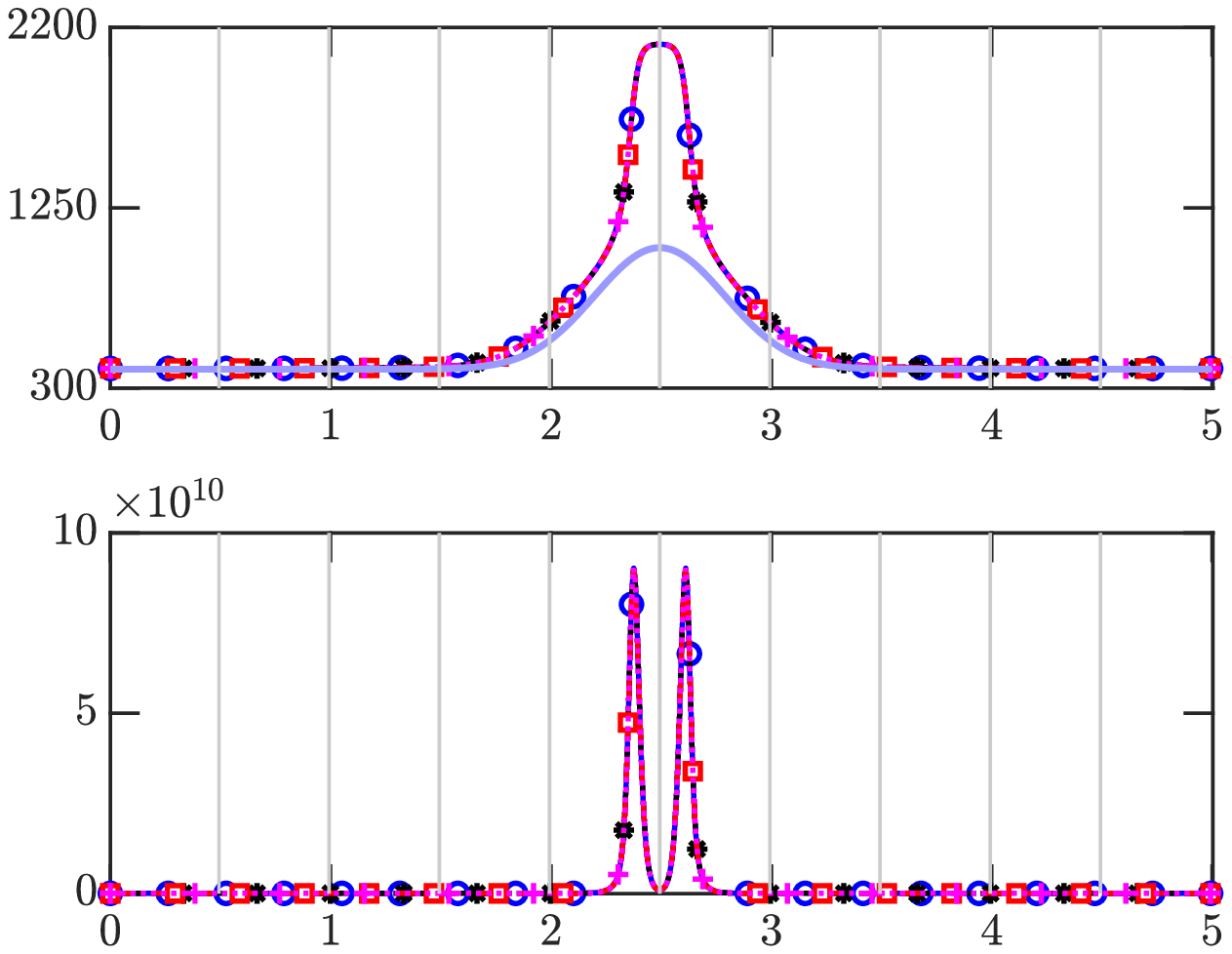}}
\hspace{0.3cm}
\subfigure{\includegraphics[trim={0cm 0cm 0cm 0cm},clip,width=0.31\textwidth]{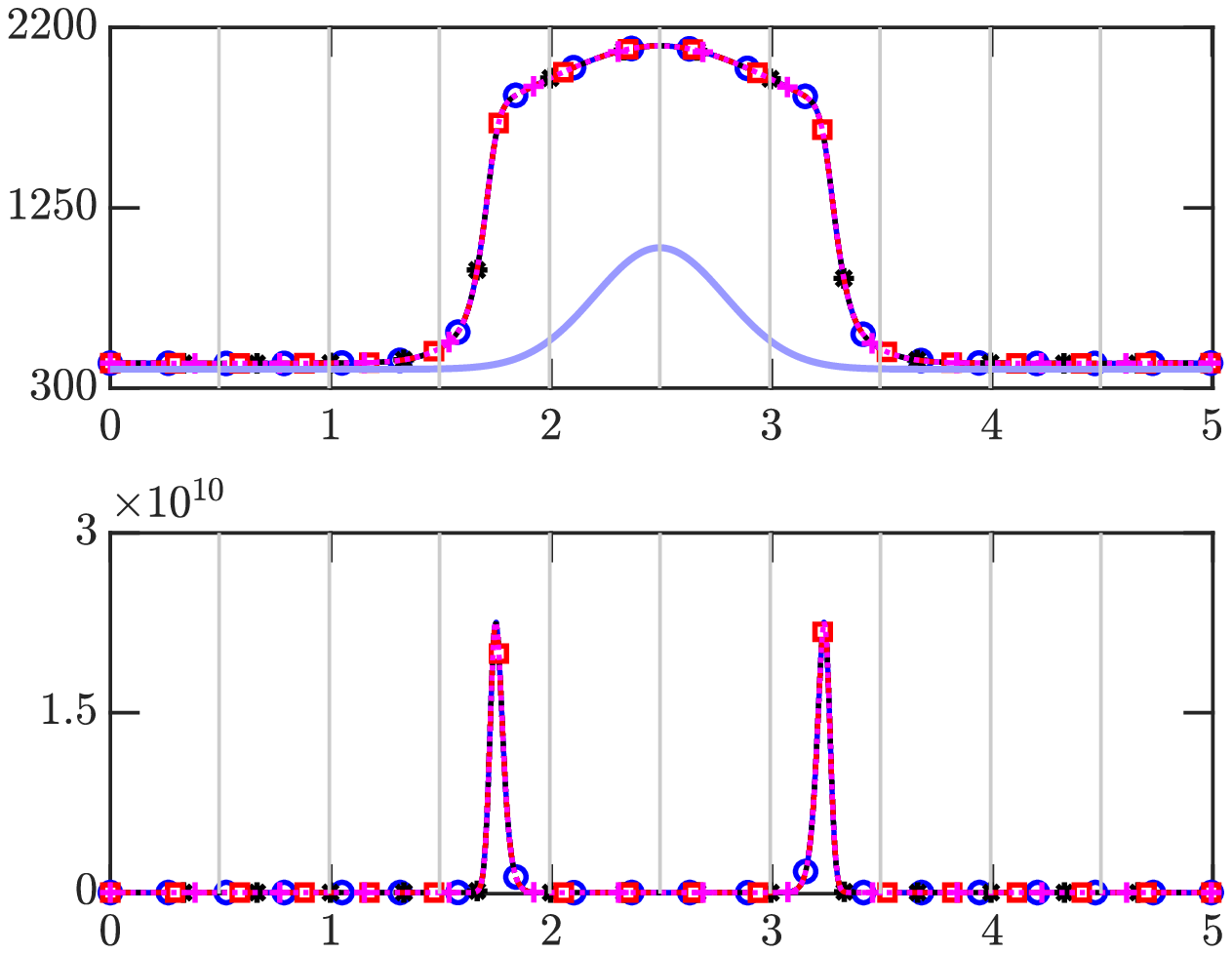}}
\hspace{0.3cm}\\
\subfigure{\includegraphics[trim={0cm 0cm 0cm 0cm},clip,width=0.31\textwidth]{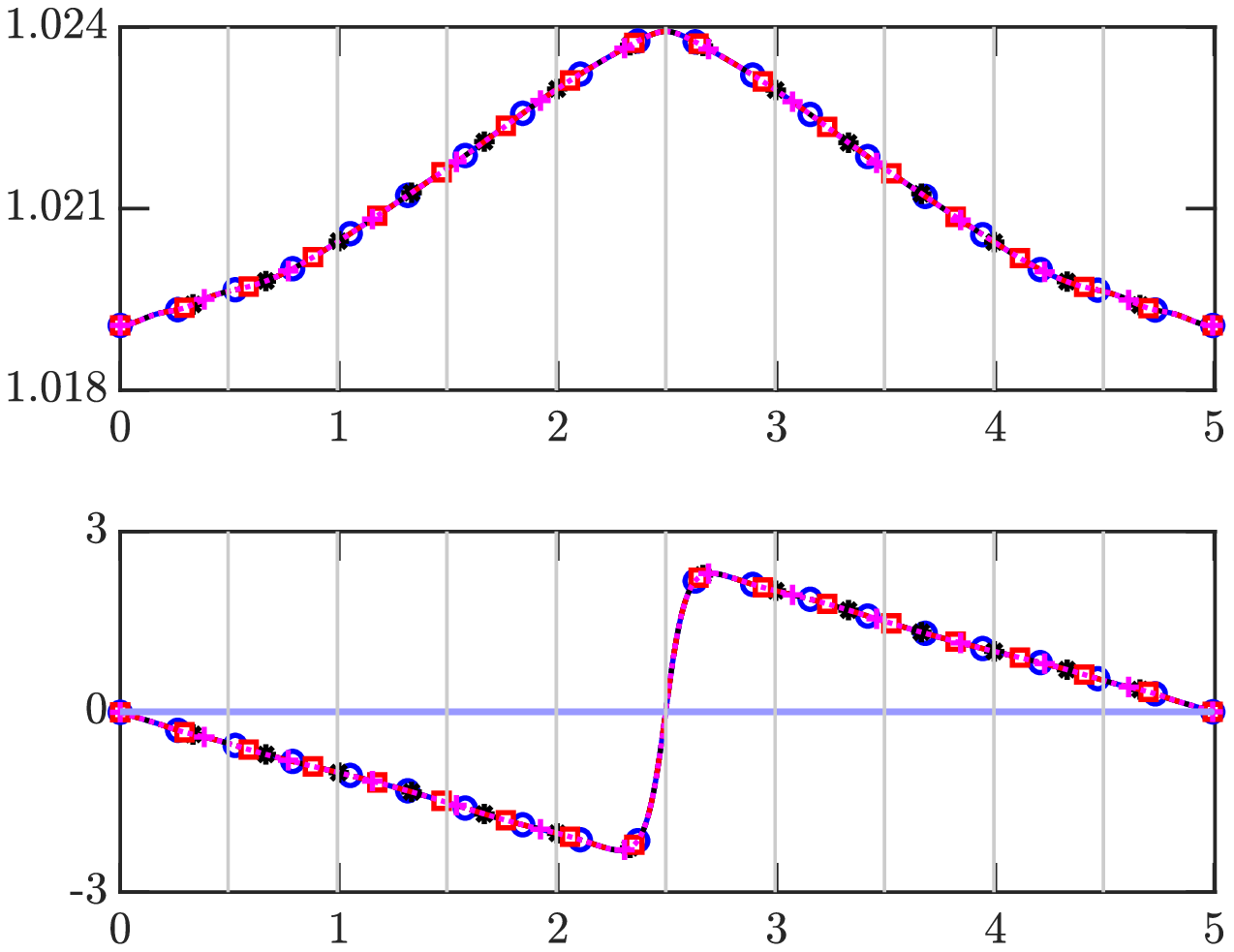}}
\hspace{0.3cm}
\subfigure{\includegraphics[trim={0cm 0cm 0cm 0cm},clip,width=0.31\textwidth]{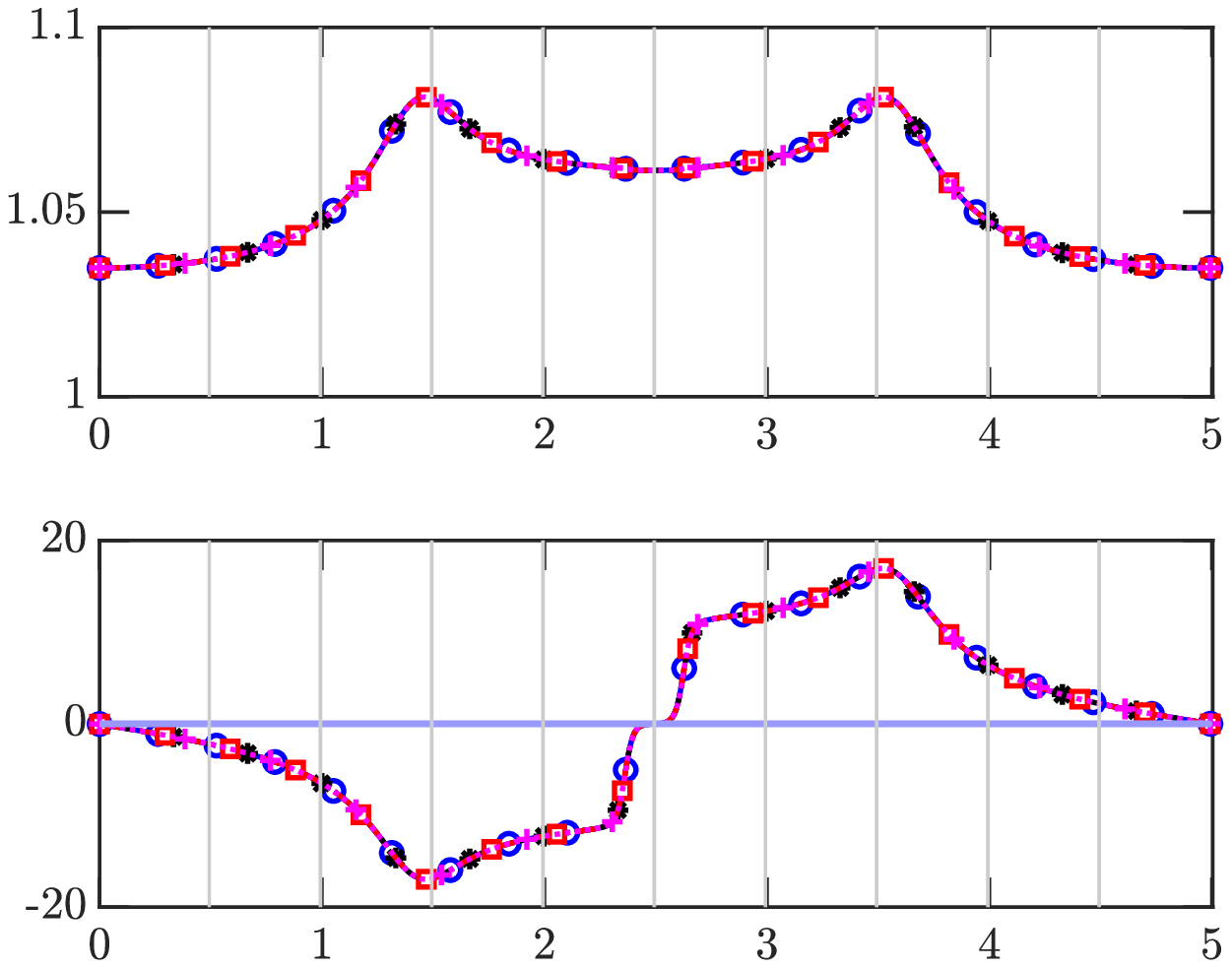}}
\hspace{0.3cm}
\subfigure{\includegraphics[trim={0cm 0cm 0cm 0cm},clip,width=0.31\textwidth]{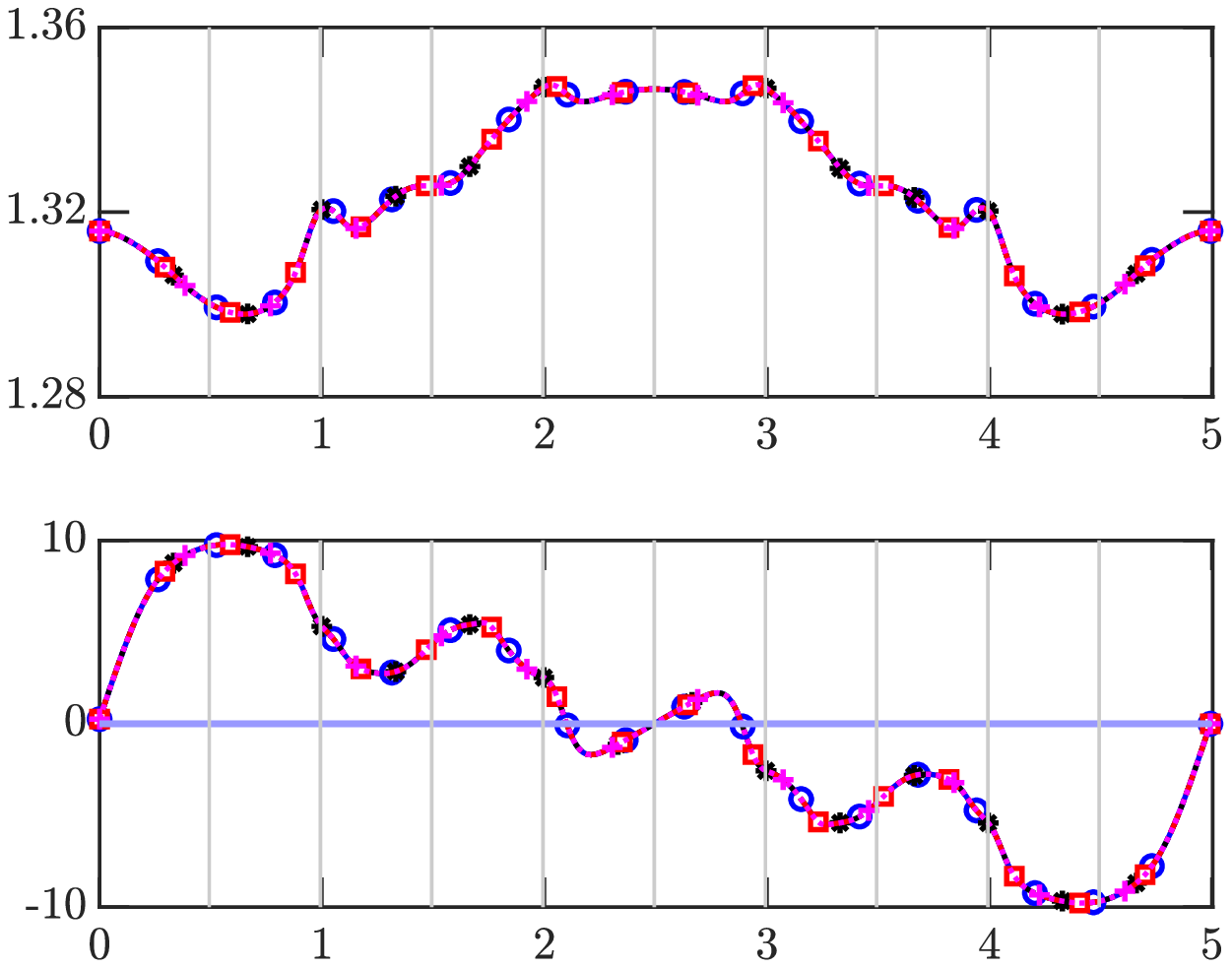}}
\hspace{0.3cm}\\
\subfigure{\includegraphics[trim={0cm 0cm 0cm 0cm},clip,width=0.31\textwidth]{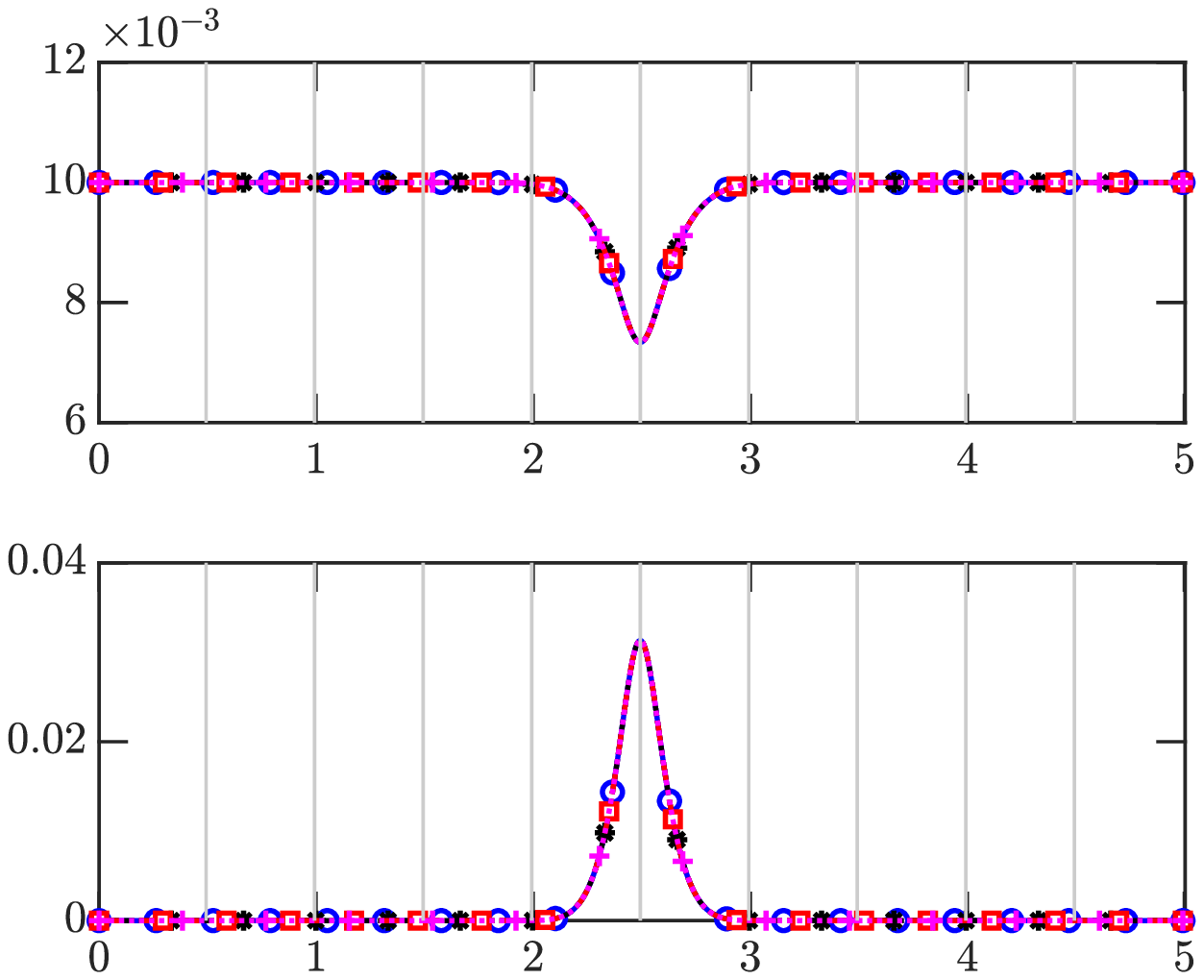}}
\hspace{0.3cm}
\subfigure{\includegraphics[trim={0cm 0cm 0cm 0cm},clip,width=0.31\textwidth]{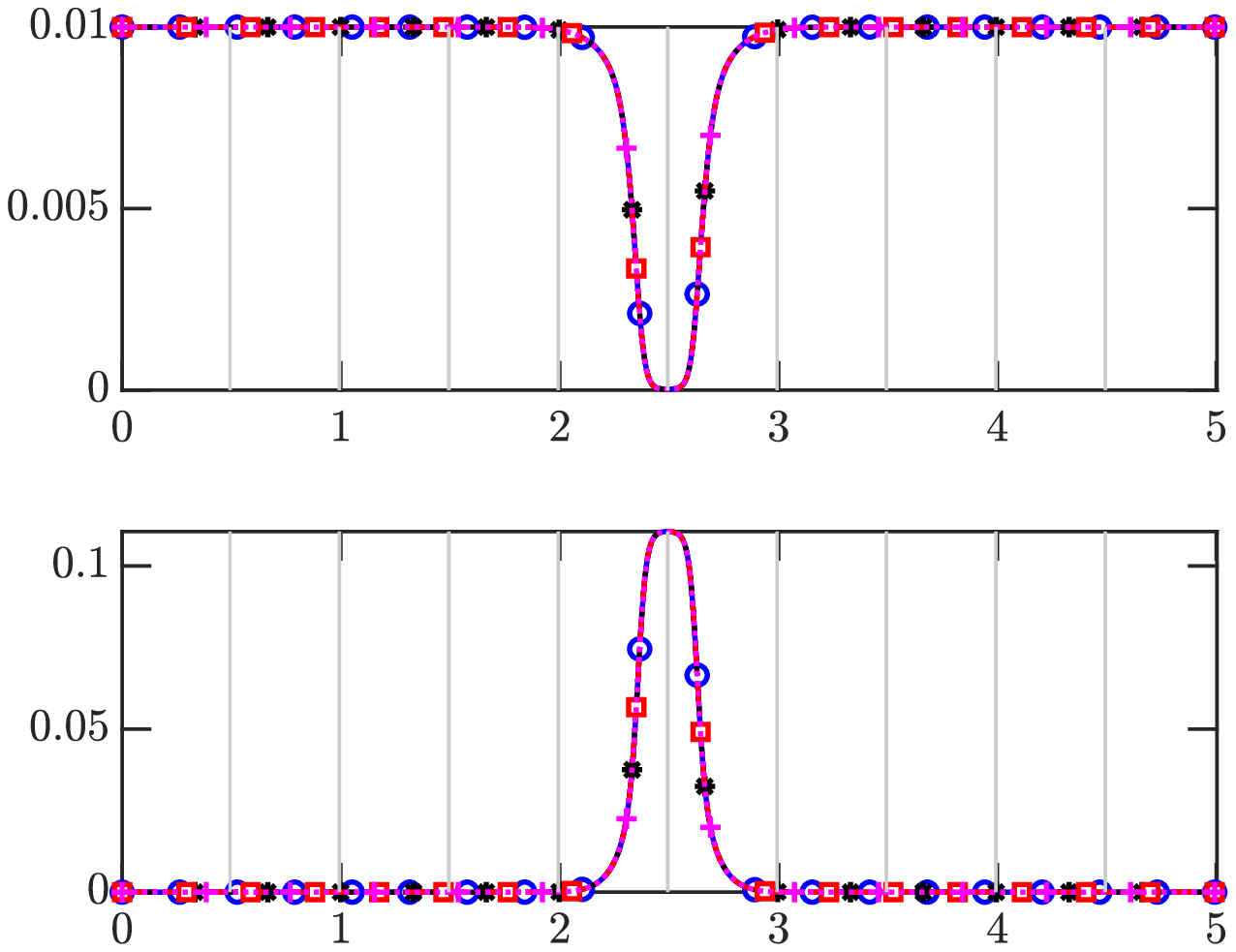}}
\hspace{0.3cm}
\subfigure{\includegraphics[trim={0cm 0cm 0cm 0cm},clip,width=0.31\textwidth]{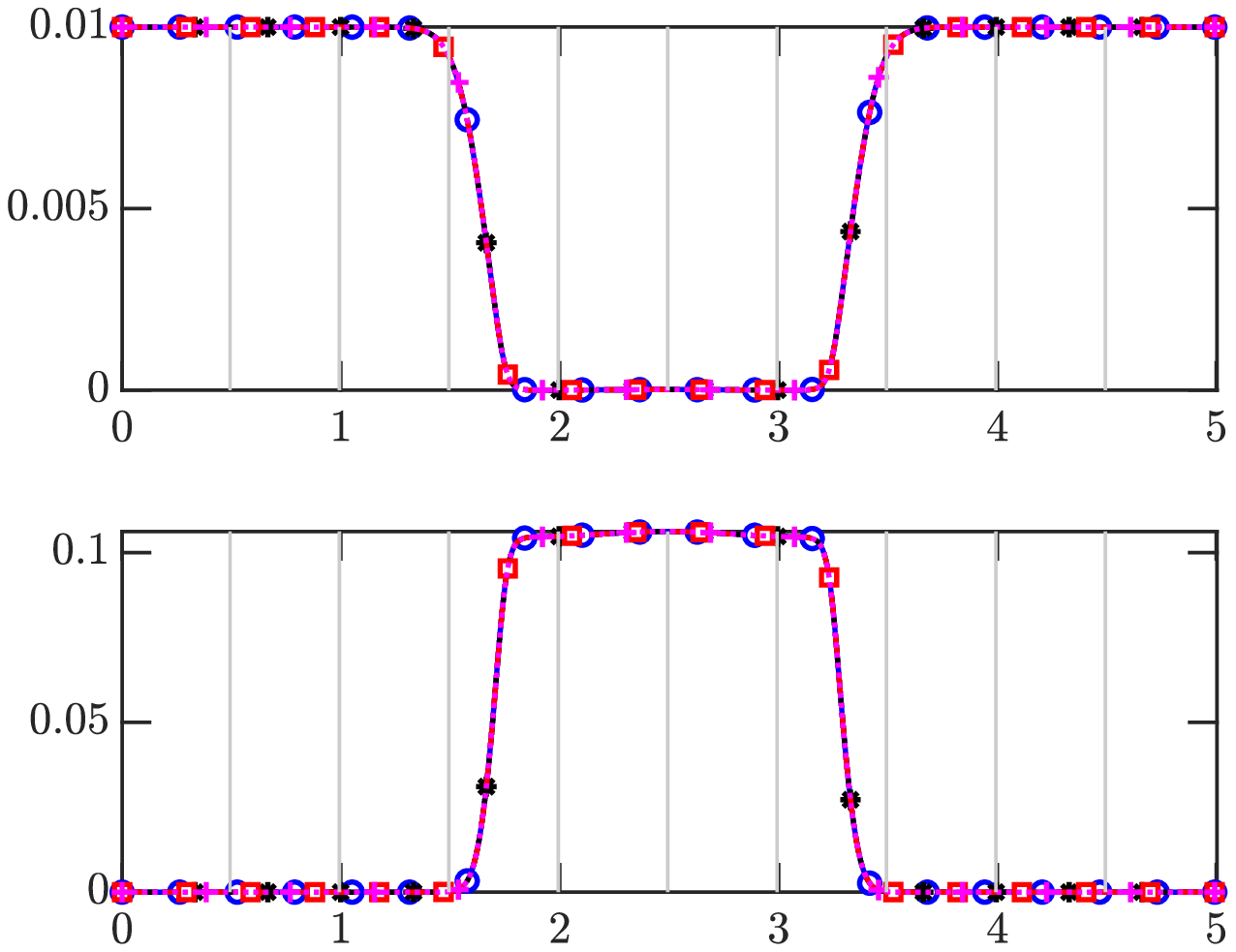}}
\hspace{0.3cm}\\
\begin{picture}(0,0)
        \put(-180,380){$t=4.96\times 10^{-5}s$}
        \put(-25,380){$t=5.52\times 10^{-5}s$}
        \put(135,380){$t=1.5\times 10^{-4}s$}
        \put(-255,344){$T$}
        \put(-255,284){$Q$}
        \put(-255,220){$P$}
        \put(-255,162){$U$}
        \put(-255,98){$Y_{H_2}$}
        \put(-260,40){$Y_{H_2O}$}
          \put(-160,0){$x~(mm)$}
          \put(-10,0){$x~(mm)$}
          \put(150,0){$x~(mm)$}
        \end{picture}
\caption{Asynchrony-tolerant (AT): auto-ignition of a premixed $H_2$/air mixture: instantaneous profiles of
temperature ($K$), heat-release rate ($Jm^{-3}s^{-3}$), pressure ($atm$), velocity ($ms^{-1}$), and mass fractions of
$Y_{H_2}$ and $Y_{H_2O}$.
The different lines: blue (Set-1), black (Set-2), red (Set-3) and magenta (Set-4) are defined in \rtab{dels}. The light blue line indicates the initial condition and the gray vertical lines represent processor boundaries.}
\figlabel{AIpbc}
\end{center}
\end{figure}

To investigate the effect of delays at processor boundaries on important quantities of interest
both before and after
spontaneous ignition,
the spatial evolution of temperature,
velocity, pressure, heat release, and
mass fractions of hydrogen and water are
shown in \rfig{AIpbc}. The time instants considered are indicated with dashed-black lines in \rfig{tpmaxAI}. Qualitatively, no difference exists between the instantaneous
values of these quantities for synchronous simulation and asynchronous simulations with AT schemes.
The $L_1$ and $L_\infty$ norm of error computed with respect to
the synchronous simulation is listed in \rtab{aiL1} and \rtab{aiLinf}, respectively, for
three time instants. Also listed are the errors
when the standard schemes are used
asynchronously. Note that in this case the errors are several orders of magnitude larger than AT schemes, irrespective
of the quantity. Moreover, for the former case, the maximum error
in temperature is approximately 9K which can trigger reactions and
result in nonphysical ignition in otherwise quiescent flow due to the strong temperature
dependence of reaction rates. Such numerical errors are not observed
for AT schemes even when the gradients, for example, in temperature
or species mass fractions exist at the processing element boundaries.

\begin{figure}[h!]
\begin{center}
\subfigure{\includegraphics[trim={0cm 0cm 0cm 0cm},clip,width=0.31\textwidth]{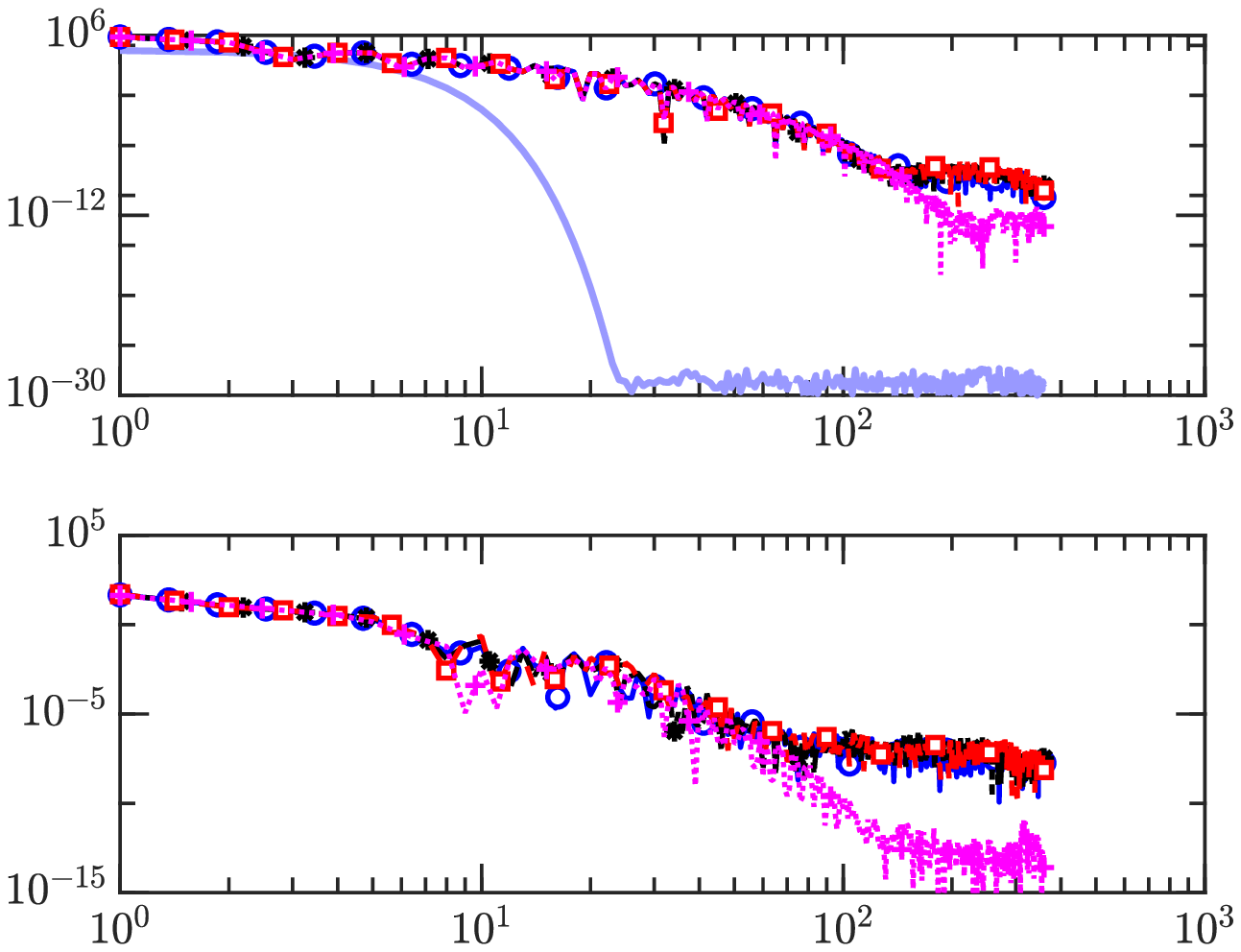}}
\hspace{1cm}
\subfigure{\includegraphics[trim={0cm 0cm 0cm 0cm},clip,width=0.31\textwidth]{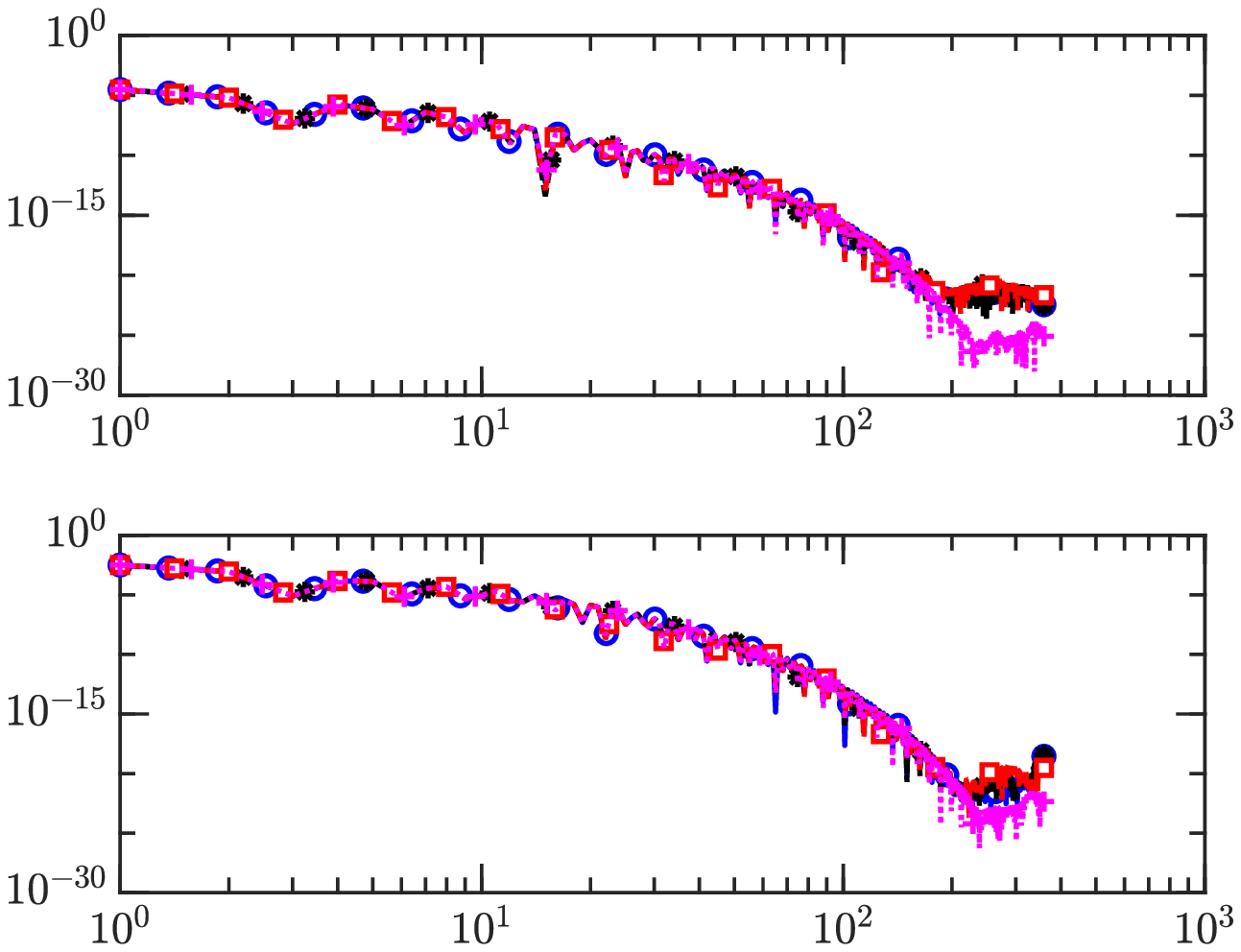}}
\begin{picture}(0,0)
         \put(-250,-10){$\ka$}
         \put(-75,-10){$\ka$}
        \put(-360,85){$\hat{T}(\ka)$}
        \put(-360,25){$\hat{U}(\ka)$}
        \put(-180,85){$\hat{Y}_{H_2}(\ka)$}
        \put(-180,15){$\hat{Y}_{H_2O}(\ka)$}
\end{picture}
\caption{AS-SFD: spectra of temperature ($\hat{T}(\ka)$),
velocity ($\hat{U}(\ka)$), and mass fractions $\hat{Y}_{H_2}(\ka)$ and $\hat{Y}_{H_2O}(\ka)$ for auto-ignition of a premixed $H_2$/air mixture. The different lines: blue (Set-1), black (Set-2), red (Set-3) and magenta (Set-4) are defined in \rtab{dels}.}
\figlabel{AIpbcSpecAS}
\end{center}
\end{figure}

\begin{figure}[h!]
\begin{center}
\subfigure{\includegraphics[trim={0cm 0cm 0cm 0cm},clip,width=0.31\textwidth]{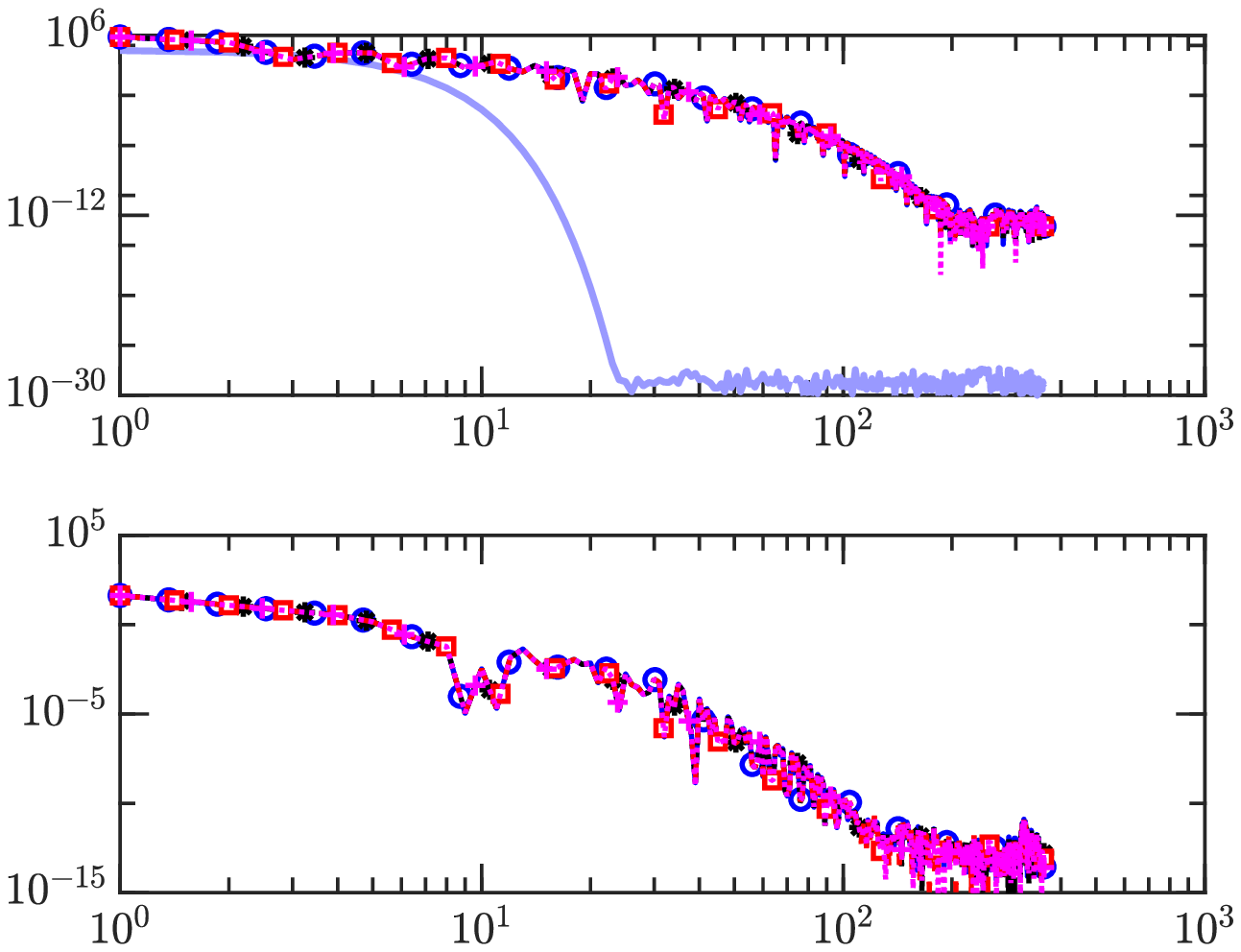}}
\hspace{1cm}
\subfigure{\includegraphics[trim={0cm 0cm 0cm 0cm},clip,width=0.31\textwidth]{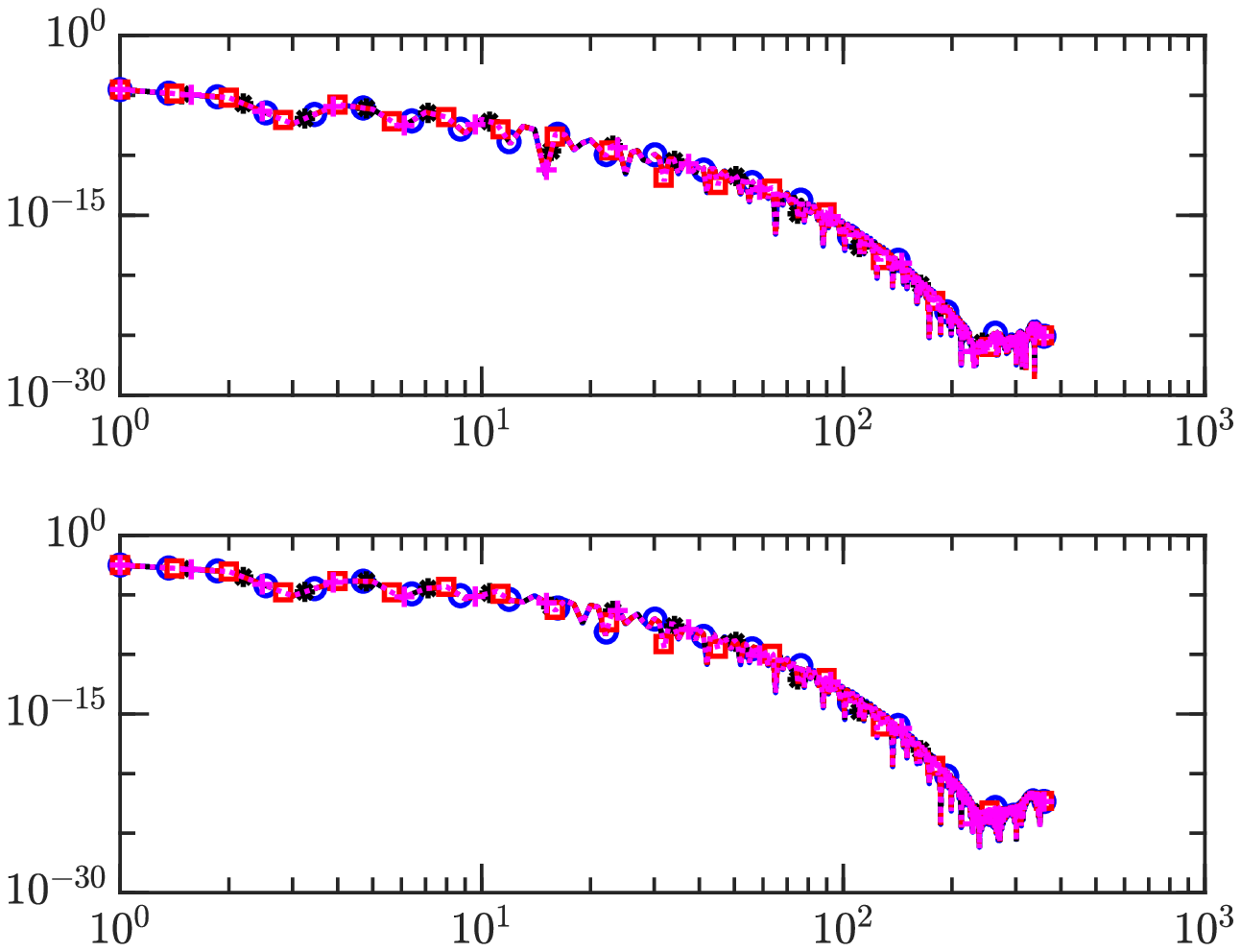}}
\begin{picture}(0,0)
         \put(-250,-10){$\ka$}
         \put(-75,-10){$\ka$}
        \put(-360,85){$\hat{T}(\ka)$}
        \put(-360,25){$\hat{U}(\ka)$}
        \put(-180,85){$\hat{Y}_{H_2}(\ka)$}
        \put(-180,15){$\hat{Y}_{H_2O}(\ka)$}
\end{picture}
\caption{Asynchrony-tolerant (AT): spectra of temperature ($\hat{T}(\ka)$),
velocity ($\hat{U}(\ka)$), and mass fractions $\hat{Y}_{H_2}(\ka)$ and $\hat{Y}_{H_2O}(\ka)$ for auto-ignition of premixed $H_2$. The different lines: blue (Set-1), black (Set-2), red (Set-3) and magenta (Set-4) are defined in \rtab{dels}.}
\figlabel{AIpbcSpecAT}
\end{center}
\end{figure}

The spectra of
temperature, velocity and mass fraction of reactants and products,
are shown in \rfig{AIpbcSpecAS} and \rfig{AIpbcSpecAT}.
The AT schemes exhibit
excellent agreement with the synchronous simulations at
all wavenumbers ($\kappa$). Moreover, non-physical accumulation of energy at high wavenumbers
due to numerical errors does not exist which is in fact observed in \rfig{AIpbcSpecAS} when the standard
schemes are used with asynchrony.
Hence, AT schemes demonstrate an excellent resolving efficiency in both physical and
spectral space despite delayed data being used for the computation of
derivatives at processing element boundaries.

\begin{table}
\centering
\begin{tabular}{ |c|c|c|c|c|c|c| }
 \hline
  Time & Case & $U~(ms^{-1})$& $P~(atm)$&$T~(K)$ & $H_2$ & $H_2O$ \\
  \hline
  \hline
 \multicolumn{7}{|c|}{AS-SFD}\\
 \hline
        &1 & 2.4294e-01 &  8.1737e-04  & 3.2713e-01 &  1.2700e-06  & 2.5152e-05\\
  1.5e-4  &2 & 3.9893e-01 &  1.3420e-03 &  5.3992e-01 &  2.1059e-06 &  4.1778e-05\\
        &3 & 5.6643e-01  & 1.9061e-03  & 7.6244e-01 &  2.9054e-06  & 5.8324e-05  \\
        \hline
 \hline
   \multicolumn{7}{|c|}{Asynchrony-tolerant (AT)}\\
  \hline
        &1 & 1.07e-07 &  3.16e-10  & 8.50e-08 &  1.89e-13 & 2.29e-12\\
4.96e-5  &2 &  1.23e-07 &  3.68e-10  & 8.19e-08 &  2.96e-13&   2.31e-12\\
        &3 & 1.33e-07 &  4.21e-10 &  1.11e-07  &4.22e-13&   3.66e-12 \\
 \hline
           &1 & 6.80e-07  & 2.18e-09 &  7.14e-07&   2.49e-12 &  2.53e-11 \\
  7.2e-5 &2 & 6.46e-07  & 2.20e-09 &  9.74e-07  & 3.77e-12  & 4.89e-11  \\
          &3 &  1.07e-06 &  3.52e-09  & 1.09e-06  & 2.79e-12 &  3.67e-11
 \\
 \hline
        &1 &  1.17e-04 &  2.70e-07  & 7.67e-04  & 4.56e-09  & 4.87e-08  \\
  1.5e-4  &2 &  5.67e-05&   1.49e-07&   2.52e-04 &  1.54e-09&   1.61e-08 \\
        &3 &1.10e-04  & 2.70e-07 &  7.17e-04  & 4.33e-09   &4.62e-08  \\
 \hline
\end{tabular}
\caption{$L_1$ norm of error in temperature, velocity, and mass fraction of
$H_2$ and $H_2O$ for auto-ignition of premixed hydrogen.}
\tablabel{aiL1}
\end{table}

\begin{table}
\centering
\begin{tabular}{ |c|c|c|c|c|c|c| }
 \hline
  Time & Case & $U~(ms^{-1})$& $P~(atm)$&$T~(K)$ & $H_2$ & $H_2O$ \\
  \hline
   \hline
 \multicolumn{7}{|c|}{AS-SFD}\\
 \hline
        &1 & 8.73e-01  & 2.72e-03 &  3.50e+00 &  1.80e-05 &  3.19e-04  \\
  1.5e-4  &2 &1.51e+00 &  4.46e-03  & 5.78e+00 &  2.96e-05 &  5.25e-04  \\
        &3 &  2.14e+00  & 6.23e-03  & 8.79e+00 &  4.41e-05 &  7.88e-04 \\
  \hline
  \hline
   \multicolumn{7}{|c|}{Asynchrony-tolerant (AT)}\\
  \hline
        &1 &1.30e-06  & 2.44e-09  & 1.48e-06  & 4.61e-12   &6.60e-11\\
  2e-5  &2 & 1.64e-06 &  2.30e-09 &  2.07e-06 &  1.07e-11  & 9.12e-11   \\
        &3 & 1.83e-06 &  2.80e-09 &  1.18e-06 &  1.19e-11  & 9.79e-11 \\
 \hline
       &1 & 7.55e-06  & 2.64e-08  & 9.52e-06&   7.27e-11  & 8.87e-10   \\
  5.52e-5 &2 & 5.70e-06  & 1.95e-08 &  1.88e-05 &  1.36e-10 &  1.90e-09 \\
          &3 &  9.45e-06 &  3.26e-08&   1.19e-05&   9.30e-11  & 1.10e-09  \\
 \hline
        &1 &  1.08e-03  & 2.44e-06  & 1.14e-02  & 5.51e-08  & 9.58e-07 \\
  1.5e-4  &2 &  4.08e-04 &  7.97e-07&   3.60e-03 &  1.77e-08&   3.05e-07   \\
        &3 & 8.06e-04   &1.75e-06  & 1.07e-02   &5.12e-08 &  8.93e-07 \\
 \hline
\end{tabular}
\caption{$L_{\infty}$ norm of error in temperature, velocity, and mass fraction of
$H_2$ and $H_2O$ for auto-ignition of premixed hydrogen.}
\tablabel{aiLinf}
\end{table}

\subsection{Auto-ignition: temperature fluctuations at the inflow boundary}
The effect of turbulent
fluctuations on fuel-air mixing, wrinkling of flames or spontaneous ignition front propagation is central to turbulent combustion. To simulate this effect in one-dimension and to ensure that the AT schemes can propagate  fluctuations accurately across processing element boundaries, a temperature perturbation is forced at the left inflow boundary. The initial condition, shown by the light blue line
in \rfig{AI2}, is the steady state solution obtained from the auto-ignition of premixed $C_2H_4$/air mixture at a pressure of $2~atm$. A 22 species 18-step reduced mechanism describing the oxidation kinetics of ethylene/air~\cite{luo2012chemical} is used.
The steady solution is perturbed with sinusoidal temperature
fluctuations with a magnitude of $180K$ and frequency $20Khz$
at the left boundary using an oscillatory inflow boundary condition
\cite{yooCharacteristicBoundaryConditions2005,yooCharacteristicBoundaryConditions2007}.
As the perturbations approach the igniting front, the adjacent temperature increases as is evident from the increase in the mass fraction of $CH_2O$ radical
in \rfig{AI2}. This induces a secondary ignition kernel to the left of the
initial front as can be seen in \rfig{AI2} (left column).
The two kernels eventually interact and as time progresses,
a steady ignition front develops to the left of the initial kernel
as shown in the plots in the right column of \rfig{AI2}.
This front oscillates about
a mean location, with the peak temperature and pressure also oscillating
in response to the incoming sinusoidal fluctuations.
The AT schemes accurately capture the transient
and steady state evolution of
both the temperature and intermediate species even in the
presence of
delays at the processor boundaries. An excellent qualitative
agreement between the instantaneous profiles for AT and synchronous
simulations in \rfig{AI2} is observed. Both $L_1$ and $L_{\infty}$ norms of the errors tabulated in \rtab{l1AI2} and \rtab{linfAI2} are also
reasonably small at both times.

\begin{table}
\centering
\begin{tabular}{ |c|c|c|c|c|c|c|c| }
 \hline
  Time & Case & $U~(ms^{-1})$& $P~(atm)$&$T~(K)$& $OH$ & $CH_2O$ &$H_2O$\\
  \hline
   \hline
   \multicolumn{8}{|c|}{Asynchrony-tolerant (AT)}\\
  \hline
          &1 &   3.20e-08 & 1.10e-10 & 1.32e-07 & 6.67e-13 & 3.53e-13 & 4.48e-12\\
  4.48e-4 &2 &   3.89e-08 & 1.27e-10 & 1.70e-07 & 8.75e-13 & 4.82e-13 & 5.91e-12 \\
          &3 &   4.97e-08 & 1.66e-10 & 1.75e-07 & 9.06e-13 & 4.90e-13 & 6.08e-12 \\
 \hline
            &1 & 3.09e-09 & 1.09e-11 & 3.32e-08 & 1.50e-13 & 9.12e-14 & 1.16e-12\\
 9.98e-4       &2 & 3.99e-09 & 1.38e-11 & 1.93e-08 & 1.20e-13 & 4.42e-14 & 6.33e-13  \\
            &3 & 4.77e-09 & 1.69e-11 & 2.82e-08 & 1.69e-13 & 7.69e-14 & 8.95e-13 \\
 \hline
\end{tabular}
\caption{Asynchrony-tolerant (AT): $L_{1}$ norm of error in temperature, velocity, and mass fraction of
$OH$ and $CH_2O$ for auto-ignition of $C_2H_4/air$ flame with temperature
fluctuations at the inflow.}
\tablabel{l1AI2}
\end{table}

\begin{table}
\centering
\begin{tabular}{ |c|c|c|c|c|c|c|c| }
 \hline
  Time & Case & $U~(ms^{-1})$& $P~(atm)$&$T~(K)$ & $OH$ & $CH_2O$& $H_2O$ \\
  \hline
   \hline
   \multicolumn{8}{|c|}{Asynchrony-tolerant (AT)}\\
  \hline
        &1 &  2.10e-07 & 8.22e-10 & 1.08e-06 & 5.76e-12 & 4.36e-12 & 5.12e-11\\
  4.48e-4 &2 & 2.20e-07 & 9.66e-10 & 1.53e-06 & 9.46e-12 & 6.58e-12 & 7.44e-11 \\
        &3 &  3.46e-07 & 1.30e-09 & 1.33e-06 & 8.23e-12 & 5.76e-12 & 6.54e-11 \\
 \hline
         &1 & 2.05e-08 & 7.70e-11 & 5.33e-07 & 2.81e-12 & 2.15e-12 & 2.58e-11\\
1.9e-3   &2 & 2.78e-08 & 1.01e-10 & 3.13e-07 & 1.68e-12 & 1.26e-12 & 1.50e-11 \\
         &3 & 3.57e-08 & 1.28e-10 & 4.41e-07 & 2.42e-12 & 1.83e-12 & 2.14e-11 \\
 \hline
\end{tabular}
\caption{Asynchrony-tolerant (AT): $L_{\infty}$ norm of error in temperature, velocity, and mass fraction of
$OH$ and $CH_2O$ for auto-ignition of $C_2H_4/air$ flame with temperature fluctuations
at the inflow.}
\tablabel{linfAI2}
\end{table}

\begin{figure}[h!]
\begin{center}
\subfigure{\includegraphics[trim={0cm 0cm 0cm 0cm},clip,width=0.32\textwidth]{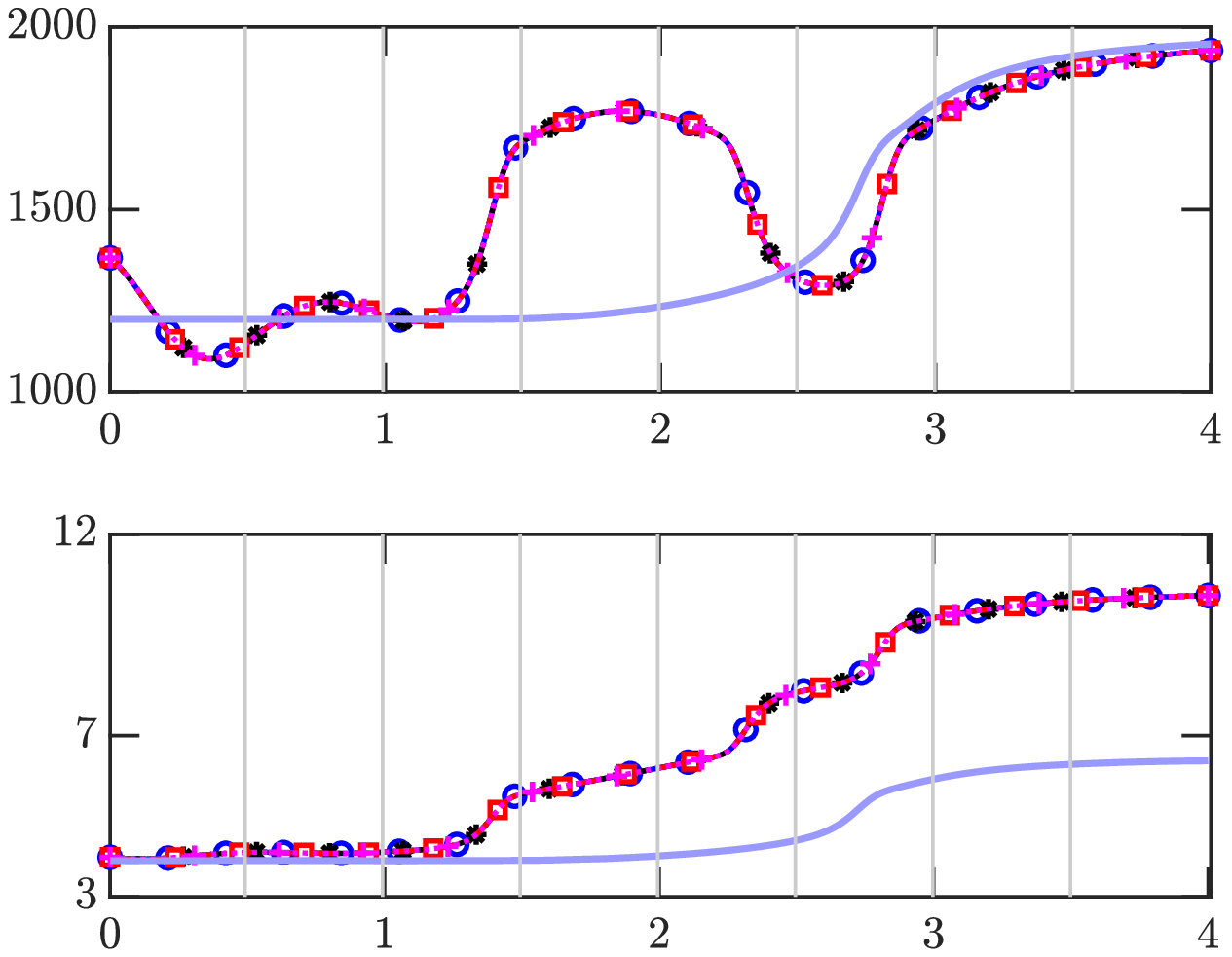}}
\hspace{1cm}\subfigure{\includegraphics[trim={0cm 0cm 0cm 0cm},clip,width=0.32\textwidth]{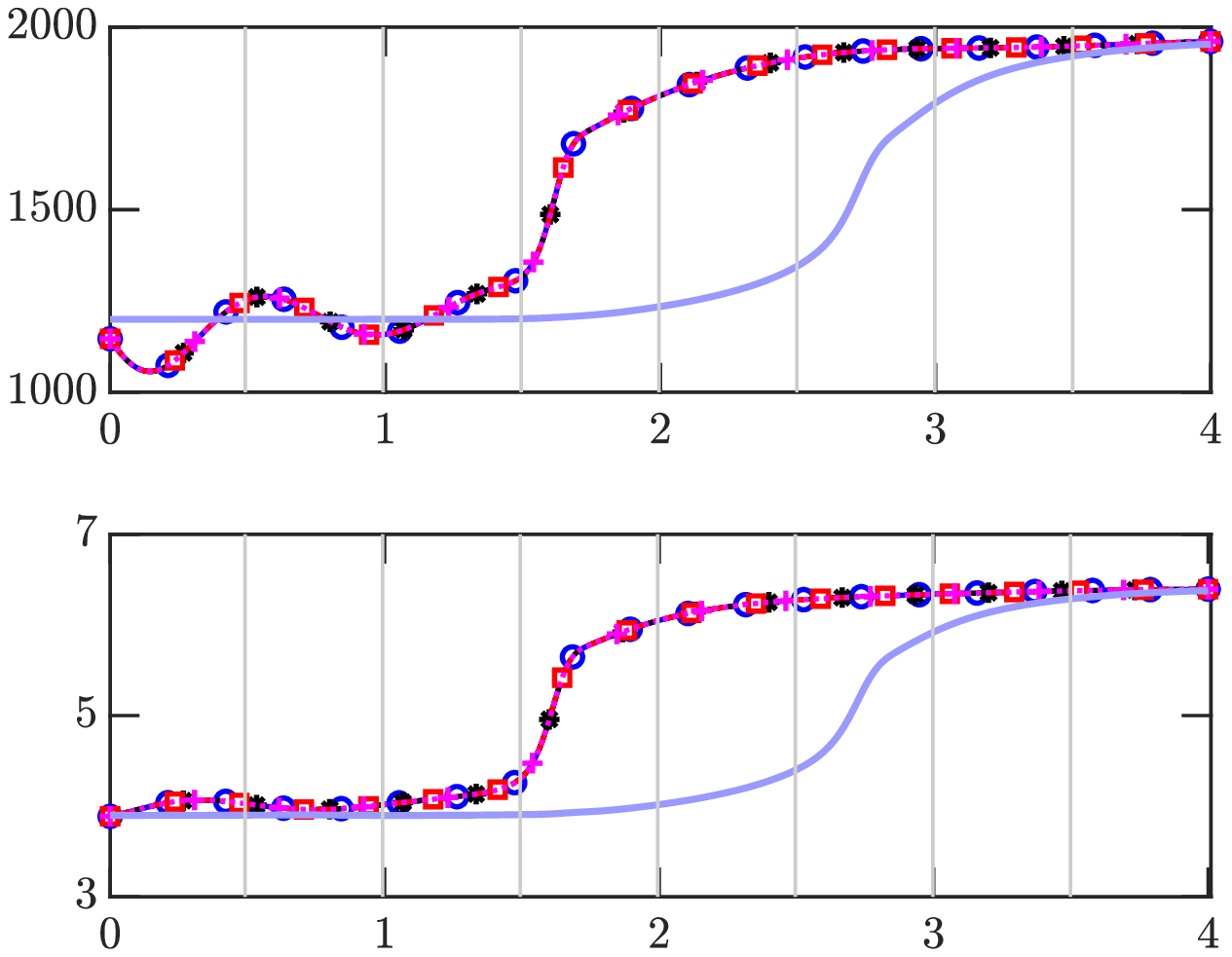}}
\hspace{1cm}
\\
\hspace{2mm}
\subfigure{\includegraphics[trim={0cm 0cm 0cm 0cm},clip,width=0.3\textwidth]{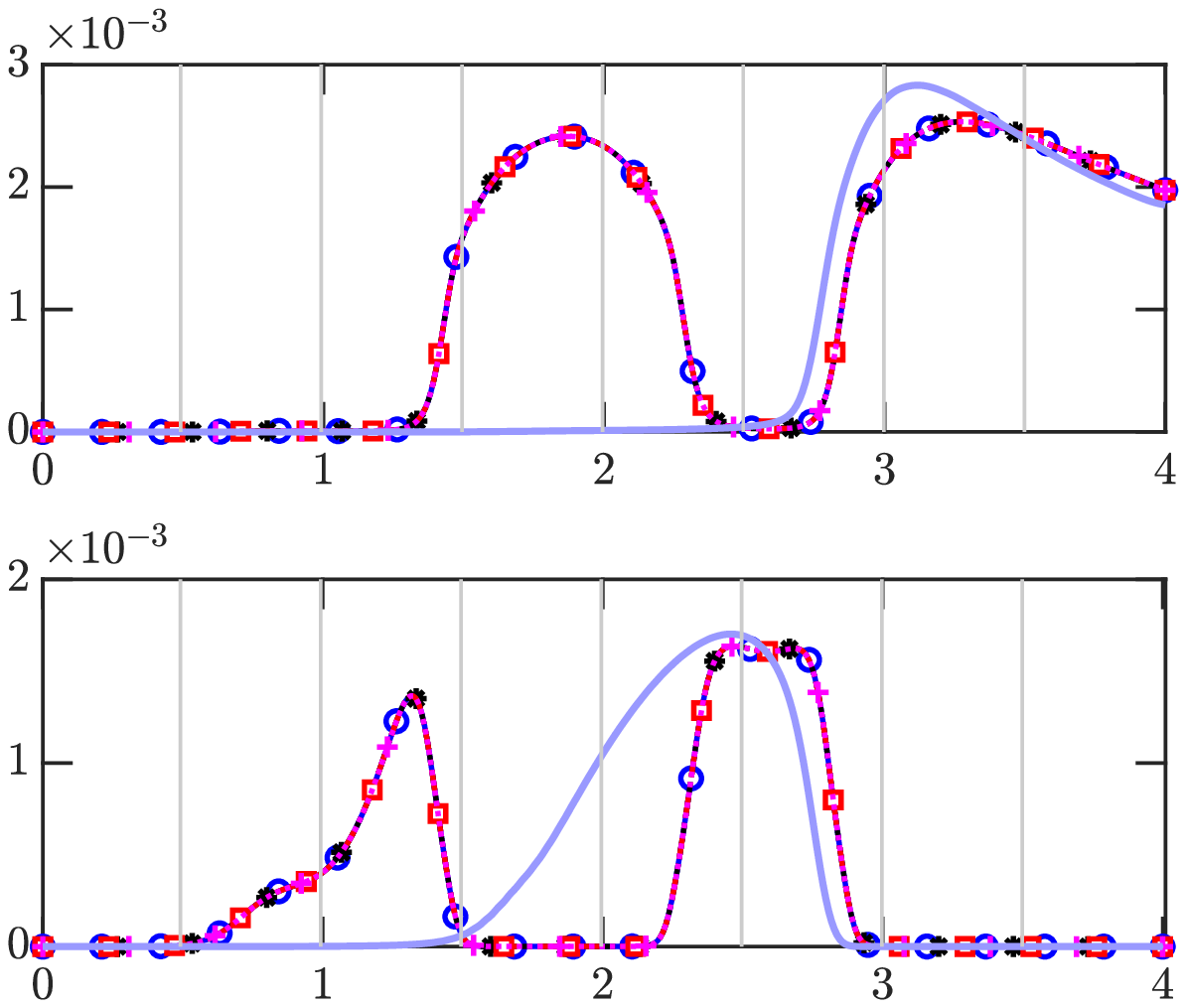}}
\hspace{1.4cm}\subfigure{\includegraphics[trim={0cm 0cm 0cm 0cm},clip,width=0.3\textwidth]{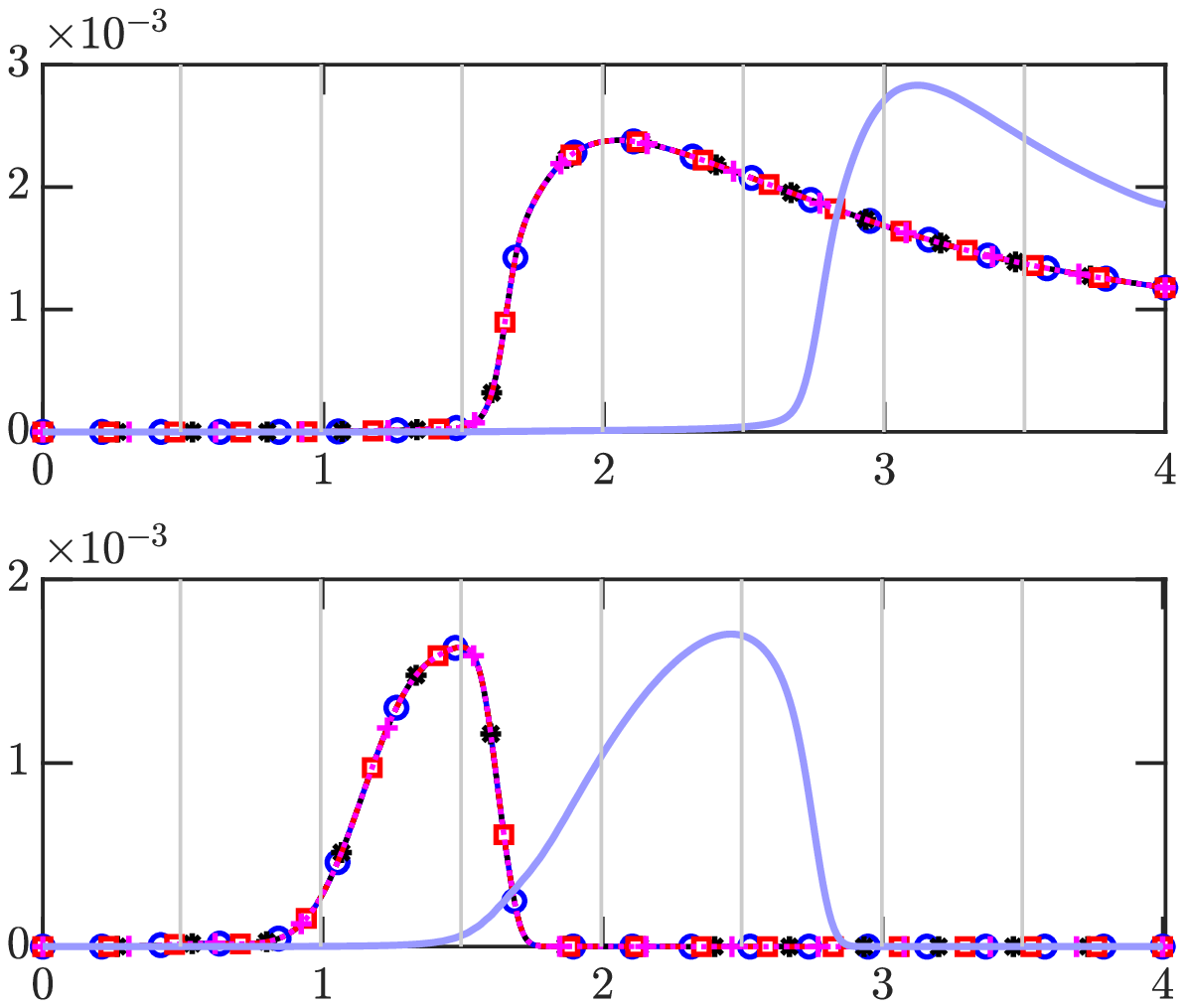}}
\begin{picture}(0,0)
         \put(-260,-10){$x~(mm)$}
         \put(-85,-10){$x~(mm)$}
        \put(-360,210){$T$}
        \put(-360,150){$U$}
        \put(-360,85){$Y_{OH}$}
        \put(-360,25){$Y_{CH_2O}$}
        \put(-285,245){$t=4.48\times 10^{-4}s$}
        \put(-105,245){$t=9.98\times 10^{-4}s$}
        \end{picture}
\caption{Asynchrony-tolerant (AT): instantaneous profiles of temperature ($K$),
velocity ($ms^{-1}$), and mass fractions of $OH$ and $CH_2O$ radicals for autoignition of
premixed $C_2H_4$/air mixtures with temperature fluctuations at the inflow. The different lines: blue (Set-1), black (Set-2), red (Set-3) and magenta (Set-4) are defined in \rtab{dels}. The light blue line indicates the initial condition and gray vertical lines represent processor boundaries.}
\figlabel{AI2}
\end{center}
\end{figure}

\subsection{Premixed flame propagation}
\begin{figure}[h!]
\begin{center}
\subfigure{\includegraphics[trim={0cm 0cm 0cm 0cm},clip,width=0.31\textwidth]{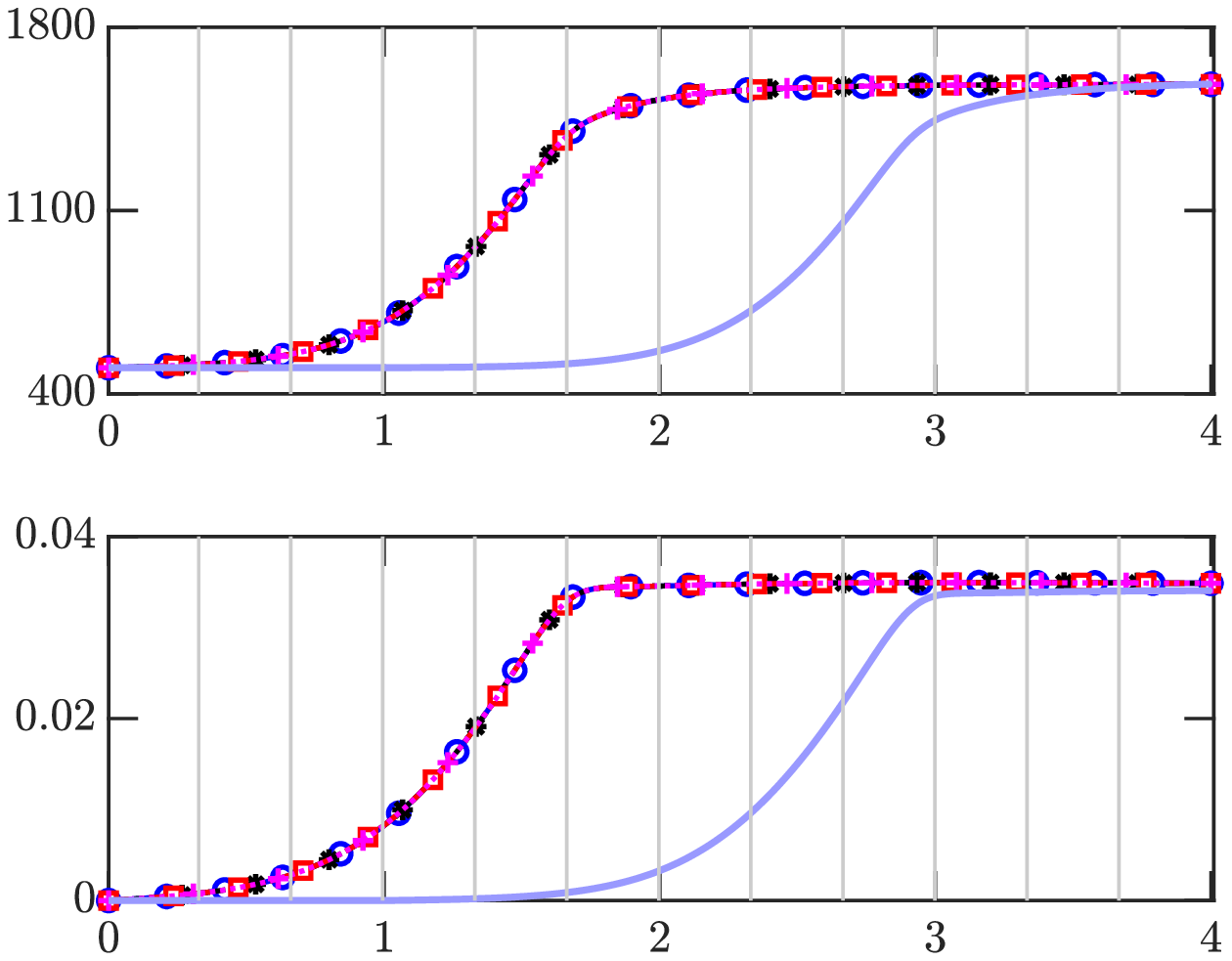}}
\hspace{0.3cm}
\subfigure{\includegraphics[width=0.31\textwidth]{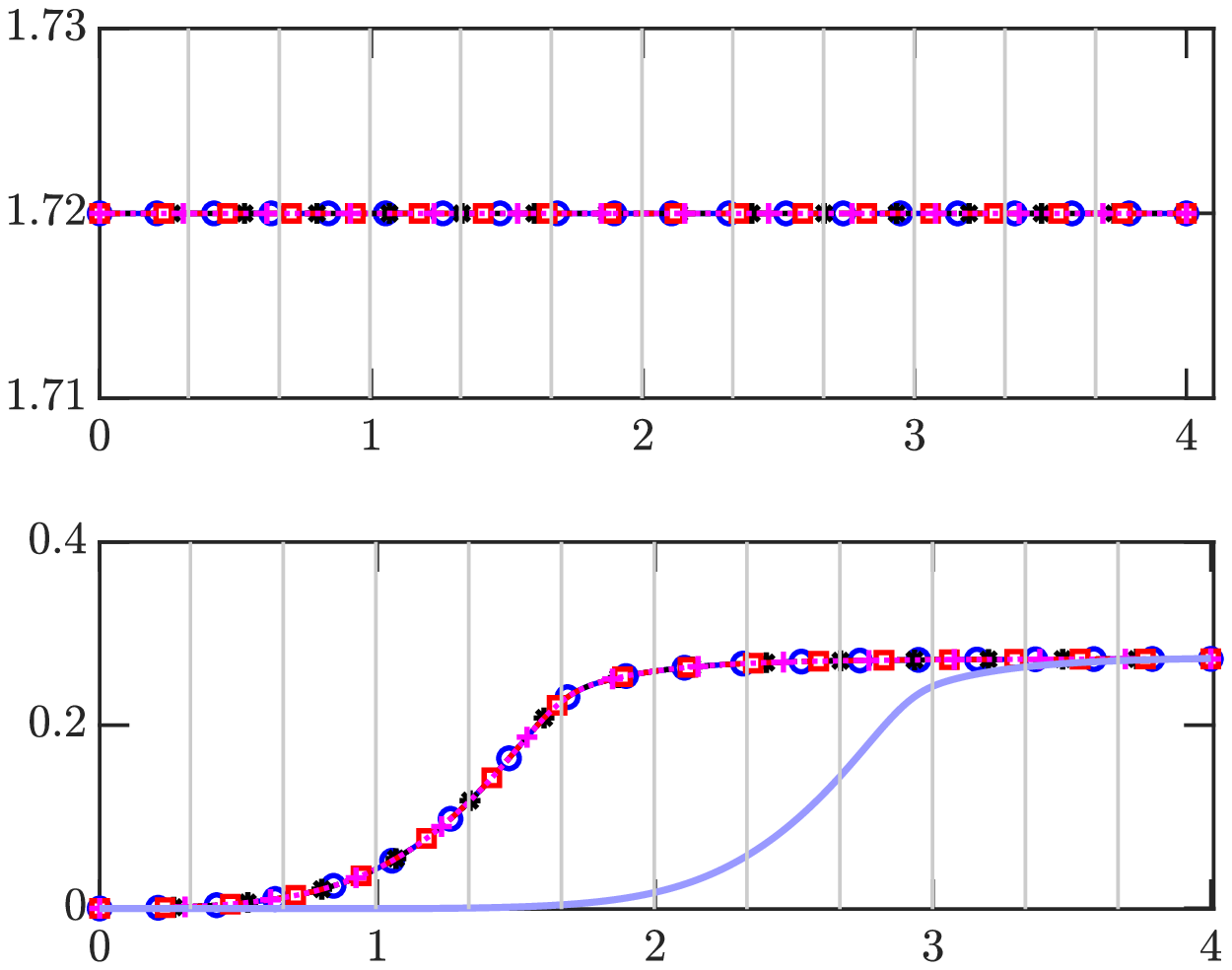}}
\hspace{0.3cm}
\subfigure{\includegraphics[width=0.30\textwidth]{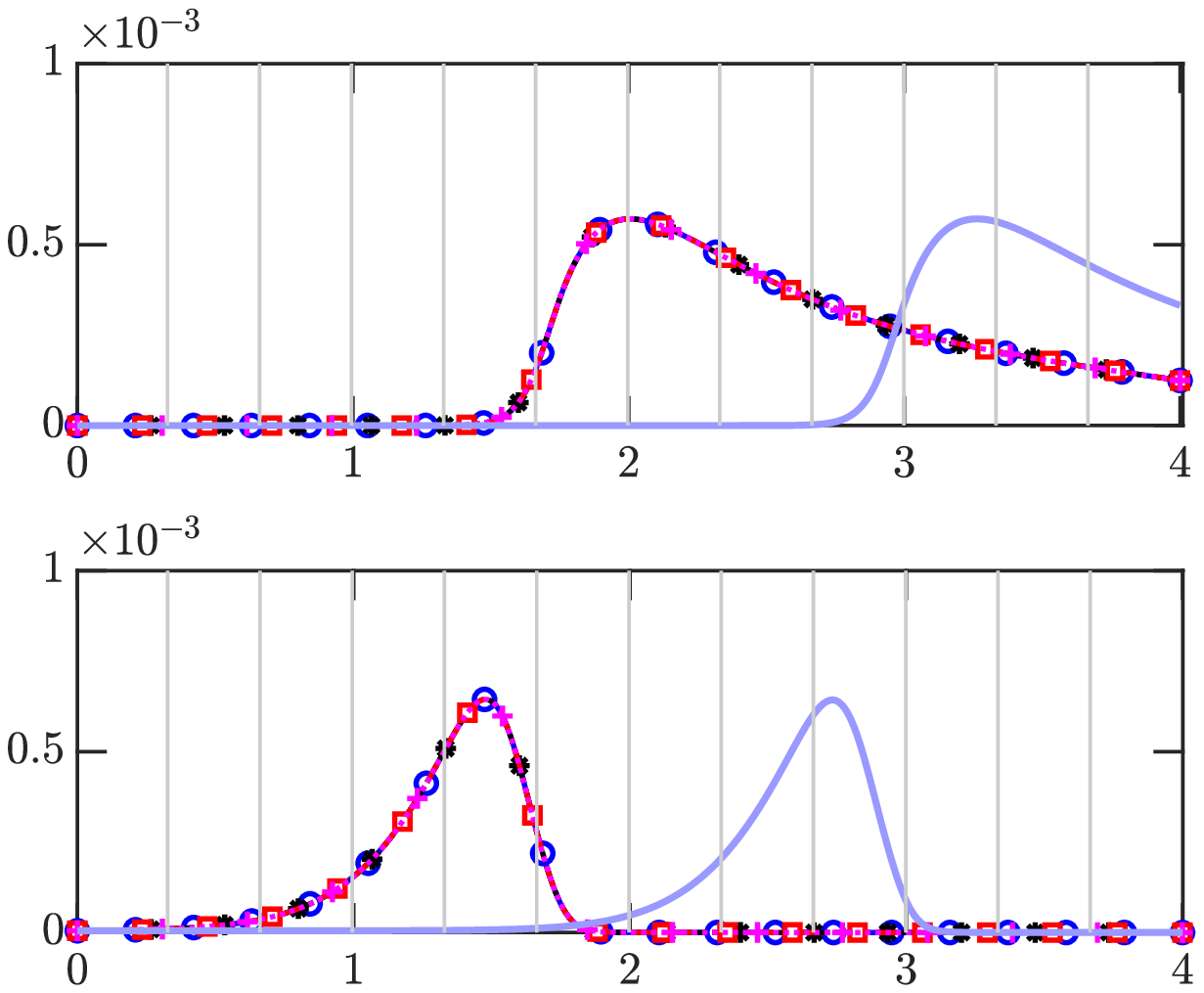}}
\begin{picture}(0,0)
        \put(-80,115){$Y_{OH}$}
        \put(-230,115){$P$}
        \put(-390,115){$T$}
        \put(-397,53){$Y_{H_2O}$}
        \put(-230,53){$U$}
        \put(-85,53){$Y_{CH_2O}$}
         \put(-88,-10){$x~(mm)$}
        \put(-238,-10){$x~(mm)$}
        \put(-398,-10){$x~(mm)$}
        \end{picture}
\caption{Asynchrony-tolerant (AT): instantaneous profiles of temperature ($K$), pressure ($atm$), velocity ($ms^{-1}$), and mass fractions of product, $H_2O$, and radicals, $OH$ and $CH_2O$, at $t=1\times10^{-2}s$
for premixed $C_2H_4$ flame propagation. The different lines: blue (Set-1), black (Set-2), red (Set-3) and magenta (Set-4) are defined in \rtab{dels}. The light blue line indicates the initial condition and gray vertical lines represent processor boundaries.
}
\figlabel{pm1}
\end{center}
\end{figure}

The previous cases
focused on investigation of the effect of
asynchrony on spontaneous ignition dominated
by unsteadiness and advection-reaction balance.
The present test case is intended to study the effect of
delayed data on laminar
premixed flame propagation. The flame
propagates at a subsonic velocity and is characterized by
a balance between the reactive and diffusive
terms in the
steady species conservation equations.
A mixture of $C_2H_4$/air at an equivalence ratio, $\phi=0.42$, $T=500~K$ and
$P=1.72~atm$ is considered. The one-dimensional domain is $4~mm$ and
is discretized with $576$ grid points that
are distributed across 12 processors. The 22 species, 18-step reduced mechanism from \cite{luo2012chemical} for ethylene/air used in the earlier case is also used here. The initial condition is generated using an auto-ignition case, and non-reflecting inflow and subsonic outflow are imposed at the left and right boundaries, respectively.
The flame is initially located close to the right edge of the domain and in time propagates to the left while consuming the reactants mixture.
The initial flame front, comprised of preheat and reaction zones,
spans across three PEs as shown by the light blue line
\rfig{pm1}. This flame
traverses across  multiple processing element
boundaries and the errors are computed to assess the effect of
the asynchronous data encountered at the boundaries on flow and flame quantities.

The flame structure is invariant with time, therefore,
only the instantaneous
profiles and errors at the time instant when the
flame approaches near the left boundary is considered
for accuracy analysis.
There is excellent agreement between the spatial
profiles for the synchronous and the AT simulations, as shown in \rfig{pm1}
for temperature, velocity as well as major and minor species.
The drop in pressure in the reaction zone
is negligibly small and thus the pressure values
depicted in \rfig{pm1} are nearly constant. As is
evident from \rtab{pml}, both $L_1$ and $L_{\infty}$
norms of the error between synchronous and AT simulations are
negligibly small and as low as $\mathcal{O}(10^{-14})$
for some of the species.
Furthermore, the flame speed computed from the time evolution of the location of
the peak heat release rate is equal to $12.59~cm/s$ for both synchronous
and asynchronous simulations. The thermal flame
thickness is equal to $\delta_T=\frac{T_2-T_1}{\max(\partial T/\partial x)}=7.81\times10^{-4}m$, irrespective of the delays.

\begin{table}
\centering
\begin{tabular}{ |c|c|c|c|c|c|c|c| }
 \hline
  Time & Case &$U~(ms^{-1})$& $P~(atm)$&$T~(K)$ & $OH$ & $CH_2O$ & $H_2O$\\
  \hline
   \hline
   \multicolumn{8}{|c|}{$L_1$ error}\\
  \hline
        & 1 & 6.22e-11 & 2.32e-13 & 2.72e-09 & 2.51e-15 & 3.21e-15 & 8.60e-14  \\
  1e-2  & 2 & 9.96e-11 & 3.81e-13 & 7.12e-10 & 8.73e-16 & 8.04e-16 & 2.04e-14\\
        & 3 & 9.42e-11 & 3.61e-13 & 1.27e-09 & 1.16e-15 & 1.35e-15 & 3.42e-14  \\
 \hline
 \hline
    \multicolumn{8}{|c|}{$L_{\infty}$ error}\\
  \hline
        & 1 & 5.72e-10 & 2.07e-12 & 1.42e-08 & 3.02e-14 & 4.66e-13 & 2.38e-14 \\
  1e-2  & 2 & 1.06e-09 & 3.07e-12 & 4.18e-09 & 8.81e-15 & 1.23e-13 & 6.31e-15  \\
        & 3 & 6.52e-10 & 3.36e-12 & 6.61e-09 & 1.42e-14 & 2.09e-13 & 1.19e-14  \\
 \hline
\end{tabular}
\caption{Asynchrony-tolerant (AT): $L_{1}$ and $L_{\infty}$ norm of error in velocity, pressure, temperature, and mass fractions of
$OH$, $CH_2O$ and $H_2O$ for premixed $C_2H_4$ flame propagation.}
\tablabel{pml}
\end{table}


\subsection{Non-premixed ignition}
\begin{figure}[h!]
\begin{center}
\includegraphics[width=0.32\textwidth]{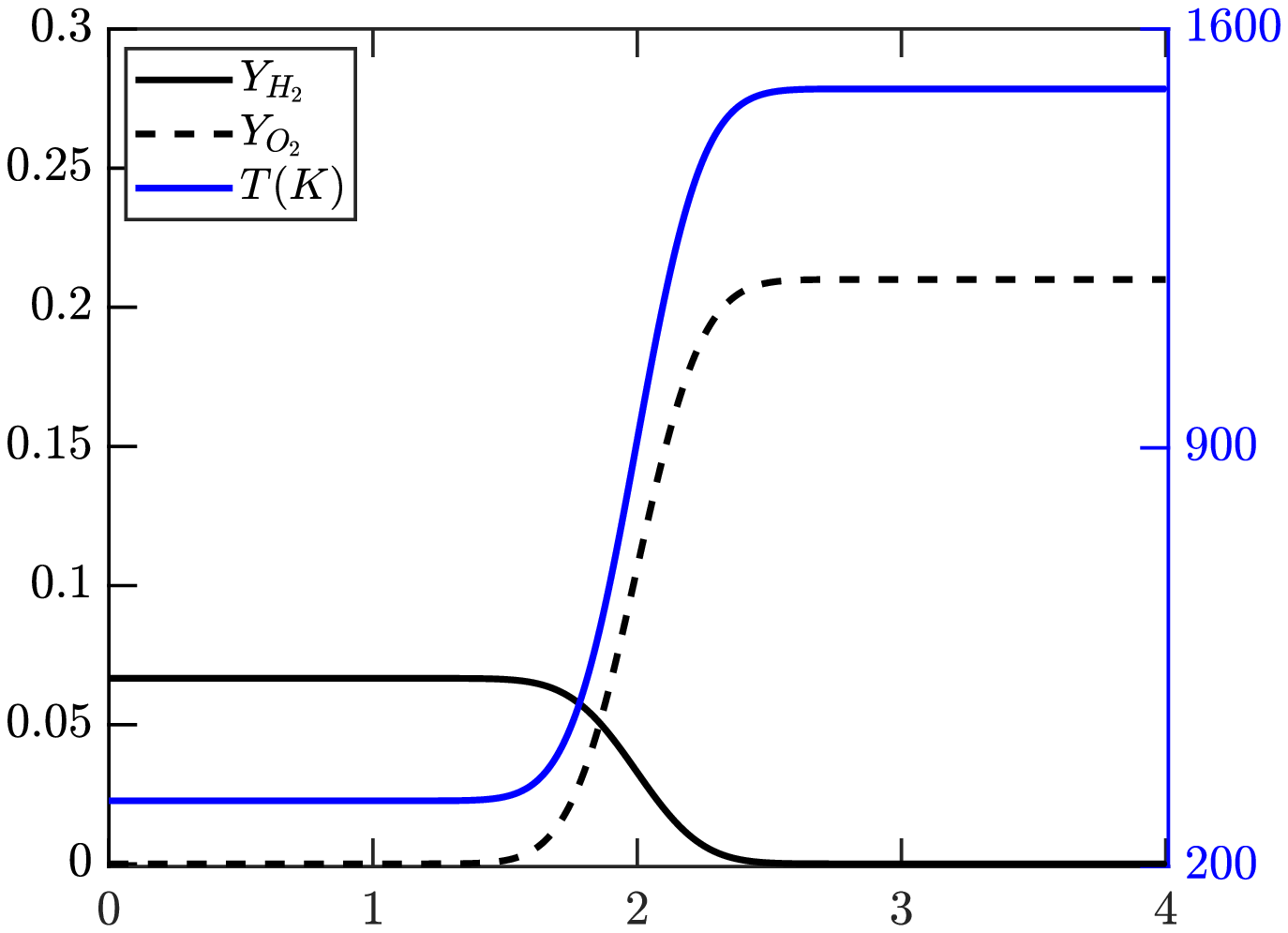}
\begin{picture}(0,0)
        \put(-170,55){$Y$}
        \put(5,55){$T$}
        \end{picture}
\caption{Initial condition for non-premixed ignition with diluted fuel on the left and vitiated air on the right.}
\figlabel{diffIC}
\end{center}
\end{figure}
An ignition of non-premixed $H_2$ using
Burke's mechanism~\cite{burke2012comprehensive} with 9 species is considered next. The setup is similar to the one used in~\cite{baum1993} and~\cite{sutherlandImprovedBoundaryConditions2003}
with fuel diluted with nitrogen on the left and
air heated to $1500K$ on the right, as shown in \rfig{diffIC}.
An inflow boundary condition
from~\cite{yooCharacteristicBoundaryConditions2005} is used on the left boundary with
zero velocity and an outflow boundary condition is used on the right boundary
with appropriate viscous conditions on each side. The
one-dimensional domain is 4 $mm$ in length,  discretized into 576 grid-points that are
distributed across 12 PEs.

As time progresses and diffusion process tends
to homogenize the gradients, there is ignition close to the
stoichiometric mixture fraction, resulting in a rise in
peak temperature and concentration of
intermediate species.
The instantaneous profiles of key quantities are shown in
\rfig{diff} at two times. The
first series of plots on the left column in
\rfig{diff} are at an earlier time when ignition is localized.
With time both fuel and intermediate
species diffuse across processing element boundaries
and the flame expands, which is shown in the plots on
the right column in
\rfig{diff}. The evolution of both major and
minor species as well as temperature and
velocity is accurately resolved by the AT schemes. The
average and maximum errors, tabulated in \rtab{diffL1} and
\rtab{diffLinf}, respectively, are both negligibly small.
The errors were similarly negligible for other
species that are not shown here.

\begin{table}
\begin{center}
\begin{tabular}{ |c|c|c|c|c|c|c|c|c| }
 \hline
  Time & Case &$U~(ms^{-1})$& $P~(atm)$&$T~(K)$ & $H_2$& $O$ & $H$   &$H_2O$ \\
  \hline
   \hline
   \multicolumn{9}{|c|}{Asynchrony-tolerant (AT)}\\
  \hline
          &1 &1.94e-08 & 4.43e-11 & 1.52e-07 & 6.04e-12 & 3.63e-12 & 4.85e-13 & 2.54e-11 \\
7.92e-5   &2 & 2.59e-08 & 5.94e-11 & 1.97e-07 & 6.72e-12 & 5.40e-12 & 6.60e-13 & 3.74e-11\\
          &3 & 3.14e-08 & 6.44e-11 & 3.13e-07 & 1.12e-11 & 7.77e-12 & 9.38e-13 & 5.42e-11\\
            \hline
          &1 & 5.94e-10 & 1.14e-12 & 2.57e-08 & 6.04e-13 & 2.69e-13 & 1.95e-14 & 4.43e-12  \\
3.19e-4   &2 & 5.25e-10 & 1.06e-12 & 3.45e-08 & 6.98e-13 & 2.55e-13 & 2.22e-14 & 5.91e-12  \\
          &3 & 6.18e-10 & 1.30e-12 & 5.01e-08 & 1.61e-12 & 2.56e-13 & 4.91e-14 & 4.80e-12 \\
 \hline
\end{tabular}
\caption{Asynchrony-tolerant (AT): $L_{1}$ norm of error in temperature, velocity, and mass fractions of
$H_2$, $O$, $H$ and $H_2O$ for non-premixed $H_2/air$ ignition.}
\tablabel{diffL1}
\end{center}
\end{table}

\begin{table}
\begin{center}
\begin{tabular}{ |c|c|c|c|c|c|c|c|c| }
 \hline
  Time & Case &$U~(ms^{-1})$& $P~(atm)$&$T~(K)$ & $H_2$& $O$ & $H$   &$H_2O$ \\
  \hline
   \hline
   \multicolumn{9}{|c|}{Asynchrony-tolerant (AT)}\\
  \hline
          &1 & 1.09e-07 & 1.91e-10 & 6.10e-07 & 3.89e-11 & 1.31e-10 & 2.63e-11 & 4.05e-12 \\
7.92e-5   &2 & 1.11e-07 & 2.25e-10 & 8.52e-07 & 4.83e-11 & 1.85e-10 & 3.69e-11 & 5.24e-12 \\
          &3 &1.21e-07 & 2.14e-10 & 1.25e-06 & 7.16e-11 & 2.73e-10 & 5.27e-11 & 7.10e-12  \\
     \hline
          &1 & 5.08e-09 & 9.80e-12 & 9.89e-08 & 2.38e-12 & 1.65e-11 & 1.50e-12 & 7.53e-14 \\
7.92e-5   &2 & 4.01e-09 & 8.79e-12 & 8.94e-08 & 2.49e-12 & 1.64e-11 & 1.29e-12 & 6.47e-14 \\
          &3 & 4.93e-09 & 7.91e-12 & 1.52e-07 & 5.15e-12 & 1.41e-11 & 1.17e-12 & 1.38e-13  \\
 \hline
\end{tabular}
\caption{Asynchrony-tolerant (AT): $L_{\infty}$ norm of error in temperature, velocity, and mass fractions of
$H_2$, $O$, $H$ and $H_2O$ for non-premixed $H_2/air$ ignition.}
\tablabel{diffLinf}
\end{center}
\end{table}

\begin{figure}[h!]
\begin{center}
\subfigure{\includegraphics[trim={0cm 0cm 0cm 0cm},clip,width=0.31\textwidth]{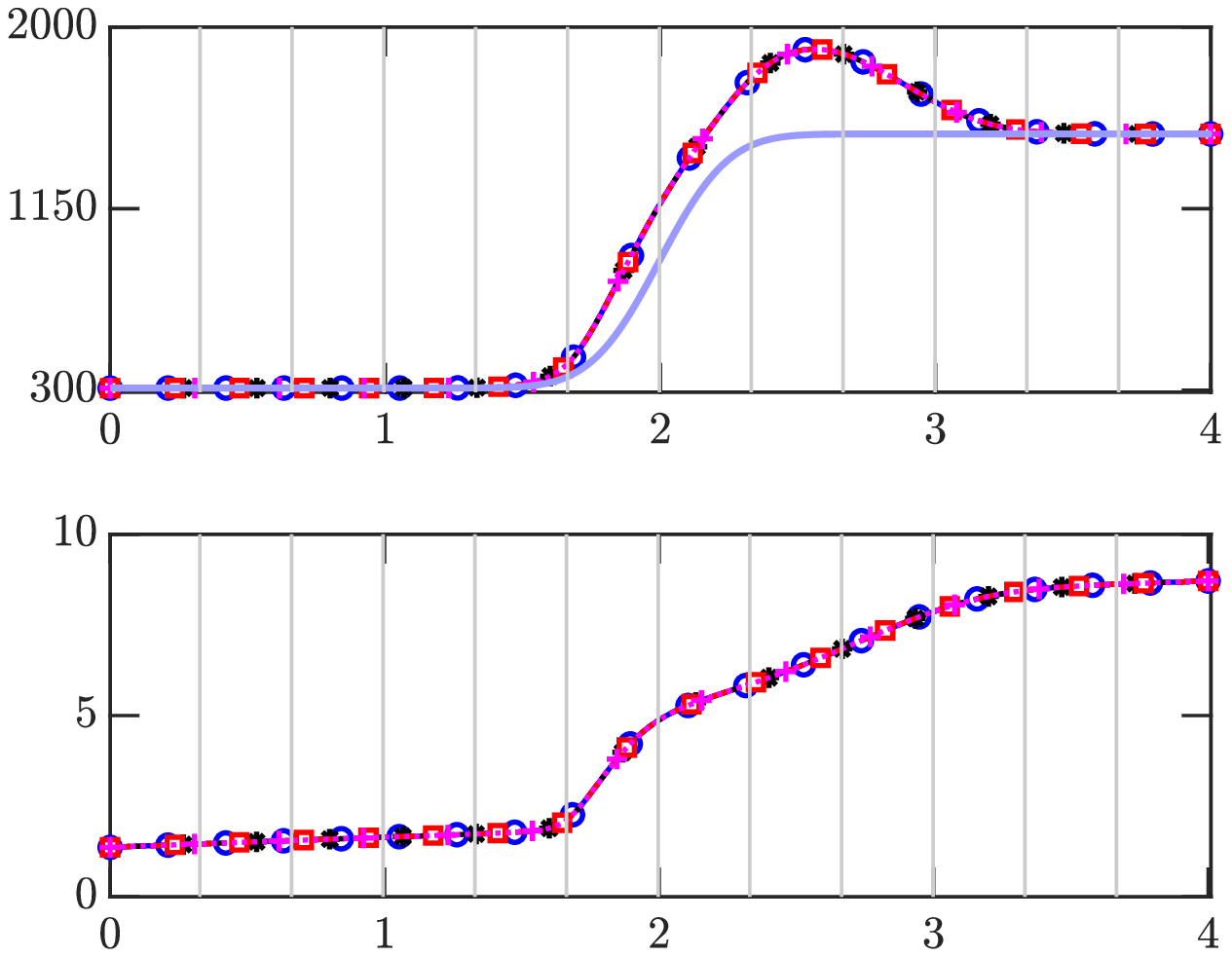}}
\hspace{1cm}
\subfigure{\includegraphics[width=0.31\textwidth]{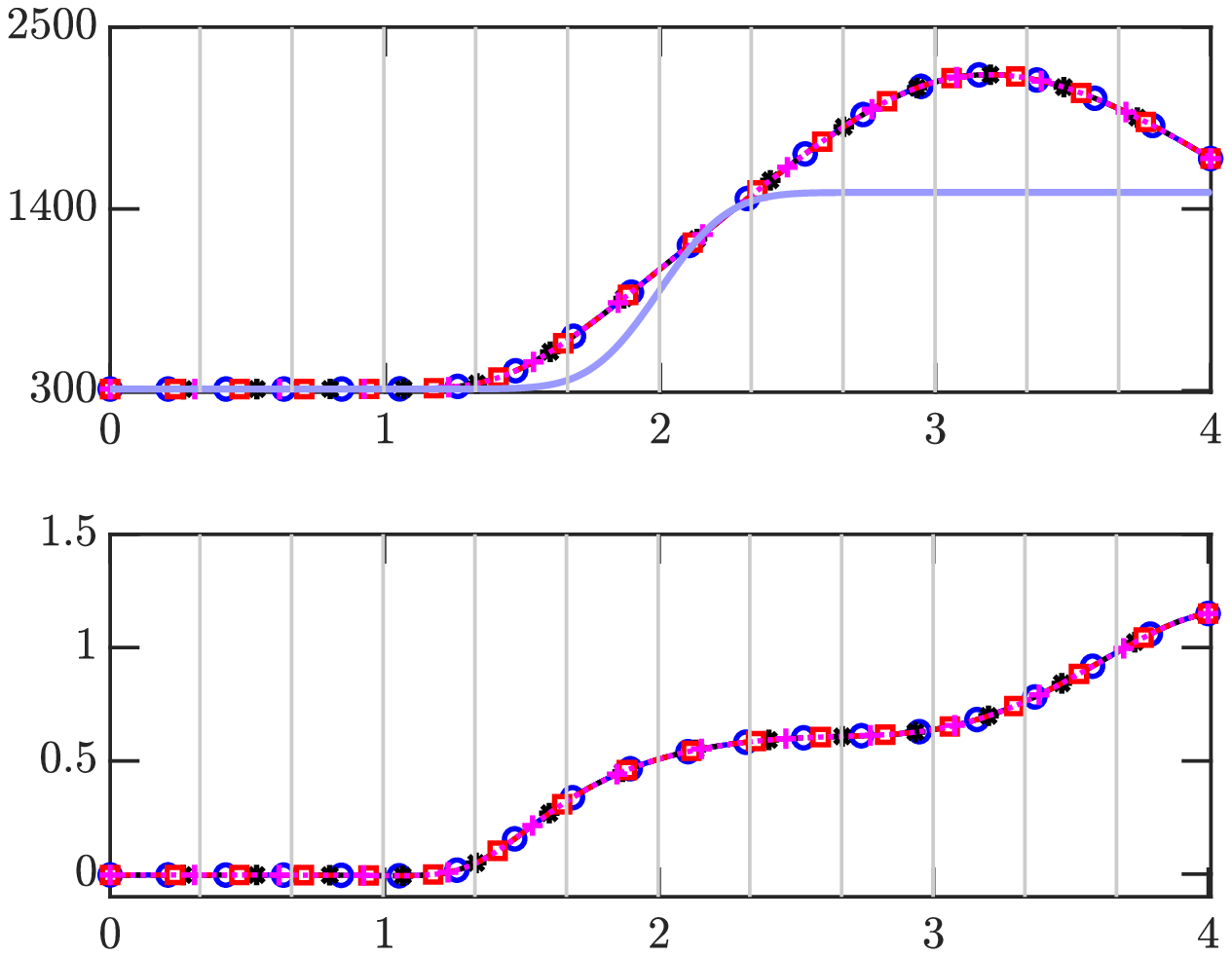}}
 \\
\subfigure{\includegraphics[trim={0cm 0cm 0cm 0cm},clip,width=0.31\textwidth]{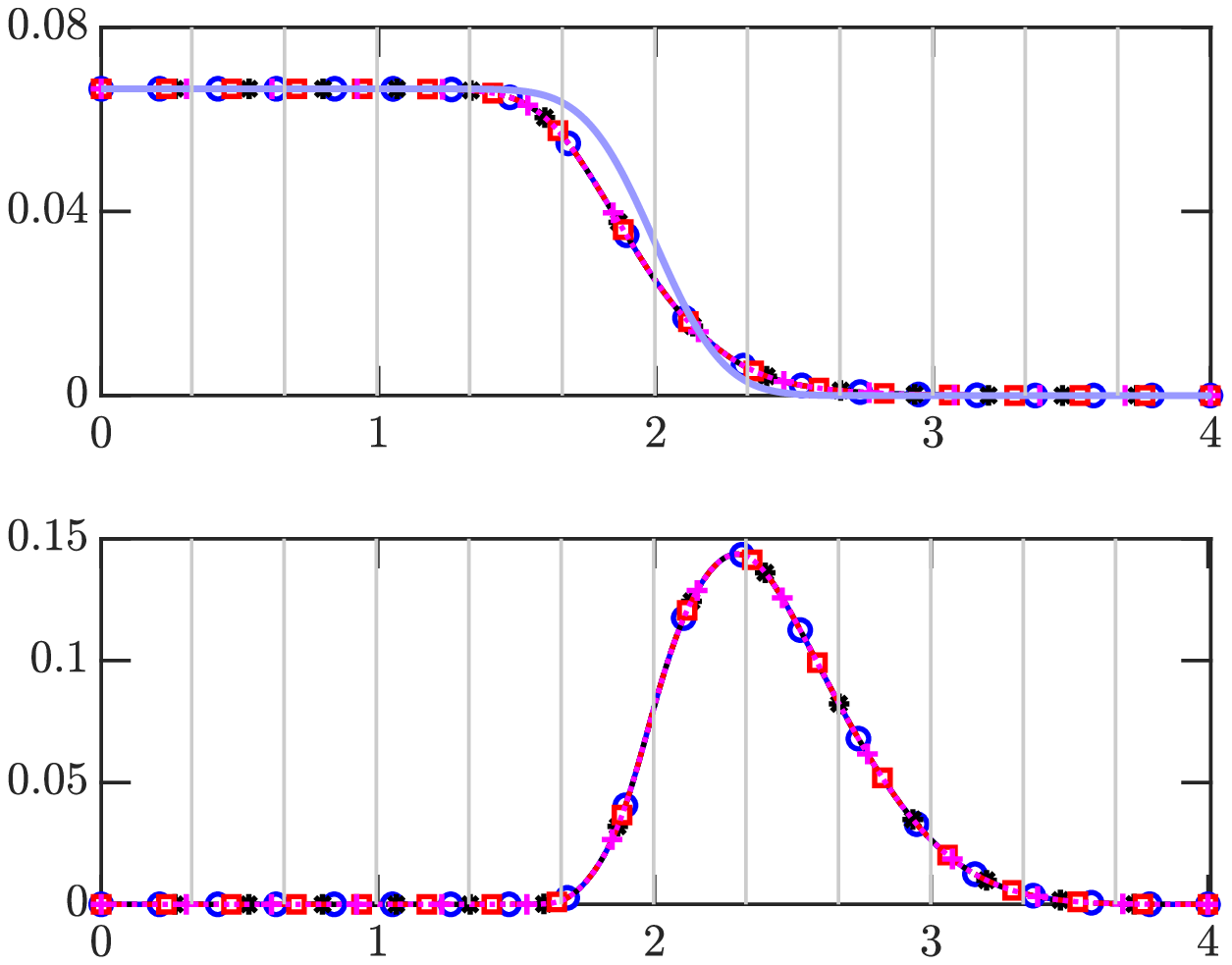}}
\hspace{1cm}
\subfigure{\includegraphics[width=0.31\textwidth]{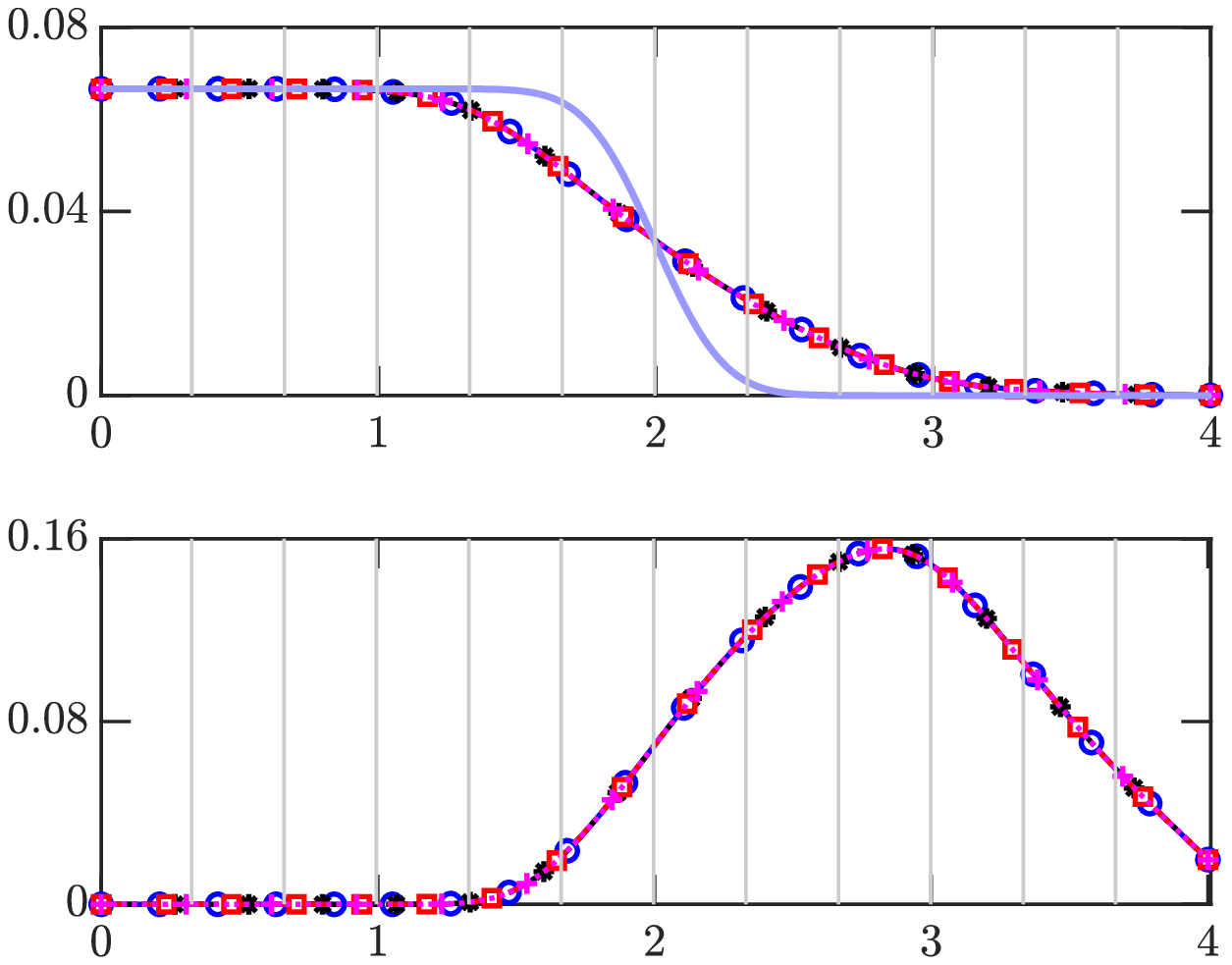}}
 \\
\subfigure{\includegraphics[trim={0cm 0cm 0cm 0cm},clip,width=0.31\textwidth]{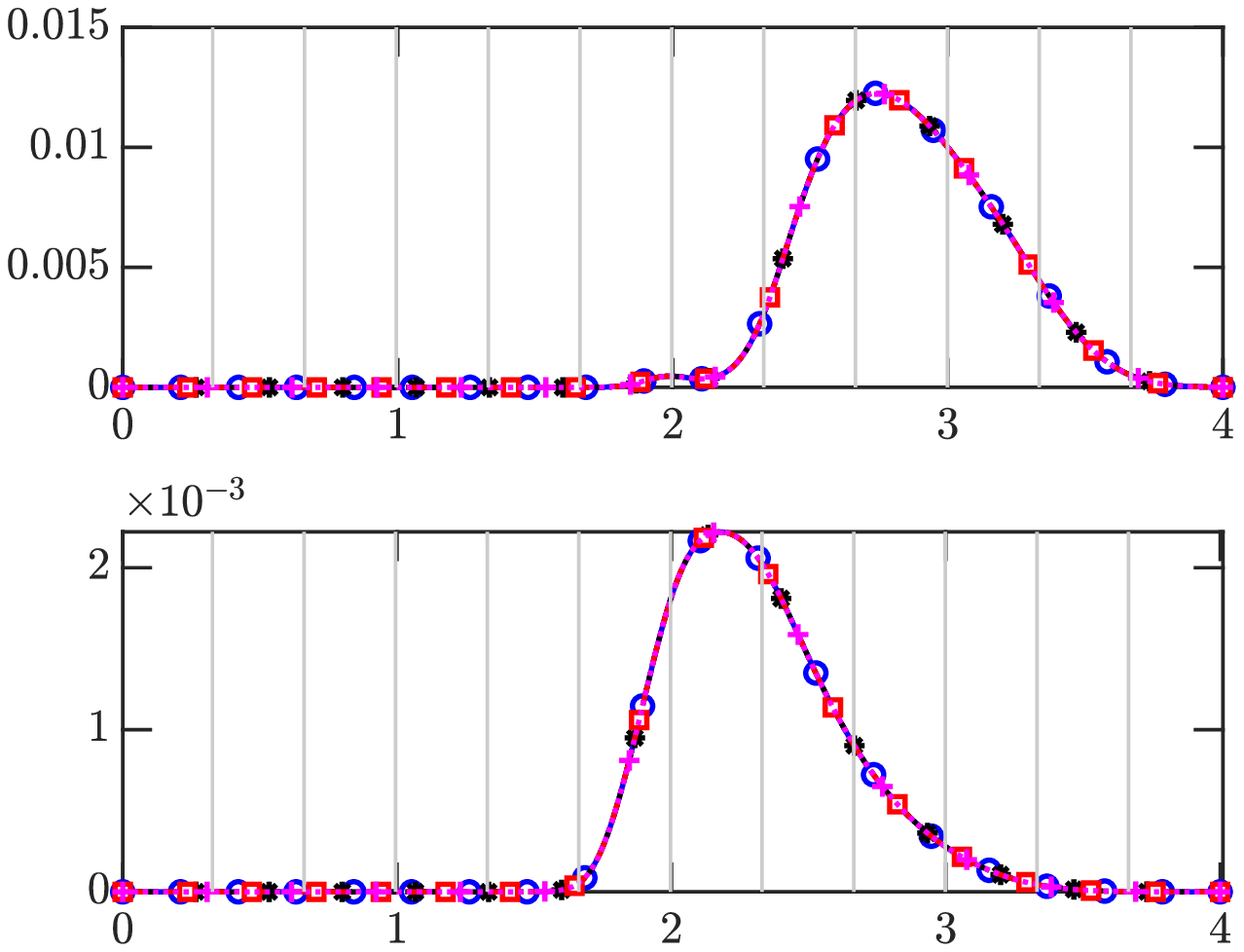}}
\hspace{1cm}
\subfigure{\includegraphics[width=0.31\textwidth]{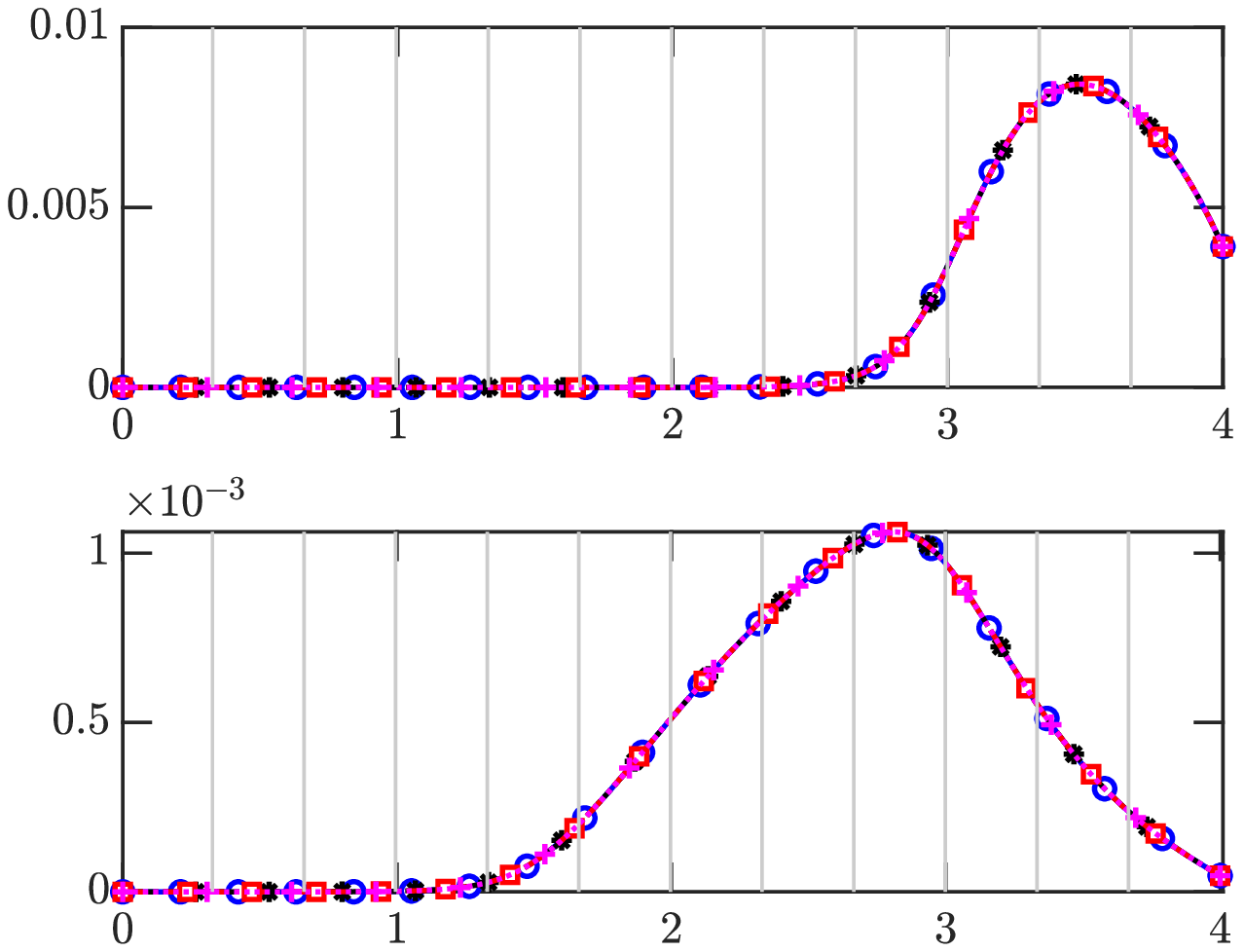}}
\begin{picture}(0,0)
        \put(-280,360){$t=7.92\times 10^{-5}s$}
        \put(-95,360){$t=3.19\times 10^{-4}s$}
        \put(-260,-10){$x~(mm)$}
        \put(-65,-10){$x~(mm)$}

         \put(-350,330){$T$}
        \put(-350,270){$U$}
         \put(-350,200){$Y_{H_2}$}
        \put(-350,150){$Y_{H_2O}$}
          \put(-350,85){$Y_{O}$}
        \put(-350,25){$Y_{H}$}
        \end{picture}
\caption{Asynchrony-tolerant (AT): Instantaneous profiles and error in temperature ($K$), velocity ($ms^{-1}$)
and mass fractions of $H_2$, $H_2O$, $O$ and $H$ for non-premixed ignition of $H_2$. The different lines: blue (Set-1), black (Set-2), red (Set-3) and magenta (Set-4) are defined in \rtab{dels}. The light blue line indicates the initial condition and the gray vertical lines represent processor boundaries.
}
\figlabel{diff}
\end{center}
\end{figure}

\subsection{Propagation of a detonation wave}
To test the numerical accuracy of AT-WENO
schemes for reacting flows, a detonation
of stoichiometric hydrogen/oxygen mixture
diluted with argon is considered. The mixture with molar ratio $H_2:O_2:Ar=2:1:7$ is at an initial temperature of 305K and
pressure of 6670 Pa. A detailed hydrogen/air mechanism~\cite{westbrook1982} comprised of 9 species and 34 reversible
reactions is used. A similar setup has also been used as
a test case in previous studies to investigate the
efficacy of numerical schemes~\cite{deiterdingParallelAdaptiveSimulation2003,lvDiscontinuousGalerkinMethod2014, houimLowdissipationTimeaccurateMethod2011}.
A supersonic outflow condition is applied at the left boundary
while a subsonic inlet condition is applied at the right boundary. The simulation is initialized with a
shock that ignites the mixture at the left end of the computational domain. The domain length is $30~cm$ which is discretized into 6000
grid-points that are distributed across 100 PEs.

The simulation results are shown in~\rfig{det} with pressure
at different time instants in the leftmost plot.
There is a leading shock front that compresses the fuel/air mixture
followed by the induction and the reaction zones.
The pressure obtained with the synchronous WENO scheme is shown in solid magenta. It is in excellent agreement with the pressure
computed with the AT-WENO scheme shown in black circles and dashed lines.
A zoomed-in view of the detonation structure depicted by temperature and pressure
is shown in the centre plots of \rfig{det}. The mixture reacts after
the induction zone leading to an increase in temperature and a corresponding decrease in pressure. The mass fractions
for different radicals including fuel $H_2$ are plotted on the rightmost plot in~\rfig{det}. The results of this simulation show that the detonation structure is captured accurately with the AT-WENO scheme. Furthermore, the detonation velocity obtained from standard WENO and AT-WENO simulations is $1619.7\ ms^{-1}$ and $1620.1\ ms^{-1}$, respectively. These values are in good agreement with the detonation velocity reported in
previous studies~\cite{deiterdingParallelAdaptiveSimulation2003, lvDiscontinuousGalerkinMethod2014,houimLowdissipationTimeaccurateMethod2011}. The peak von-Neumann pressure is $165.35\ kPa$ and $165.34\ kPa$ for WENO and AT-WENO which are also similar to the previous studies. Overall, the AT-WENO schemes derived in the present study exhibit excellent numerical accuracy despite the use of delayed data at processor boundaries. Since multidimensional simulations of detonation phenomena in combustion devices face highly intensive computational resource requirements~\cite{luong2020statistical,luong2020effects}, AT-WENO schemes hold the potential of offsetting a certain portion of communication overhead associated with standard WENO schemes at large node counts, thereby leading to improved scalability.

\begin{figure}[h!]
\begin{center}
\subfigure{\includegraphics[trim={0cm 0cm 0cm 0cm},clip,width=0.3\textwidth]{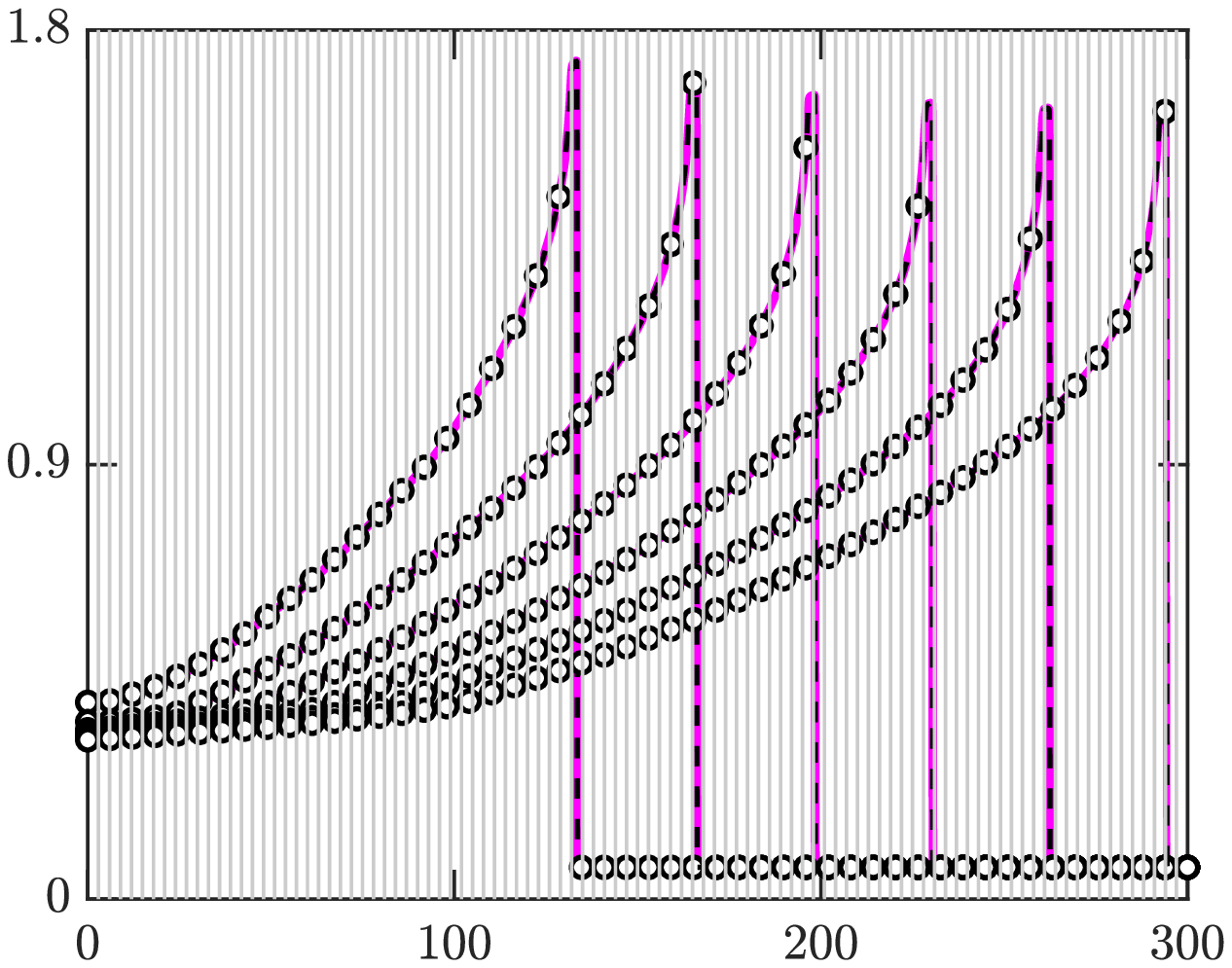}}
\hspace{0.3cm}
\subfigure{\includegraphics[width=0.31\textwidth]{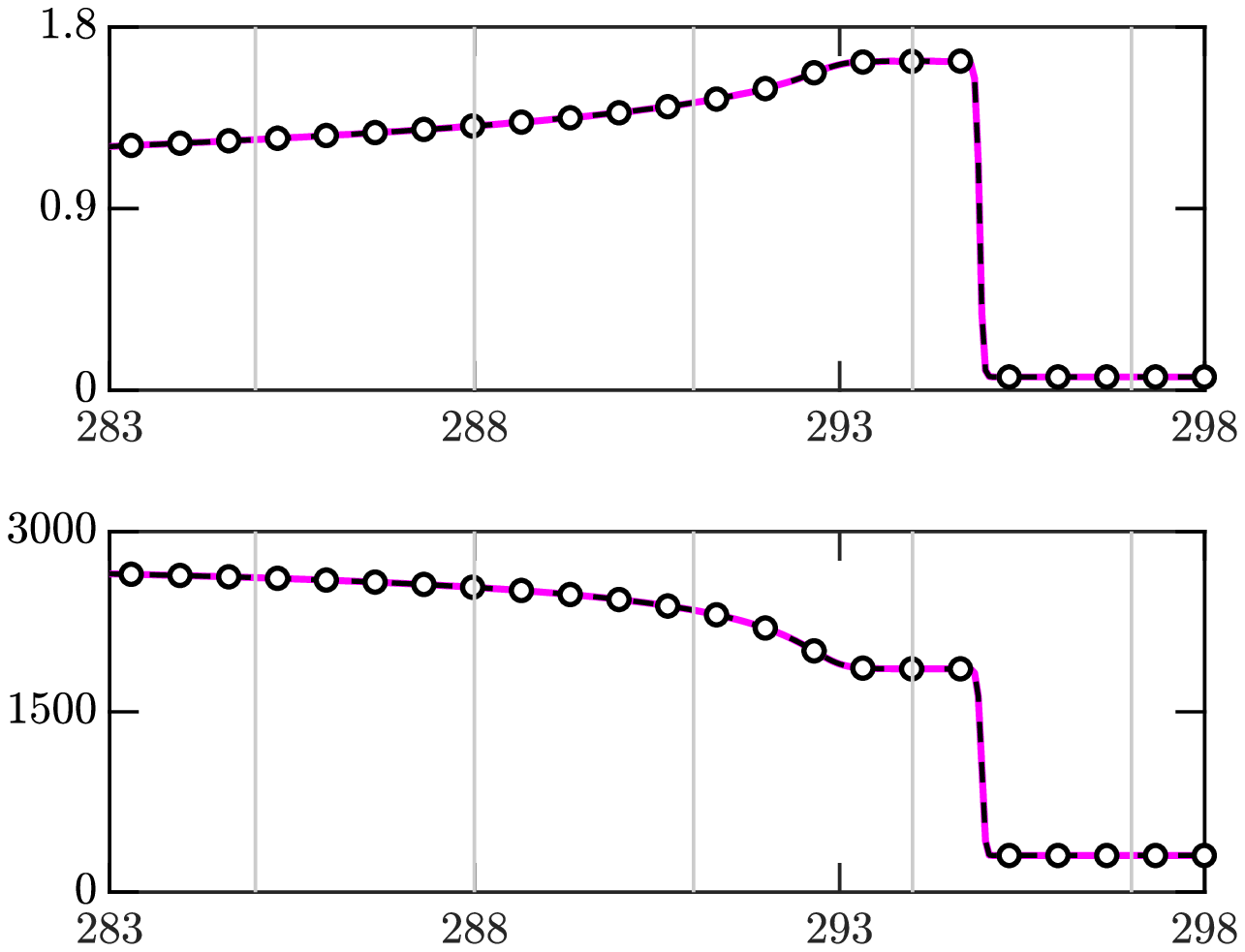}}
\hspace{0.3cm}
\subfigure{\includegraphics[width=0.3\textwidth]{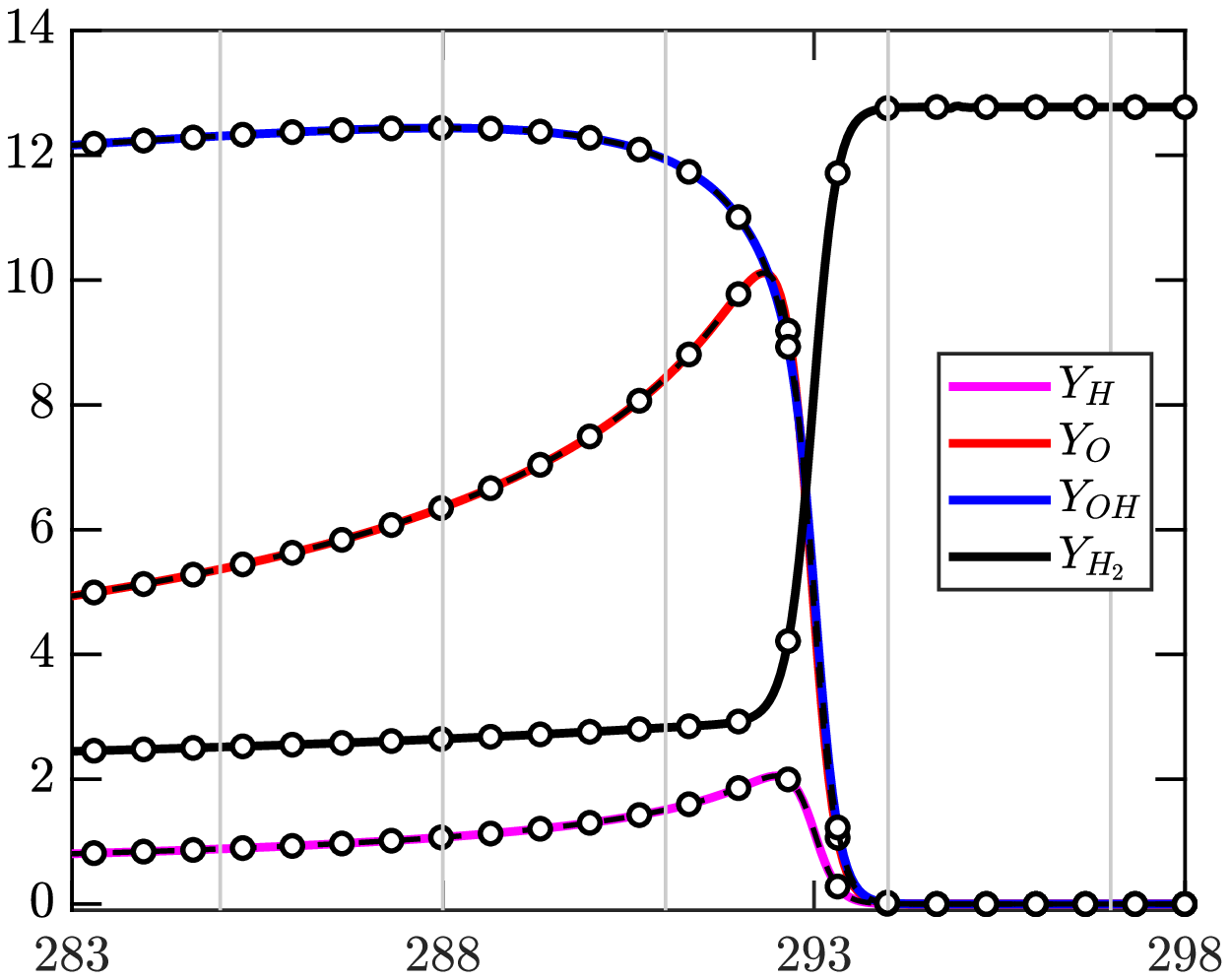}}
\begin{picture}(0,0)

        \put(-400,-10){$x~(mm)$}
        \put(-245,-10){$x~(mm)$}
        \put(-80,-10){$x~(mm)$}

         \put(-390,112){$P$}
         \put(-230,112){$P$}
         \put(-230,53){$T$}
         \put(-90,112){$Y\times 1000$}
        \end{picture}
\caption{Asynchrony-tolerant (AT):
Pressure (atm) profiles (left) at different times,
structure of detonation wave (middle)
for  pressure (atm) and temperature ($K$) and
mass fractions of $H,O,OH,H_2$ (right). The solid lines are
standard WENO and black-dashed line with black symbols (Set-3 in \rtab{dels}) corresponds to AT-WENO. gray vertical lines
represent processor boundaries.}
\figlabel{det}
\end{center}
\end{figure}

\section{Discussion}
DNS of turbulent combustion at higher Reynolds numbers
with detailed chemical mechanisms and at
conditions relevant for practical devices will
require efficient utilization of massive
computing resources anticipated on  next generation exascale
machines. These simulations at scales and conditions
that are currently infeasible will provide fine-grained details into
the interactions between turbulence and chemistry and will provide fundamental insights into the physical processes that result in pollutant formation, flame stabilization, blow-off, higher efficiency, etc.
This in turn
will not only aid in the design of fuel-flexible, low-emission combustion
engines but also in the development of physics-based
models for engineering simulations that will be used to design advanced clean engines. However, a successful transition of
DNS codes to exascale machines and beyond is possible only through
redesign and development of
new computational algorithms in order to overcome some of the obvious bottlenecks and challenges.

The anticipated computing power of a quintillion ($10^{18}$)
double precision floating point
operations per second on the exascale machine will be
realized through extreme levels of parallelism with millions of processing elements (PEs), including CPUs, GPUs and FPGAs.
For example, Frontier at the Oak Ridge Leadership Computing Facility (OCLF) will have 1:4 CPU to GPU ratio with
unified memory. Despite the use
of state-of-the-art low-latency, high-bandwidth
communication links, the sheer
number of communications will significantly hinder scaling to
large node counts. To this end, efficient
strategies to mask communications and data movement will be needed to improve scalability.
The use of AT schemes at the PE boundaries for computation
of derivatives accurately, despite delayed data, will
facilitate further overlap between communications
and computations. However, a challenge in
extending the AT schemes to three-dimensions is that the
derivatives of convective and diffusive terms
need to be expanded in terms of
derivatives of primitive variables by applying the chain rule.
For example, $\partial (\rho u e) /\partial x:= (\rho u e)_x =
\rho u e_x + \rho e u_x + e u \rho_x$. This expansion
increases the total number of derivatives that need to be computed,
especially when a large
number of species are involved.
Alternatively, one can compute
the convective and diffusive terms
(for example, $\rho u e, ~\rho Y_i u,~ J_i,~ \tau$)
from the primitive variables at different time levels as required
by the AT schemes. These computations can be performed
on the fly in order to avoid
storing such product terms at multiple time-levels.
Essentially there is a trade-off between memory (data movement) and computation. Re-design of data
structures for better memory coherency will also be extremely
helpful since AT schemes use data from multiple time levels
and can thus incur high cache miss rates.
Besides these challenges, a critical
first step in using AT schemes for scalable simulations of turbulent
combustion on massive supercomputers
is to ascertain that the numerical artifacts
due to delayed data at PE boundary are significantly
small and do not affect the underlying physics. The comprehensive set of
numerical experiments presented in the present study specifically highlights the excellent numerical accuracy of AT schemes
for compressible reacting unsteady flows.

A key component of simulations of reacting flows
even for simple geometries
includes the prescription of nonperiodic boundary conditions.
The NSCBC \cite{thompsonTimeDependentBoundary1987,
poinsotBoundaryConditionsDirect} with
improvements for reacting flows
\cite{yooCharacteristicBoundaryConditions2005,yooCharacteristicBoundaryConditions2007,sutherlandImprovedBoundaryConditions2003,lodatoCharacteristicOutflowsOptimal} has
been successfully used for these simulations.
In the normal direction, the terms in the NSCBC boundary conditions
are not affected by AT schemes. For example, at the outflow, one-sided
schemes are used, which require only local information within a
PE. At the inflow, relaxation terms are used which again do not
require PEs to communicate. Since transverse and
viscous terms are also included for reacting flows
\cite{yooCharacteristicBoundaryConditions2005,yooCharacteristicBoundaryConditions2007,sutherlandImprovedBoundaryConditions2003},
the AT schemes can be utilized in the
transverse direction for computation of all the
 spatial derivatives involved.
Extension to corners can be done following
\cite{lodatoCharacteristicOutflowsOptimal} with AT schemes used
wherever necessary.

Another advantage for using asynchronous computations on
exascale machines is to leverage the
high flop-rate of GPUs, i.e. throughput computing. In hybrid computing architectures, GPUs are expected to
handle most of the computations, while CPUs will facilitate communications
between PEs. With AT schemes, GPUs will not have to wait on the
CPUs for the most updated
data from the neighboring PEs. This would enhance utilization of GPUs without
affecting numerical accuracy or introducing idling penalties. A similar approach has been utilized in \cite{PK2020}
where asynchronous copies between CPUs and GPUs are used to overlap computation and data movement, but delayed data with asynchronous computations have not been used. At a compiler level, new asynchronous run-time systems that are capable of dynamic task parallelism are being developed to improve the computation-communication overlap \cite{regent2015,charm2014,charm2019}. The AT schemes, which relax synchronization at a mathematical level, can be coupled with such data-centric programming models to create highly scalable PDE solvers.
Asynchrony has also been utilized
in \cite{kollaImprovingScalabilitySilentError}
for scalable resilience to soft faults.
However, in \cite{kollaImprovingScalabilitySilentError}
all derivatives are still computed
with the most updated data but the computations are
re-arranged to ensure maximum overlap
between communication and computation.
The asynchronization approach utilized in \cite{kollaImprovingScalabilitySilentError} coupled with
mathematical level asynchrony with AT schemes can be
effective in pushing the scaling wall.
Furthermore, since AT schemes do not require updated data,
these schemes can also be used to recover from node failures without halting
the simulation altogether.


\section{Conclusions}
In this paper, a
series of numerical simulations that utilize delayed data at
processor boundaries are presented
for canonical reacting flows with single-step and detailed
reaction mechanisms.
It is shown that large numerical errors are incurred
in physical as well as spectral space when standard schemes are used asynchronously, even for the simplest problems.
To overcome this loss of accuracy, novel asynchrony-tolerant (AT) schemes are
used for accurate computation of spatial derivatives with
delayed data at the boundaries. These asynchronous simulations using AT schemes are compared with standard synchronous simulations and are shown
to exhibit excellent qualitative and quantitative agreement.
Both spatial and temporal evolution of hydrodynamic and
thermo-chemical quantities are accurately captured by the AT schemes.

For simulations of high-speed flows with shocks and jump
discontinuities, the standard shock-resolving WENO
schemes have degraded numerical accuracy. Thus, following the
procedure listed in \cite{adityaHighorderAsynchronytolerantFinite2017}
AT-WENO schemes are derived. However, the schemes so derived
have non-convex linear weights which can lead to
numerical instabilities and oscillations \cite{shiTechniqueTreatingNegative2002}. To overcome this,
the asynchronous stencil
is modified to include an additional grid-point. With this degree-of-freedom, an additional low-order truncation
error term is selectively eliminated while preserving the
convexity of the ideal weights.
For evaluation of non-linear weights
in AT-WENO schemes, asynchrony-tolerant smoothness indicators are also derived.
The superior accuracy of AT-WENO schemes in comparison to the
standard WENO schemes is exhibited through the order-of-accuracy tests
for linear and non-linear equations.
These schemes are also used for simulation of a detonation wave which is found to have
excellent agreement with its synchronous counterpart.

A common conclusion one can arrive at
through these simulations is
that the AT schemes have excellent accuracy despite relaxed synchronizations at  processor boundaries. Thus, AT schemes
provide a potential path for highly scalable
asynchronous simulations of turbulent combustion at the extreme scales promised by exascale and beyond.

\section*{Acknowledgements}
The first author gratefully acknowledges support from the NSF-INTERN program through Grant No. 1605914.
The work at IISc was supported by the Start-up Research Grant, SERB, India.
The work at Sandia was supported by the US Department of Energy, Office of Basic Energy Sciences, Division of Chemical Sciences, Geosciences, and Biosciences. Sandia National Laboratories is a multi-mission laboratory managed and operated by National Technology and Engineering Solutions of Sandia, LLC., a wholly owned subsidiary of Honeywell International, Inc., for the U.S. Department of Energy's National Nuclear Security Administration under contract DE-NA-0003525. The views expressed in the article do not necessarily represent the views of the U.S. Department of Energy or the United States Government.





%



\appendix
\section{AT schemes}
\begin{table}[h]
\begin{center}
{\tabulinesep=0.5mm
\begin{tabu}{|c|c|c|}
\hline
(Derivative, & Boundary & Scheme\\
Order) & &\\
\hline
\hline
(2,4) & Left & {$\scriptsize\begin{array}{c}
\frac{1}{2}(\k{}^2+3\k{}+2)\left({-\U{j+2}{n} + 16 \U{j+1}{n} - 30 \U{j}{n} + 16\U{j-1}{n-\k{}} - \U{j-2}{n-\k{}}}\right)/{12\dx^2}  \\
 -(\k{}^2+2\k{}) \left({-\U{j+2}{n} + 16 \U{j+1}{n} - 30 \U{j}{n} + 16\U{j-1}{n-\k{}-1} - \U{j-2}{n-\k{}-1}}\right)/{12\dx^2} \\
 +\frac{1}{2}(\k{}^2+\k{}) \left({-\U{j+2}{n} + 16 \U{j+1}{n} - 30 \U{j}{n} + 16\U{j-1}{n-\k{}-2} - \U{j-2}{n-\k{}-2}}\right)/{12\dx^2}
\end{array}$} \\
\hline
(2,4) & Right &{$\scriptsize\begin{array}{c}
\frac{1}{2}(\k{}^2+3\k{}+2) \left({-\U{j+2}{n-\k{}} + 16 \U{j+1}{n-\k{}} - 30 \U{j}{n} + 16\U{j-1}{n} - \U{j-2}{n}}\right)/{12\dx^2}  \\
 -(\k{}^2+2\k{}) \left({-\U{j+2}{n-\k{}-1} + 16 \U{j+1}{n-\k{}-1} - 30 \U{j}{n} + 16\U{j-1}{n} - \U{j-2}{n}}\right)/{12\dx^2}  \\
 +\frac{1}{2}(\k{}^2+\k{}) \left({-\U{j+2}{n-\k{}-2} + 16 \U{j+1}{n-\k{}-2} - 30 \U{j}{n} + 16\U{j-1}{n} - \U{j-2}{n}}\right)/{12\dx^2}
\end{array}$}\\
\hline

(1,4) & Left & {$\scriptsize\begin{array}{c}
\frac{1}{2}(\k{}^2+3\k{}+2)\left({-\U{j+2}{n} + 8 \U{j+1}{n} - 8\U{j-1}{n-\k{}} + \U{j-2}{n-\k{}}}\right)/{12\dx}  \\
 -(\k{}^2+2\k{})\left({-\U{j+2}{n} + 8 \U{j+1}{n} - 8 \U{j-1}{n-\k{}-1} + \U{j-2}{n-\k{}-1}}\right)/{12\dx} \\
 +\frac{1}{2}(\k{}^2+\k{})\left({-\U{j+2}{n} + 8 \U{j+1}{n} - 8 \U{j-1}{n-\k{}-2} + \U{j-2}{n-\k{}-2}}\right)/{12\dx}
\end{array}$}\\
\hline
(1,4) & Right &
{ $ \scriptsize\begin{array}{c}
\frac{1}{2}(\k{}^2+3\k{}+2)\left({-\U{j+2}{n-\k{}} + 8 \U{j+1}{n-\k{}} - 8\U{j-1}{n} + \U{j-2}{n}}\right)/{12\dx}  \\
 -(\k{}^2+2\k{})\left({-\U{j+2}{n-\k{}-1} + 8 \U{j+1}{n-\k{}-1} - 8 \U{j-1}{n} + \U{j-2}{n}}\right)/{12\dx} \\
 +\frac{1}{2}(\k{}^2+\k{})\left({-\U{j+2}{n-\k{}-2} + 8 \U{j+1}{n-\k{}-2} - 8 \U{j-1}{n} + \U{j-2}{n}}\right)/{12\dx}
\end{array}$}\\
\hline
\end{tabu}}
\caption{Asynchrony-tolerant (AT) schemes for left and right boundary
used in numerical simulations for first and second derivative.}
\label{tab:atschemes}
\end{center}
\end{table}

\section{Reconstruction approximation at the right boundary}
For points located at the right processor boundary
the modified smaller asynchronous stencil is
given by,
$\tilde{S}^{(0)}_R=\{ u_{j-2}^n, u_{j-1}^n,u_{j}^n\}$,
$\tilde{S}^{(1)}_R=\{ u_{j-1}^{n}, u_j^n,u_{j}^{n-\kt},u_{j+1}^{n-\kt}u_{j+1}^{n-\kt-1} \}$,
and $\tilde{S}^{(2)}_R=
\{ u_{j}^{n}, u_{j+1}^{n-\kt},u_{j+1}^{n-\kt-1},u_{j+2}^{n-\kt}u_{j+2}^{n-\kt-1} \}
$.
Here the delay appears at the rightmost gird points.
We derive the following reconstruction approximation at
the right boundary,
\begin{equation}
    \begin{aligned}
    &\tilde{u}_{j+\frac{1}{2}}^{n,(0)}=
\frac{u_{j-2}^n}{3}-\frac{7 u_{j-1}^n}{6}+\frac{11 u_j^n}{6}\\
     &\tilde{u}_{j+\frac{1}{2}}^{n,(1)}=
         -\frac{1}{3} \tilde{k} u_{j+1}^{-\tilde{k}+n-1}+\frac{1}{6} \left(2 \tilde{k}+2\right) u_{j+1}^{n-\tilde{k}}-\frac{1}{6} u_{j-1}^n+\frac{5 u_j^n}{6}
     \\
   &\tilde{u}_{j+\frac{1}{2}}^{n,(2)}=
   -\frac{5}{6} \tilde{k} u_{j+1}^{-\tilde{k}+n-1}+\frac{1}{6} \tilde{k} u_{j+2}^{-\tilde{k}+n-1}+\frac{1}{6} \left(5 \tilde{k}+5\right) u_{j+1}^{n-\tilde{k}}+\frac{1}{6} \left(-\tilde{k}-1\right) u_{j+2}^{n-\tilde{k}}+\frac{u_j^n}{3}
    \end{aligned}
    \eqnlabel{atReconR}
\end{equation}
Similarly, the approximation at point $u_{j-1/2}$ on
the right boundary at each of the
candidate stencils is listed below
\begin{equation}
    \begin{aligned}
    &\tilde{u}_{j-\frac{1}{2}}^{n,(0)}=
-\frac{1}{6} u_{j-2}^n+\frac{5 u_{j-1}^n}{6}+\frac{u_j^n}{3}\\
     &\tilde{u}_{j-\frac{1}{2}}^{n,(1)}=
\frac{1}{6} \tilde{k} u_{j+1}^{-\tilde{k}+n-1}+\frac{1}{6} \left(-\tilde{k}-1\right) u_{j+1}^{n-\tilde{k}}+\frac{u_{j-1}^n}{3}+\frac{5 u_j^n}{6}
     \\
   &\tilde{u}_{j-\frac{1}{2}}^{n,(2)}=
\frac{7}{6} \tilde{k} u_{j+1}^{-\tilde{k}+n-1}-\frac{1}{3} \tilde{k} u_{j+2}^{-\tilde{k}+n-1}+\frac{1}{6} \left(-7 \tilde{k}-7\right) u_{j+1}^{n-\tilde{k}}+\frac{1}{6} \left(2 \tilde{k}+2\right) u_{j+2}^{n-\tilde{k}}+\frac{11 u_j^n}{6}
    \end{aligned}
    \eqnlabel{atReconRm}
\end{equation}

Following the procedure described in the previous sections, the
smoothness indicator at the right-boundary can be computed to be
\begin{equation}
    \begin{aligned}
         \tilde{\beta}^{(0)}=\frac{1}{3} &
         \Bigg(
   4 \left(u^n\right)_{j-2}^2+\left(11 u^n_j-19 u^n_{j-1}\right) u^n_{j-2}+25 \left(u^n\right)_{j-1}^2+10 \left(u^n\right)_j^2-31 u^n_{j-1} u^n_j
         \Bigg)
    \end{aligned}
    \eqnlabel{beta2AT0}
\end{equation}
\begin{equation}
    \begin{aligned}
        \tilde{ \beta
        }^{(1)}=\frac{1}{3} &
         \Bigg(
         u_{j-1}^n \left(-5 \tilde{k} u_{j+1}^{n-\tilde{k}-1}+5 \left(\tilde{k}+1\right) u_{j+1}^{n-\tilde{k}}-13 u_j^n\right)+4 \left(\tilde{k} u_{j+1}^{n-\tilde{k}-1}-\left(\tilde{k}+1\right) u_{j+1}^{n-\tilde{k}}\right)^2\\
         &+13 u_j^n \Big(\tilde{k} u_{j+1}^{n-\tilde{k}-1}-\left(\tilde{k}+1\right) u_{j+1}^{n-\tilde{k}}\Big)+4 \left(u_{j-1}^n\right)^2+13 \left(u_j^n\right)^2
         \Bigg)
    \end{aligned}
    \eqnlabel{beta2AT1}
\end{equation}
\begin{equation}
    \begin{aligned}
        \tilde{ \beta}^{(2)}=\frac{1}{3} &
         \Bigg(
      \tilde{k}^2 \Bigg(25 \left(u_{j+1}^{n-\tilde{k}-1}\right)^2+\left(19 \left(u_{j+2}^{n-\tilde{k}}-u_{j+2}^{n-\tilde{k}-1}\right)-50 u_{j+1}^{n-\tilde{k}}\right) u_{j+1}^{n-\tilde{k}-1}+25 \left(u_{j+1}^{n-\tilde{k}}\right)^2\\
      &+4 \left(u_{j+2}^{n-\tilde{k}-1}-u_{j+2}^{n-\tilde{k}}\right)^2+19 u_{j+1}^{n-\tilde{k}} \left(u_{j+2}^{n-\tilde{k}-1}-u_{j+2}^{n-\tilde{k}}\right)\Bigg)+\tilde{k} \Big(50 \left(u_{j+1}^{n-\tilde{k}}\right)^2\\
      &+19 \left(u_{j+2}^{n-\tilde{k}-1}-2 u_{j+2}^{n-\tilde{k}}\right) u_{j+1}^{n-\tilde{k}}+8 u_{j+2}^{n-\tilde{k}} \left(u_{j+2}^{n-\tilde{k}}-u_{j+2}^{n-\tilde{k}-1}\right)+u_{j+1}^{n-\tilde{k}-1} \left(19 u_{j+2}^{n-\tilde{k}}-50 u_{j+1}^{n-\tilde{k}}\right)\Big)\\
      &+25 \left(u_{j+1}^{n-\tilde{k}}\right)^2+4 \left(u_{j+2}^{n-\tilde{k}}\right)^2-19 u_{j+1}^{n-\tilde{k}} u_{j+2}^{n-\tilde{k}}+u_j^n \Big(-31 u_{j+1}^{n-\tilde{k}}+11 u_{j+2}^{n-\tilde{k}}\\
      &+\tilde{k} \left(31 u_{j+1}^{n-\tilde{k}-1}-31 u_{j+1}^{n-\tilde{k}}-11 u_{j+2}^{n-\tilde{k}-1}+11 u_{j+2}^{n-\tilde{k}}\right)\Big)+10 \left(u_j^n\right)^2
         \Bigg)
    \end{aligned}
    \eqnlabel{beta2AT2R}
\end{equation}
Lastly, we note that the approximation
of $u_{j-1/2}$ at the left-boundary is
\begin{equation}
    \begin{aligned}
    &\tilde{u}_{j-\frac{1}{2}}^{n,(0)}=
\frac{1}{6} \tilde{k} u_{j-2}^{-\tilde{k}+n-1}-\frac{5}{6} \tilde{k} u_{j-1}^{-\tilde{k}+n-1}+\frac{1}{6} \left(-\tilde{k}-1\right) u_{j-2}^{n-\tilde{k}}+\frac{1}{6} \left(5 \tilde{k}+5\right) u_{j-1}^{n-\tilde{k}}+\frac{u_j^n}{3}
\\
     &\tilde{u}_{j-\frac{1}{2}}^{n,(1)}=
-\frac{1}{3} \tilde{k} u_{j-1}^{-\tilde{k}+n-1}+\frac{1}{6} \left(2 \tilde{k}+2\right) u_{j-1}^{n-\tilde{k}}+\frac{5 u_j^n}{6}-\frac{u_{j+1}^n}{6}
     \\
   &\tilde{u}_{j-\frac{1}{2}}^{n,(2)}=
   \frac{11}{6}{u_j^n}-\frac{7}{6} u_{j+1}^n+\frac{1}{3}u_{j+2}^n.
    \end{aligned}
    \eqnlabel{atReconM}
\end{equation}
 \bibliographystyle{model1-num-names}

\bibliography{main}

\begin{thebibliography}{62}
\expandafter\ifx\csname natexlab\endcsname\relax\def\natexlab#1{#1}\fi
\providecommand{\bibinfo}[2]{#2}
\ifx\xfnm\relax \def\xfnm[#1]{\unskip,\space#1}\fi
\bibitem[{Moin and Mahesh(1998)}]{moinDIRECTNUMERICALSIMULATION1998}
\bibinfo{author}{P.~Moin}, \bibinfo{author}{K.~Mahesh},
\newblock \bibinfo{title}{{{DIRECT NUMERICAL SIMULATION}}: {{A Tool}} in
  {{Turbulence Research}}},
\newblock \bibinfo{journal}{Annu. Rev. Fluid Mech.} \bibinfo{volume}{30}
  (\bibinfo{year}{1998}) \bibinfo{pages}{539--578}.
\bibitem[{Visbal et~al.(2001)Visbal, Gaitonde, and
  Rizzetta}]{visbalHighOrderSchemesDNS}
\bibinfo{author}{M.~Visbal}, \bibinfo{author}{D.~Gaitonde},
  \bibinfo{author}{D.~Rizzetta},
\newblock \bibinfo{title}{High-order schemes for {{DNS}}/{{LES}} and {{CAA}} on
  curvilinear dynamic meshes},
\newblock \bibinfo{journal}{DNS/LES Progress and Challenges, Greyden Press}
  (\bibinfo{year}{2001}).
\bibitem[{Chen et~al.(2009)Chen, Choudhary, {de Supinski}, DeVries, Hawkes,
  Klasky, Liao, Ma, {Mellor-Crummey}, Podhorszki, Sankaran, Shende, and
  Yoo}]{chenTerascaleDirectNumerical2009}
\bibinfo{author}{J.~H. Chen}, \bibinfo{author}{A.~Choudhary},
  \bibinfo{author}{B.~{de Supinski}}, \bibinfo{author}{M.~DeVries},
  \bibinfo{author}{E.~R. Hawkes}, \bibinfo{author}{S.~Klasky},
  \bibinfo{author}{W.~K. Liao}, \bibinfo{author}{K.~L. Ma},
  \bibinfo{author}{J.~{Mellor-Crummey}}, \bibinfo{author}{N.~Podhorszki},
  \bibinfo{author}{R.~Sankaran}, \bibinfo{author}{S.~Shende},
  \bibinfo{author}{C.~S. Yoo},
\newblock \bibinfo{title}{Terascale direct numerical simulations of turbulent
  combustion using {{S3D}}},
\newblock \bibinfo{journal}{Comput. Sci. Disc.} \bibinfo{volume}{2}
  (\bibinfo{year}{2009}) \bibinfo{pages}{015001}.
\bibitem[{Attili et~al.(2014)Attili, Bisetti, Mueller, and
  Pitsch}]{attiliFormationGrowthTransport2014}
\bibinfo{author}{A.~Attili}, \bibinfo{author}{F.~Bisetti},
  \bibinfo{author}{M.~E. Mueller}, \bibinfo{author}{H.~Pitsch},
\newblock \bibinfo{title}{Formation, growth, and transport of soot in a
  three-dimensional turbulent non-premixed jet flame},
\newblock \bibinfo{journal}{Combustion and Flame} \bibinfo{volume}{161}
  (\bibinfo{year}{2014}) \bibinfo{pages}{1849--1865}.
\bibitem[{Aditya et~al.(2019)Aditya, Gruber, Xu, Lu, Krisman, Bothien, and
  Chen}]{adityaDirectNumericalSimulation2019}
\bibinfo{author}{K.~Aditya}, \bibinfo{author}{A.~Gruber},
  \bibinfo{author}{C.~Xu}, \bibinfo{author}{T.~Lu},
  \bibinfo{author}{A.~Krisman}, \bibinfo{author}{M.~R. Bothien},
  \bibinfo{author}{J.~H. Chen},
\newblock \bibinfo{title}{Direct numerical simulation of flame stabilization
  assisted by autoignition in a reheat gas turbine combustor},
\newblock \bibinfo{journal}{Proceedings of the Combustion Institute}
  \bibinfo{volume}{37} (\bibinfo{year}{2019}) \bibinfo{pages}{2635--2642}.
\bibitem[{Savard et~al.(2019)Savard, Hawkes, Aditya, Wang, and
  Chen}]{savard2019}
\bibinfo{author}{B.~Savard}, \bibinfo{author}{E.~R. Hawkes},
  \bibinfo{author}{K.~Aditya}, \bibinfo{author}{H.~Wang},
  \bibinfo{author}{J.~H. Chen},
\newblock \bibinfo{title}{Regimes of premixed turbulent spontaneous ignition
  and deflagration under gas-turbine reheat combustion conditions},
\newblock \bibinfo{journal}{Combustion and Flame} \bibinfo{volume}{208}
  (\bibinfo{year}{2019}) \bibinfo{pages}{402--419}.
\bibitem[{Gruber et~al.(2018)Gruber, Richardson, Aditya, and
  Chen}]{gruberDirectNumericalSimulations2018}
\bibinfo{author}{A.~Gruber}, \bibinfo{author}{E.~S. Richardson},
  \bibinfo{author}{K.~Aditya}, \bibinfo{author}{J.~H. Chen},
\newblock \bibinfo{title}{Direct numerical simulations of premixed and
  stratified flame propagation in turbulent channel flow},
\newblock \bibinfo{journal}{Phys. Rev. Fluids} \bibinfo{volume}{3}
  (\bibinfo{year}{2018}) \bibinfo{pages}{110507}.
\bibitem[{Zhang et~al.(2021)Zhang, Luong, P{\'e}rez, Han, Im, and
  Huang}]{zhangExergyLossCharacteristics2021}
\bibinfo{author}{J.~Zhang}, \bibinfo{author}{M.~B. Luong},
  \bibinfo{author}{F.~E.~H. P{\'e}rez}, \bibinfo{author}{D.~Han},
  \bibinfo{author}{H.~G. Im}, \bibinfo{author}{Z.~Huang},
\newblock \bibinfo{title}{Exergy loss characteristics of {{DME}}/air and
  ethanol/air mixtures with temperature and concentration fluctuations under
  {{HCCI}}/{{SCCI}} conditions: {{A DNS}} study},
\newblock \bibinfo{journal}{Combustion and Flame} \bibinfo{volume}{226}
  (\bibinfo{year}{2021}) \bibinfo{pages}{334--346}.
\bibitem[{Berger et~al.(2020)Berger, Hesse, Kleinheinz, Hegetschweiler, Attili,
  Beeckmann, Linteris, and Pitsch}]{bergerDNSStudyImpact2020}
\bibinfo{author}{L.~Berger}, \bibinfo{author}{R.~Hesse},
  \bibinfo{author}{K.~Kleinheinz}, \bibinfo{author}{M.~J. Hegetschweiler},
  \bibinfo{author}{A.~Attili}, \bibinfo{author}{J.~Beeckmann},
  \bibinfo{author}{G.~T. Linteris}, \bibinfo{author}{H.~Pitsch},
\newblock \bibinfo{title}{A {{DNS}} study of the impact of gravity on
  spherically expanding laminar premixed flames},
\newblock \bibinfo{journal}{Combustion and Flame} \bibinfo{volume}{216}
  (\bibinfo{year}{2020}) \bibinfo{pages}{412--425}.
\bibitem[{Kim et~al.(2021)Kim, Lee, and
  Im}]{kimEffectsDifferentialDiffusion2020}
\bibinfo{author}{Y.~J. Kim}, \bibinfo{author}{B.~J. Lee},
  \bibinfo{author}{H.~G. Im},
\newblock \bibinfo{title}{Effects of {{Differential Diffusion}} on the
  {{Stabilization}} of {{Unsteady Lean Premixed Flames Behind}} a
  {{Bluff}}-{{Body}}},
\newblock \bibinfo{journal}{Flow Turbulence Combust}  (\bibinfo{year}{2021})
  \bibinfo{pages}{1125-- 1141}.
\bibitem[{Nivarti and Cant(2017)}]{Nivarti2017}
\bibinfo{author}{G.~Nivarti}, \bibinfo{author}{S.~Cant},
\newblock \bibinfo{title}{Direct numerical simulation of the bending effect in
  turbulent premixed flames},
\newblock \bibinfo{journal}{Proceedings of the Combustion Institute}
  \bibinfo{volume}{36} (\bibinfo{year}{2017}) \bibinfo{pages}{1903--1910}.
\bibitem[{Desai et~al.(2021)Desai, Kim, Song, Luong, P{\'e}rez, Sankaran, and
  Im}]{desai2021direct}
\bibinfo{author}{S.~Desai}, \bibinfo{author}{Y.~J. Kim},
  \bibinfo{author}{W.~Song}, \bibinfo{author}{M.~B. Luong},
  \bibinfo{author}{F.~E.~H. P{\'e}rez}, \bibinfo{author}{R.~Sankaran},
  \bibinfo{author}{H.~G. Im},
\newblock \bibinfo{title}{Direct numerical simulations of turbulent reacting
  flows with shock waves and stiff chemistry using many-core/gpu acceleration},
\newblock \bibinfo{journal}{Computers \& Fluids} \bibinfo{volume}{215}
  (\bibinfo{year}{2021}) \bibinfo{pages}{104787}.
\bibitem[{Beardsell and Blanquart(2021)}]{BEARDSELL2021}
\bibinfo{author}{G.~Beardsell}, \bibinfo{author}{G.~Blanquart},
\newblock \bibinfo{title}{Fully compressible simulations of the impact of
  acoustic waves on the dynamics of laminar premixed flames for engine-relevant
  conditions},
\newblock \bibinfo{journal}{Proceedings of the Combustion Institute}
  \bibinfo{volume}{38} (\bibinfo{year}{2021}) \bibinfo{pages}{1923--1931}.
\bibitem[{Aditya and Donzis(2012)}]{konduri2012async}
\bibinfo{author}{K.~Aditya}, \bibinfo{author}{D.~A. Donzis},
\newblock \bibinfo{title}{Poster: Asynchronous computing for partial
  differential equations at extreme scales},
\newblock in: \bibinfo{booktitle}{Proceedings of the 2012 SC Companion: High
  Performance Computing, Networking Storage and Analysis}, SCC '12,
  \bibinfo{publisher}{IEEE Computer Society}, \bibinfo{address}{Washington, DC,
  USA}, \bibinfo{year}{2012}, p. \bibinfo{pages}{1444}.
\bibitem[{Donzis and
  Aditya(2014)}]{donzisAsynchronousFinitedifferenceSchemes2014}
\bibinfo{author}{D.~A. Donzis}, \bibinfo{author}{K.~Aditya},
\newblock \bibinfo{title}{Asynchronous finite-difference schemes for partial
  differential equations},
\newblock \bibinfo{journal}{J. Comp. Phys.} \bibinfo{volume}{274}
  (\bibinfo{year}{2014}) \bibinfo{pages}{370--392}.
\bibitem[{Dongarra et~al.(2011)Dongarra, Beckman, Moore, Aerts, Aloisio, Andre,
  Barkai, Berthou, Boku, Braunschweig, Cappello, Chapman, {Xuebin Chi},
  Choudhary, Dosanjh, Dunning, Fiore, Geist, Gropp, Harrison, Hereld, Heroux,
  Hoisie, Hotta, {Zhong Jin}, Ishikawa, Johnson, Kale, Kenway, Keyes, Kramer,
  Labarta, Lichnewsky, Lippert, Lucas, Maccabe, Matsuoka, Messina, Michielse,
  Mohr, Mueller, Nagel, Nakashima, Papka, Reed, Sato, Seidel, Shalf, Skinner,
  Snir, Sterling, Stevens, Streitz, Sugar, Sumimoto, Tang, Taylor, Thakur,
  Trefethen, Valero, {van der Steen}, Vetter, Williams, Wisniewski, and
  Yelick}]{dongarraInternationalExascaleSoftware2011}
\bibinfo{author}{J.~Dongarra}, \bibinfo{author}{P.~Beckman},
  \bibinfo{author}{T.~Moore}, \bibinfo{author}{P.~Aerts},
  \bibinfo{author}{G.~Aloisio}, \bibinfo{author}{J.-C. Andre},
  \bibinfo{author}{D.~Barkai}, \bibinfo{author}{J.-Y. Berthou},
  \bibinfo{author}{T.~Boku}, \bibinfo{author}{B.~Braunschweig},
  \bibinfo{author}{F.~Cappello}, \bibinfo{author}{B.~Chapman},
  \bibinfo{author}{{Xuebin Chi}}, \bibinfo{author}{A.~Choudhary},
  \bibinfo{author}{S.~Dosanjh}, \bibinfo{author}{T.~Dunning},
  \bibinfo{author}{S.~Fiore}, \bibinfo{author}{A.~Geist},
  \bibinfo{author}{B.~Gropp}, \bibinfo{author}{R.~Harrison},
  \bibinfo{author}{M.~Hereld}, \bibinfo{author}{M.~Heroux},
  \bibinfo{author}{A.~Hoisie}, \bibinfo{author}{K.~Hotta},
  \bibinfo{author}{{Zhong Jin}}, \bibinfo{author}{Y.~Ishikawa},
  \bibinfo{author}{F.~Johnson}, \bibinfo{author}{S.~Kale},
  \bibinfo{author}{R.~Kenway}, \bibinfo{author}{D.~Keyes},
  \bibinfo{author}{B.~Kramer}, \bibinfo{author}{J.~Labarta},
  \bibinfo{author}{A.~Lichnewsky}, \bibinfo{author}{T.~Lippert},
  \bibinfo{author}{B.~Lucas}, \bibinfo{author}{B.~Maccabe},
  \bibinfo{author}{S.~Matsuoka}, \bibinfo{author}{P.~Messina},
  \bibinfo{author}{P.~Michielse}, \bibinfo{author}{B.~Mohr},
  \bibinfo{author}{M.~S. Mueller}, \bibinfo{author}{W.~E. Nagel},
  \bibinfo{author}{H.~Nakashima}, \bibinfo{author}{M.~E. Papka},
  \bibinfo{author}{D.~Reed}, \bibinfo{author}{M.~Sato},
  \bibinfo{author}{E.~Seidel}, \bibinfo{author}{J.~Shalf},
  \bibinfo{author}{D.~Skinner}, \bibinfo{author}{M.~Snir},
  \bibinfo{author}{T.~Sterling}, \bibinfo{author}{R.~Stevens},
  \bibinfo{author}{F.~Streitz}, \bibinfo{author}{B.~Sugar},
  \bibinfo{author}{S.~Sumimoto}, \bibinfo{author}{W.~Tang},
  \bibinfo{author}{J.~Taylor}, \bibinfo{author}{R.~Thakur},
  \bibinfo{author}{A.~Trefethen}, \bibinfo{author}{M.~Valero},
  \bibinfo{author}{A.~{van der Steen}}, \bibinfo{author}{J.~Vetter},
  \bibinfo{author}{P.~Williams}, \bibinfo{author}{R.~Wisniewski},
  \bibinfo{author}{K.~Yelick},
\newblock \bibinfo{title}{The {{International Exascale Software Project}}
  roadmap},
\newblock \bibinfo{journal}{The International Journal of High Performance
  Computing Applications} \bibinfo{volume}{25} (\bibinfo{year}{2011})
  \bibinfo{pages}{3--60}.
\bibitem[{Bertsekas and Tsitsiklis(1989)}]{bert1989}
\bibinfo{author}{D.~P. Bertsekas}, \bibinfo{author}{J.~N. Tsitsiklis},
  \bibinfo{title}{Parallel and Distributed Computation: Numerical Methods},
  \bibinfo{publisher}{Prentice-Hall, Inc.}, \bibinfo{address}{USA},
  \bibinfo{year}{1989}.
\bibitem[{Frommer and Szyld(2000)}]{frommerAsynchronousIterations2000}
\bibinfo{author}{A.~Frommer}, \bibinfo{author}{D.~B. Szyld},
\newblock \bibinfo{title}{On asynchronous iterations},
\newblock \bibinfo{journal}{Journal of Computational and Applied Mathematics}
  \bibinfo{volume}{123} (\bibinfo{year}{2000}) \bibinfo{pages}{201--216}.
\bibitem[{Lee et~al.(2016)Lee, Bhattacharya, Dass, Sakuru, and
  Mahapatra}]{leeRelaxedSynchronizationApproach2016}
\bibinfo{author}{K.~Lee}, \bibinfo{author}{R.~Bhattacharya},
  \bibinfo{author}{J.~Dass}, \bibinfo{author}{V.~N. S.~P. Sakuru},
  \bibinfo{author}{R.~N. Mahapatra},
\newblock \bibinfo{title}{A {{Relaxed Synchronization Approach}} for {{Solving
  Parallel Quadratic Programming Problems}} with {{Guaranteed Convergence}}},
\newblock in: \bibinfo{booktitle}{2016 {{IEEE International Parallel}} and
  {{Distributed Processing Symposium}} ({{IPDPS}})}, pp.
  \bibinfo{pages}{182--191}.
\bibitem[{Lee et~al.(2015)Lee, Bhattacharya, and
  Gupta}]{leeSwitchedDynamicalSystem2015}
\bibinfo{author}{K.~Lee}, \bibinfo{author}{R.~Bhattacharya},
  \bibinfo{author}{V.~Gupta},
\newblock \bibinfo{title}{A switched dynamical system framework for analysis of
  massively parallel asynchronous numerical algorithms},
\newblock in: \bibinfo{booktitle}{2015 {{American Control Conference}}
  ({{ACC}})}, pp. \bibinfo{pages}{1095--1100}.
\bibitem[{Amitai et~al.(1992)Amitai, Averbuch, Itzikowitz, and
  Israeli}]{amitai1992}
\bibinfo{author}{D.~Amitai}, \bibinfo{author}{A.~Averbuch},
  \bibinfo{author}{S.~Itzikowitz}, \bibinfo{author}{M.~Israeli},
\newblock \bibinfo{title}{Parallel adaptive and time-stabilizing schemes for
  constant-coefficient parabolic pde's},
\newblock \bibinfo{journal}{Computers \& Mathematics with Applications}
  \bibinfo{volume}{24} (\bibinfo{year}{1992}) \bibinfo{pages}{33 -- 53}.
\bibitem[{Amitai et~al.(1994)Amitai, Averbuch, Itzikowitz, and
  Turkel}]{amitaiAsynchronousCorrectedasynchronousFinite1994}
\bibinfo{author}{D.~Amitai}, \bibinfo{author}{A.~Averbuch},
  \bibinfo{author}{S.~Itzikowitz}, \bibinfo{author}{E.~Turkel},
\newblock \bibinfo{title}{Asynchronous and corrected-asynchronous finite
  difference solutions of {{PDEs}} on {{MIMD}} multiprocessors},
\newblock \bibinfo{journal}{Numer. Algorithms} \bibinfo{volume}{6}
  (\bibinfo{year}{1994}) \bibinfo{pages}{275--296}.
\bibitem[{Mittal and Girimaji(2017)}]{mittalProxyequationParadigmStrategy2017}
\bibinfo{author}{A.~Mittal}, \bibinfo{author}{S.~Girimaji},
\newblock \bibinfo{title}{Proxy-equation paradigm: {{A}} strategy for massively
  parallel asynchronous computations},
\newblock \bibinfo{journal}{Phys. Rev. E} \bibinfo{volume}{96}
  (\bibinfo{year}{2017}) \bibinfo{pages}{033304}.
\bibitem[{Aditya and
  Donzis(2017)}]{adityaHighorderAsynchronytolerantFinite2017}
\bibinfo{author}{K.~Aditya}, \bibinfo{author}{D.~A. Donzis},
\newblock \bibinfo{title}{High-order asynchrony-tolerant finite difference
  schemes for partial differential equations},
\newblock \bibinfo{journal}{Journal of Computational Physics}
  \bibinfo{volume}{350} (\bibinfo{year}{2017}) \bibinfo{pages}{550--572}.
\bibitem[{Mudigere et~al.(2014)Mudigere, Sherlekar, and
  Ansumali}]{mudigereDelayedDifferenceScheme2014}
\bibinfo{author}{D.~Mudigere}, \bibinfo{author}{S.~D. Sherlekar},
  \bibinfo{author}{S.~Ansumali},
\newblock \bibinfo{title}{Delayed {{Difference Scheme}} for {{Large Scale
  Scientific Simulations}}},
\newblock \bibinfo{journal}{Phys. Rev. Lett.} \bibinfo{volume}{113}
  (\bibinfo{year}{2014}) \bibinfo{pages}{218701}.
\bibitem[{Aditya et~al.(2019)Aditya, Gysi, Kwasniewski, Hoefler, Donzis, and
  Chen}]{aditya2019}
\bibinfo{author}{K.~Aditya}, \bibinfo{author}{T.~Gysi},
  \bibinfo{author}{G.~Kwasniewski}, \bibinfo{author}{T.~Hoefler},
  \bibinfo{author}{D.~A. Donzis}, \bibinfo{author}{J.~H. Chen},
  \bibinfo{title}{A scalable weakly-synchronous algorithm for solving partial
  differential equations}, \bibinfo{year}{2019}.
\bibitem[{Kumari and Donzis(2020)}]{kumariDirectNumericalSimulations2020}
\bibinfo{author}{K.~Kumari}, \bibinfo{author}{D.~A. Donzis},
\newblock \bibinfo{title}{Direct numerical simulations of turbulent flows using
  high-order asynchrony-tolerant schemes: {{Accuracy}} and performance},
\newblock \bibinfo{journal}{Journal of Computational Physics}
  \bibinfo{volume}{419} (\bibinfo{year}{2020}) \bibinfo{pages}{109626}.
\bibitem[{Kumari and Donzis(2021)}]{kumariGeneralizedNeumannAnalysis2021}
\bibinfo{author}{K.~Kumari}, \bibinfo{author}{D.~A. Donzis},
\newblock \bibinfo{title}{A generalized von {{Neumann}} analysis for
  multi-level schemes: {{Stability}} and spectral accuracy},
\newblock \bibinfo{journal}{Journal of Computational Physics}
  \bibinfo{volume}{424} (\bibinfo{year}{2021}) \bibinfo{pages}{109868}.
\bibitem[{Hoefler et~al.(2010)Hoefler, Schneider, and
  Lumsdaine}]{hoefler2008noise}
\bibinfo{author}{T.~Hoefler}, \bibinfo{author}{T.~Schneider},
  \bibinfo{author}{A.~Lumsdaine},
\newblock \bibinfo{title}{{Characterizing the Influence of System Noise on
  Large-Scale Applications by Simulation}},
\newblock in: \bibinfo{booktitle}{International Conference for High Performance
  Computing, Networking, Storage and Analysis (SC'10)}.
\bibitem[{Shu(1998)}]{shuEssentiallyNonoscillatoryWeighted1998}
\bibinfo{author}{C.-W. Shu},
\newblock \bibinfo{title}{Essentially non-oscillatory and weighted essentially
  non-oscillatory schemes for hyperbolic conservation laws},
\newblock in: \bibinfo{editor}{A.~Quarteroni} (Ed.),
  \bibinfo{booktitle}{Advanced {{Numerical Approximation}} of {{Nonlinear
  Hyperbolic Equations}}: {{Lectures}} given at the 2nd {{Session}} of the
  {{Centro Internazionale Matematico Estivo}} ({{C}}.{{I}}.{{M}}.{{E}}.) Held
  in {{Cetraro}}, {{Italy}}, {{June}} 23\textendash 28, 1997},
  \bibinfo{publisher}{{Springer Berlin Heidelberg}}, \bibinfo{address}{{Berlin,
  Heidelberg}}, \bibinfo{year}{1998}, pp. \bibinfo{pages}{325--432}.
\bibitem[{Shu(2009)}]{shuHighOrderWeighted2009}
\bibinfo{author}{C.-W. Shu},
\newblock \bibinfo{title}{High {{Order Weighted Essentially Nonoscillatory
  Schemes}} for {{Convection Dominated Problems}}},
\newblock \bibinfo{journal}{SIAM Rev.} \bibinfo{volume}{51}
  (\bibinfo{year}{2009}) \bibinfo{pages}{82--126}.
\bibitem[{Pirozzoli(2011)}]{pirozzoli2011numerical}
\bibinfo{author}{S.~Pirozzoli},
\newblock \bibinfo{title}{Numerical methods for high-speed flows},
\newblock \bibinfo{journal}{Annual review of fluid mechanics}
  \bibinfo{volume}{43} (\bibinfo{year}{2011}) \bibinfo{pages}{163--194}.
\bibitem[{{Bermejo-Moreno} et~al.(2013){Bermejo-Moreno}, {Bodart}, {Larsson},
  {Barney}, {Nichols}, and {Jones}}]{bermejo2013scaling}
\bibinfo{author}{I.~{Bermejo-Moreno}}, \bibinfo{author}{J.~{Bodart}},
  \bibinfo{author}{J.~{Larsson}}, \bibinfo{author}{B.~M. {Barney}},
  \bibinfo{author}{J.~W. {Nichols}}, \bibinfo{author}{S.~{Jones}},
\newblock \bibinfo{title}{Solving the compressible navier-stokes equations on
  up to 1.97 million cores and 4.1 trillion grid points},
\newblock in: \bibinfo{booktitle}{SC '13: Proceedings of the International
  Conference on High Performance Computing, Networking, Storage and Analysis},
  pp. \bibinfo{pages}{1--10}.
\bibitem[{Mosedale and Drikakis(2007)}]{mosedale2007assessment}
\bibinfo{author}{A.~Mosedale}, \bibinfo{author}{D.~Drikakis},
\newblock \bibinfo{title}{{Assessment of Very High Order of Accuracy in
  Implicit LES models}},
\newblock \bibinfo{journal}{Journal of Fluids Engineering}
  \bibinfo{volume}{129} (\bibinfo{year}{2007}) \bibinfo{pages}{1497--1503}.
\bibitem[{Ritos et~al.(2018{\natexlab{a}})Ritos, Kokkinakis, and
  Drikakis}]{ritos2018physical}
\bibinfo{author}{K.~Ritos}, \bibinfo{author}{I.~W. Kokkinakis},
  \bibinfo{author}{D.~Drikakis},
\newblock \bibinfo{title}{Physical insight into the accuracy of finely-resolved
  {iLES} in turbulent boundary layers},
\newblock \bibinfo{journal}{Computers \& Fluids} \bibinfo{volume}{169}
  (\bibinfo{year}{2018}{\natexlab{a}}) \bibinfo{pages}{309--316}.
\bibitem[{Ritos et~al.(2018{\natexlab{b}})Ritos, Kokkinakis, and
  Drikakis}]{ritos2018performance}
\bibinfo{author}{K.~Ritos}, \bibinfo{author}{I.~W. Kokkinakis},
  \bibinfo{author}{D.~Drikakis},
\newblock \bibinfo{title}{Performance of high-order implicit large eddy
  simulations},
\newblock \bibinfo{journal}{Computers \& Fluids} \bibinfo{volume}{173}
  (\bibinfo{year}{2018}{\natexlab{b}}) \bibinfo{pages}{307--312}.
\bibitem[{Shu(2020)}]{shuEssentiallyNonoscillatoryWeighted}
\bibinfo{author}{C.-W. Shu},
\newblock \bibinfo{title}{Essentially non-oscillatory and weighted essentially
  non-oscillatory schemes},
\newblock \bibinfo{journal}{Acta Numerica}  (\bibinfo{year}{2020})
  \bibinfo{pages}{63}.
\bibitem[{Jiang and Shu(1996)}]{jiangEfficientImplementationWeighted}
\bibinfo{author}{G.-S. Jiang}, \bibinfo{author}{C.-W. Shu},
\newblock \bibinfo{title}{Efficient implementation of weighted eno schemes},
\newblock \bibinfo{journal}{Journal of Computational Physics}
  \bibinfo{volume}{126} (\bibinfo{year}{1996}) \bibinfo{pages}{202--228}.
\bibitem[{Shi et~al.(2002)Shi, Hu, and Shu}]{shiTechniqueTreatingNegative2002}
\bibinfo{author}{J.~Shi}, \bibinfo{author}{C.~Hu}, \bibinfo{author}{C.-W. Shu},
\newblock \bibinfo{title}{A {{Technique}} of {{Treating Negative Weights}} in
  {{WENO Schemes}}},
\newblock \bibinfo{journal}{Journal of Computational Physics}
  \bibinfo{volume}{175} (\bibinfo{year}{2002}) \bibinfo{pages}{108--127}.
\bibitem[{Liu et~al.(2009)Liu, Shu, and Zhang}]{liu2009positivity}
\bibinfo{author}{Y.-y. Liu}, \bibinfo{author}{C.-w. Shu},
  \bibinfo{author}{M.-p. Zhang},
\newblock \bibinfo{title}{On the positivity of linear weights in weno
  approximations},
\newblock \bibinfo{journal}{Acta Mathematicae Applicatae Sinica, English
  Series} \bibinfo{volume}{25} (\bibinfo{year}{2009})
  \bibinfo{pages}{503--538}.
\bibitem[{Kee et~al.(1996)Kee, Rupley, Meeks, and Miller}]{kee1996chemkin}
\bibinfo{author}{R.~J. Kee}, \bibinfo{author}{F.~M. Rupley},
  \bibinfo{author}{E.~Meeks}, \bibinfo{author}{J.~A. Miller},
  \bibinfo{title}{CHEMKIN-III: A FORTRAN chemical kinetics package for the
  analysis of gas-phase chemical and plasma kinetics}, \bibinfo{type}{Technical
  Report}, Sandia National Labs., Livermore, CA (United States),
  \bibinfo{year}{1996}.
\bibitem[{Kee et~al.(1986)Kee, Dixon-Lewis, Warnatz, Coltrin, and
  Miller}]{kee1986fortran}
\bibinfo{author}{R.~J. Kee}, \bibinfo{author}{G.~Dixon-Lewis},
  \bibinfo{author}{J.~Warnatz}, \bibinfo{author}{M.~E. Coltrin},
  \bibinfo{author}{J.~A. Miller},
\newblock \bibinfo{title}{A fortran computer code package for the evaluation of
  gas-phase multicomponent transport properties},
\newblock \bibinfo{journal}{Sandia National Laboratories Report SAND86-8246}
  \bibinfo{volume}{13} (\bibinfo{year}{1986}) \bibinfo{pages}{80401--1887}.
\bibitem[{Baum et~al.(1995)Baum, Poinsot, and Thévenin}]{baum1993}
\bibinfo{author}{M.~Baum}, \bibinfo{author}{T.~Poinsot},
  \bibinfo{author}{D.~Thévenin},
\newblock \bibinfo{title}{Accurate boundary conditions for multicomponent
  reactive flows},
\newblock \bibinfo{journal}{Journal of Computational Physics}
  \bibinfo{volume}{116} (\bibinfo{year}{1995}) \bibinfo{pages}{247 -- 261}.
\bibitem[{Thompson(1987)}]{thompsonTimeDependentBoundary1987}
\bibinfo{author}{K.~W. Thompson},
\newblock \bibinfo{title}{Time dependent boundary conditions for hyperbolic
  systems},
\newblock \bibinfo{journal}{Journal of Computational Physics}
  \bibinfo{volume}{68} (\bibinfo{year}{1987}) \bibinfo{pages}{1--24}.
\bibitem[{Poinsot and Lele(1992)}]{poinsotBoundaryConditionsDirect}
\bibinfo{author}{T.~J. Poinsot}, \bibinfo{author}{S.~K. Lele},
\newblock \bibinfo{title}{Boundary conditions for direct simulations of
  compressible viscous flows},
\newblock \bibinfo{journal}{Journal of Computational Physics}
  \bibinfo{volume}{101} (\bibinfo{year}{1992}) \bibinfo{pages}{104--129}.
\bibitem[{Luo et~al.(2012)Luo, Yoo, Richardson, Chen, Law, and
  Lu}]{luo2012chemical}
\bibinfo{author}{Z.~Luo}, \bibinfo{author}{C.~S. Yoo}, \bibinfo{author}{E.~S.
  Richardson}, \bibinfo{author}{J.~H. Chen}, \bibinfo{author}{C.~K. Law},
  \bibinfo{author}{T.~Lu},
\newblock \bibinfo{title}{Chemical explosive mode analysis for a turbulent
  lifted ethylene jet flame in highly-heated coflow},
\newblock \bibinfo{journal}{Combustion and Flame} \bibinfo{volume}{159}
  (\bibinfo{year}{2012}) \bibinfo{pages}{265--274}.
\bibitem[{Yoo et~al.(2005)Yoo, Wang, Trouv{\'e}, and
  Im}]{yooCharacteristicBoundaryConditions2005}
\bibinfo{author}{C.~S. Yoo}, \bibinfo{author}{Y.~Wang},
  \bibinfo{author}{A.~Trouv{\'e}}, \bibinfo{author}{H.~G. Im},
\newblock \bibinfo{title}{Characteristic boundary conditions for direct
  simulations of turbulent counterflow flames},
\newblock \bibinfo{journal}{Combustion Theory and Modelling}
  \bibinfo{volume}{9} (\bibinfo{year}{2005}) \bibinfo{pages}{617--646}.
\bibitem[{Yoo and Im(2007)}]{yooCharacteristicBoundaryConditions2007}
\bibinfo{author}{C.~S. Yoo}, \bibinfo{author}{H.~G. Im},
\newblock \bibinfo{title}{Characteristic boundary conditions for simulations of
  compressible reacting flows with multi-dimensional, viscous and reaction
  effects},
\newblock \bibinfo{journal}{Combustion Theory and Modelling}
  \bibinfo{volume}{11} (\bibinfo{year}{2007}) \bibinfo{pages}{259--286}.
\bibitem[{Burke et~al.(2012)Burke, Chaos, Ju, Dryer, and
  Klippenstein}]{burke2012comprehensive}
\bibinfo{author}{M.~P. Burke}, \bibinfo{author}{M.~Chaos},
  \bibinfo{author}{Y.~Ju}, \bibinfo{author}{F.~L. Dryer},
  \bibinfo{author}{S.~J. Klippenstein},
\newblock \bibinfo{title}{Comprehensive h2/o2 kinetic model for high-pressure
  combustion},
\newblock \bibinfo{journal}{International Journal of Chemical Kinetics}
  \bibinfo{volume}{44} (\bibinfo{year}{2012}) \bibinfo{pages}{444--474}.
\bibitem[{Sutherland and
  Kennedy(2003)}]{sutherlandImprovedBoundaryConditions2003}
\bibinfo{author}{J.~C. Sutherland}, \bibinfo{author}{C.~A. Kennedy},
\newblock \bibinfo{title}{Improved boundary conditions for viscous, reacting,
  compressible flows},
\newblock \bibinfo{journal}{Journal of Computational Physics}
  \bibinfo{volume}{191} (\bibinfo{year}{2003}) \bibinfo{pages}{502--524}.
\bibitem[{Westbrook(1982)}]{westbrook1982}
\bibinfo{author}{C.~K. Westbrook},
\newblock \bibinfo{title}{Chemical kinetics of hydrocarbon oxidation in gaseous
  detonations},
\newblock \bibinfo{journal}{Combustion and Flame} \bibinfo{volume}{46}
  (\bibinfo{year}{1982}) \bibinfo{pages}{191 -- 210}.
\bibitem[{Deiterding(2003)}]{deiterdingParallelAdaptiveSimulation2003}
\bibinfo{author}{R.~Deiterding},
\newblock \bibinfo{title}{Parallel adaptive simulation of multi-dimensional
  detonation structures}  (\bibinfo{year}{2003}).
\bibitem[{Lv and Ihme(2014)}]{lvDiscontinuousGalerkinMethod2014}
\bibinfo{author}{Y.~Lv}, \bibinfo{author}{M.~Ihme},
\newblock \bibinfo{title}{Discontinuous {{Galerkin}} method for multicomponent
  chemically reacting flows and combustion},
\newblock \bibinfo{journal}{Journal of Computational Physics}
  \bibinfo{volume}{270} (\bibinfo{year}{2014}) \bibinfo{pages}{105--137}.
\bibitem[{Houim and Kuo(2011)}]{houimLowdissipationTimeaccurateMethod2011}
\bibinfo{author}{R.~W. Houim}, \bibinfo{author}{K.~K. Kuo},
\newblock \bibinfo{title}{A low-dissipation and time-accurate method for
  compressible multi-component flow with variable specific heat ratios},
\newblock \bibinfo{journal}{Journal of Computational Physics}
  \bibinfo{volume}{230} (\bibinfo{year}{2011}) \bibinfo{pages}{8527--8553}.
\bibitem[{Luong et~al.(2020{\natexlab{a}})Luong, Desai, P{\'e}rez, Sankaran,
  Johansson, and Im}]{luong2020statistical}
\bibinfo{author}{M.~B. Luong}, \bibinfo{author}{S.~Desai},
  \bibinfo{author}{F.~E.~H. P{\'e}rez}, \bibinfo{author}{R.~Sankaran},
  \bibinfo{author}{B.~Johansson}, \bibinfo{author}{H.~G. Im},
\newblock \bibinfo{title}{A statistical analysis of developing knock intensity
  in a mixture with temperature inhomogeneities},
\newblock \bibinfo{journal}{Proceedings of the Combustion Institute}
  (\bibinfo{year}{2020}{\natexlab{a}}).
\bibitem[{Luong et~al.(2020{\natexlab{b}})Luong, Desai,
  Hern{\'a}ndez~P{\'e}rez, Sankaran, Johansson, and Im}]{luong2020effects}
\bibinfo{author}{M.~B. Luong}, \bibinfo{author}{S.~Desai},
  \bibinfo{author}{F.~E. Hern{\'a}ndez~P{\'e}rez},
  \bibinfo{author}{R.~Sankaran}, \bibinfo{author}{B.~Johansson},
  \bibinfo{author}{H.~G. Im},
\newblock \bibinfo{title}{Effects of turbulence and temperature fluctuations on
  knock development in an ethanol/air mixture},
\newblock \bibinfo{journal}{Flow, Turbulence and Combustion}
  (\bibinfo{year}{2020}{\natexlab{b}}) \bibinfo{pages}{1--21}.
\bibitem[{Lodato and Pitsch(2011)}]{lodatoCharacteristicOutflowsOptimal}
\bibinfo{author}{G.~Lodato}, \bibinfo{author}{H.~Pitsch},
\newblock \bibinfo{title}{Characteristic outflows with optimal transverse
  terms: The edges and corners coupling algorithm},
\newblock \bibinfo{journal}{Center for Turbulence Research, Annual Research
  Briefs}  (\bibinfo{year}{2011}) \bibinfo{pages}{12}.
\bibitem[{Yeung and Ravikumar(2020)}]{PK2020}
\bibinfo{author}{P.~K. Yeung}, \bibinfo{author}{K.~Ravikumar},
\newblock \bibinfo{title}{Advancing understanding of turbulence through
  extreme-scale computation: Intermittency and simulations at large problem
  sizes},
\newblock \bibinfo{journal}{Phys. Rev. Fluids} \bibinfo{volume}{5}
  (\bibinfo{year}{2020}) \bibinfo{pages}{110517}.
\bibitem[{Slaughter et~al.(2015)Slaughter, Lee, Treichler, Bauer, and
  Aiken}]{regent2015}
\bibinfo{author}{E.~Slaughter}, \bibinfo{author}{W.~Lee},
  \bibinfo{author}{S.~Treichler}, \bibinfo{author}{M.~Bauer},
  \bibinfo{author}{A.~Aiken},
\newblock \bibinfo{title}{Regent: A high-productivity programming language for
  hpc with logical regions},
\newblock in: \bibinfo{booktitle}{Proceedings of the International Conference
  for High Performance Computing, Networking, Storage and Analysis}, SC '15,
  \bibinfo{publisher}{Association for Computing Machinery},
  \bibinfo{address}{New York, NY, USA}, \bibinfo{year}{2015}.
\bibitem[{Acun et~al.(2014)Acun, Gupta, Jain, Langer, Menon, Mikida, Ni,
  Robson, Sun, Totoni, Wesolowski, and Kale}]{charm2014}
\bibinfo{author}{B.~Acun}, \bibinfo{author}{A.~Gupta},
  \bibinfo{author}{N.~Jain}, \bibinfo{author}{A.~Langer},
  \bibinfo{author}{H.~Menon}, \bibinfo{author}{E.~Mikida},
  \bibinfo{author}{X.~Ni}, \bibinfo{author}{M.~Robson},
  \bibinfo{author}{Y.~Sun}, \bibinfo{author}{E.~Totoni},
  \bibinfo{author}{L.~Wesolowski}, \bibinfo{author}{L.~Kale},
\newblock \bibinfo{title}{Parallel programming with migratable objects: Charm++
  in practice},
\newblock in: \bibinfo{booktitle}{SC '14: Proceedings of the International
  Conference for High Performance Computing, Networking, Storage and Analysis},
  pp. \bibinfo{pages}{647--658}.
\bibitem[{Kulkarni and Lumsdaine(2019)}]{charm2019}
\bibinfo{author}{A.~Kulkarni}, \bibinfo{author}{A.~Lumsdaine},
\newblock \bibinfo{title}{A comparative study of asynchronous many-tasking
  runtimes: Cilk, charm++, parallex and am++},
\newblock \bibinfo{journal}{arXiv preprint arXiv:1904.00518}
  (\bibinfo{year}{2019}).
\bibitem[{Kolla et~al.(2020)Kolla, Mayo, Teranishi, and
  Armstrong}]{kollaImprovingScalabilitySilentError}
\bibinfo{author}{H.~Kolla}, \bibinfo{author}{J.~R. Mayo},
  \bibinfo{author}{K.~Teranishi}, \bibinfo{author}{R.~C. Armstrong},
\newblock \bibinfo{title}{Improving scalability of silent-error resilience for
  message-passing solvers via local recovery and asynchrony},
\newblock \bibinfo{journal}{2020 IEEE/ACM 10th Workshop on Fault Tolerance for
  HPC at eXtreme Scale (FTXS)}  (\bibinfo{year}{2020}) \bibinfo{pages}{1--10}.

\end{thebibliography}
\end{document}